\documentclass[pre,aps]{revtex4}
\usepackage{epsfig}
\usepackage{amstext}
\usepackage{amsbsy}
\usepackage{amsmath}
\usepackage{algorithm}
\usepackage{algorithmic} 

\begin{document}

\title{Adaptive cluster expansion for the inverse Ising problem:\\ convergence, algorithm and tests}
\author{S. Cocco$^{1,2}$, R. Monasson$^{1,3}$}
\affiliation{
$^1$ Simons Center for Systems Biology, Institute for Advanced Study, Einstein Drive, Princeton, NJ 08540\\
$^2$ Laboratoire de Physique Statistique de l'ENS, CNRS \& UPMC, 24 rue Lhomond, 75005 Paris, France\\
$^3$ Laboratoire de Physique Th\'eorique de l'ENS, CNRS \& UPMC, 24 rue Lhomond,
75005 Paris, France}

\begin{abstract}
We present a procedure to solve the inverse Ising problem, that is
to find  the interactions between a set of binary variables from the measure of their equilibrium correlations. The method consists in constructing and selecting specific clusters of variables, based on their contributions to the cross-entropy of the Ising model. Small contributions are discarded to avoid overfitting and to make the computation tractable. The properties of the cluster expansion and its performances on synthetic data are studied. To make the implementation easier we give the pseudo-code of the algorithm.    
\end{abstract}
\maketitle

\section{Introduction}

The Ising model is a paradigm of statistical physics, and has been extensively studied to understand the equilibrium properties and the nature of the phase transitions in various systems in condensed matter \cite{brush67}. In its usual formulation, the Ising model is defined over a set of $N$ binary variables $\sigma_i$, with $i=1,2,\ldots , N$. The variables, called spins, are submitted to a set of $N$ local fields, $h_i$, and of $\frac 12 N(N-1)$ pairwise couplings, $J_{ij}$. The observables of the model, such as the average values of the spins or of the spin-spin correlations over the Gibbs measure, 
\begin{equation}\label{mc0}
\langle \sigma _i \rangle \ , \langle \sigma_k\sigma _l\rangle \ ,
\end{equation}
are well-defined and can be calculated from the knowledge of those interaction parameters. We will refer to the task of calculating (\ref{mc0}) given the interaction parameters as to the direct Ising problem.

In many experimental cases, the interaction parameters are unknown, while the values of observables can be estimated from measurements. A natural question is to  know if and how the interaction parameters can be deduced from the data 
(\cite{bialek,marre,Pey09,Weigt,Bal10,cavagna}).  When the coupling matrix is known {\em a priori} to have a specific and simple structure, this question can be answered with an ordinary fit. For instance, in a two-dimensional and uniform ferromagnet, all couplings vanish but between neighbors on the lattice, and $J_{ij}=J$ for contiguous sites $i$ and $j$. In such a case, the observable such as the average correlation between neighboring spins, $c$, depends on a single parameter, $J$. The measurement of $c$ gives a direct access to a value of $J$. However, data coming from complex systems arising in biology, sociology, finance, ... can generally not be interpreted with such a simple Ising model, and the fit procedure is much more complicated for two reasons. First, in the absence of any {\em prior} knowledge about the interaction network, the number of interaction parameters $J_{ij}$ to be inferred scales quadratically with the system size $N$, and can be very large. Secondly, the quality of the data is a crucial issue. Experimental data are plagued by noise, coming either from the measurement apparatus or from imperfect sampling. The task of fitting a very large number of interaction parameters from 'noisy' data has received much attention in the statistics community, under the name of high-dimensional inference \cite{tibshirani}. 

To be more specific, the inverse Ising problem is defined as follows. Assume that a set of $B$ configurations $\boldsymbol\sigma ^\tau=\{\sigma_1^\tau, \sigma_2^\tau , \ldots , \sigma_N^\tau\}$, with $\tau=1,2,\ldots , B$ are available from measurements. We compute the empirical 1- and 2-point averages through
\begin{equation}
p_i = \frac 1B \sum _{\tau =1}^B \sigma_i^\tau \ , \quad
p_{kl} = \frac 1B \sum _{\tau =1}^B \sigma_k^\tau \, \sigma_l ^\tau \ .
\label{mc}
\end{equation}
The inverse Ising problem consists in finding the values of the $N$ local fields, $h_i$, and of the $\frac 12N(N-1)$ interactions, $J_{ij}$, such that the individual and pairwise frequencies of the spins (\ref{mc0}) defined from the Gibbs measure coincide with their empirical counterparts, $p_i$ and $p_{kl}$. While the Gibbs measure corresponding to the Ising model is by no means the unique measure allowing one to reproduce the data $p_i$ and $p_{kl}$, it is the distribution with the largest entropy doing so \cite{maxent}. In other words, the Ising model is the least constrained model capable of matching the empirical values of the 1- and 2-point observables. This property explains the recent surge of interest in defining and solving the inverse Ising problem in the context of the analysis of biological, {\em e.g.} neurobiological \cite{bialek,marre,Pey09,noi} and proteomic \cite{Weigt,Bal10} data. 

As a result of its generality, the inverse Ising problem has been studied in various fields under different names, such as Boltzmann machine learning in learning theory \cite{opper,ackley} or graphical model selection in statistical inference \cite{tibshirani, wain, bento}. While the research field is currently very active, the diversity of the tools and, sometimes, of the goals make somewhat difficult to compare the results obtained across the disciplines. Several variants of the inverse Ising problem can be defined:

\begin{itemize}
\item {A:} find the interaction network from a set of spin configurations
$\boldsymbol\sigma^\tau$. It is generally assumed in the graphical model community that the Ising model is exact, that is, that the underlying distribution of the data is truly an Ising model with unknown interaction parameters $\bf J$. The question is to find which interactions $J_{ij}$ are non zero (or larger than some  $J_{min}$ is absolute value), and how many configurations (value of $B$) should be sampled to achieve this goal with acceptable probability. 
\item {B:} find the interactions $J_{ij}$ and the fields $h_i$ from the frequencies $p_i,p_{ij}$ only. Those frequencies should be reproduced within a prescribed accuracy, $\epsilon$, not too small (compared to the error on the data) to avoid overfitting. Note that in general the Ising model is not the true underlying model for the data here; it is only the model with maximal entropy given the constraints on 1- and 2-point correlations.
\item{C:} same as B, but in addition we want to know the entropy (at fixed
individual and pairwise frequencies), which measures how many configurations $\boldsymbol\sigma$ really contribute to the Gibbs distribution of the Ising model. Computing the entropy is generally intractable for the direct Ising problem, unless correlations decay fast enough \cite{Sinclair}. 
\end{itemize}
Variants B and C are harder than A: full spin configurations give access to all $K$-spin correlations, a knowledge which can be used to design fast network structure inference algorithm. Recently, a procedure to solve problem C was proposed, based on ideas and techniques coming from statistical physics \cite{coc11}. The purpose of the present paper is to discuss its performances and limitations.

It is essential to be aware of the presence of noise in the data, {\em e.g.} due to the imperfect sampling (finite number $B$ of configurations). A potential risk is overfitting: the network of interactions we find at the end of the inference process could reproduce the mere noisy data, rather than the 'true' interactions. How can one disentangle noise from signal in the data? A popular approach in the statistics community is to require that the inferred interaction network be sparse. The rationale for imposing sparsity is two-fold. First, physical lattices are very sparse, and connect only close sites in the space; it is possible but not at all obvious that networks modeling other {\em e.g.} biological data enjoy a similar property. Secondly, an Ising model with a sparse interaction network reproducing a set of correlations is a sparing representation of the statistics of the data, and, in much the same spirit as the minimal message length approach \cite{wallace}, should be preferred to models with denser networks. The appeal of the approach is largely due to the fact that imposing sparsity is computationally tractable.

\begin{figure}
\begin{center}
\epsfig{file=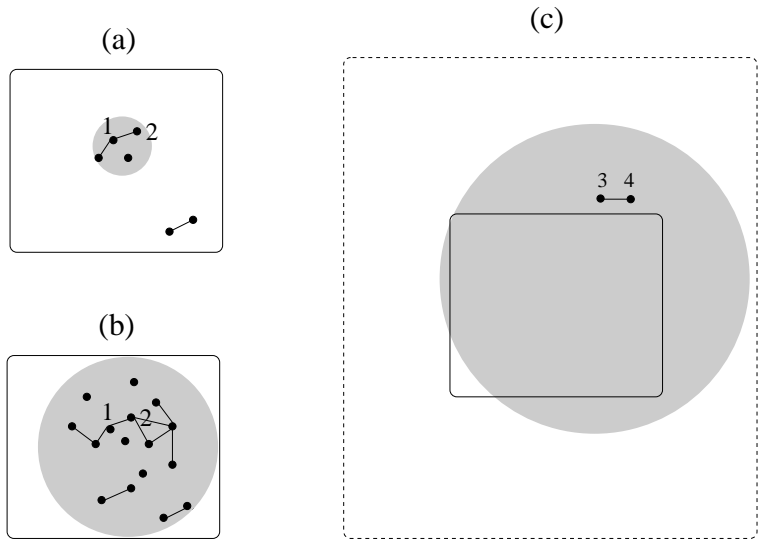,width=10cm}
\caption{Schematic representation of a well-conditioned {\bf (a)} and an ill-conditioned {\bf (b)} inverse Ising problems. The gray areas symbolize the set of spins (full dots)  whose interactions (links) and fields are affected by a change of the frequency $p_{12}$ of spins 1 and 2. The response is localized in {\bf (a)} and extended in {\bf (b)}. Experiments usually measure a restricted part of the system (dashed contour) only {\bf (c)}. Increasing the size of the measured sub-system, {\em e.g.} by including the frequencies of the extra-variables 3 and 4, will modify most of the inferred interaction parameters if the problem is ill-conditioned.}
\label{fig-concept}
\end{center}
\end{figure}

The criterion required by our procedure is not that the interaction network should be  sparse, but that the inverse Ising problem should be well-conditioned. To illustrate this notion, consider a set of data, {\em i.e.} of frequencies $p_i,p_{kl}$, and assume one has found the solution $h_i,J_{kl}$ to the corresponding inverse Ising problem. Let us now slightly modify one or a few frequencies, say, $p_{12} \to p'_{12}= p_{12} +\delta p_{12}$, and solve again the corresponding inverse Ising problem, with the results $h'_i,J'_{kl}$. Let $\delta J_{kl}=J'_{kl}-J_{kl}$ and $\delta h_i=h'_i-h_i$ measure the response of the interaction parameters to the small modification of $p_{12}$ alone. Two extreme cases are: 
\begin{itemize}
\item {\em Localized response:} the response is restricted to the parameters involving spins 1 and 2 only, {\em i.e.} $\delta h_1,\delta h_2,\delta J_{12}\ne 0$; it vanishes for all the other parameters.
\item {\em Extended response:} the response spreads all over the spin system, and all the quantities $\delta h_i,\delta J_{kl}$ are non-zero. 
\end{itemize}
Intermediate cases will generically be encountered, and are symbolized in Fig.~\ref{fig-concept}(a)\&(b). For instance, if the response is non-zero over a small number of parameters only, which define a 'neighborhood' of the spins $1,2$, we will consider it is localized. Obviously, the notion of 'smallness' cannot be rigorously defined here, unless the system size $N$ can be made arbitrarily large and sent to infinity.

Drawing our inspiration from the vocabulary of numerical analysis, we will say that the inverse Ising problem is well-conditioned if the response is localized. For a well-conditioned problem, a small change of one or a few variables essentially affects one or a few interaction parameters. On the contrary, most if not all interaction parameters of a ill-conditioned inverse Ising problem are affected by an elementary modification of the data. This notion must be distinguished from the concept of ill-posed problem. As we will see in Section \ref{secstatement}, the inverse Ising problem is always well-posed, once an appropriate regularization is introduced: given the frequencies, there exists a unique set of interaction parameters reproducing those data, regardless of how hard it is to compute.

Not all inverse Ising problems are well-conditioned. However, it is our opinion that only those ones should be solved. The reason is that, in generic experimental situations, only a (small) region of the system is accessible. Solving the inverse problem attached to this sub-system makes sense only if the problem is well-conditioned. If it is ill-conditioned, extending even by a bit the sub-system would considerably affect the values of most of the inferred parameters (Fig.~\ref{fig-concept}(c)). Hence, the interaction parameters would be very much dependent on the part of the system which is not measured! Such a possibility simply means that the inverse problem, though mathematically well-posed, is not meaningful.

Interestingly, the response of the interactions to a change of a few correlations can be localized, while the response of the correlations to a change of a few interactions is extended. An example is given by 'critical' Ising models, where correlations extend over the whole system. However, the corresponding inverse Ising problem may be well-conditioned.
 
The presence of noise in the data considerably affects the status of the inverse Ising problem. As we will see later, even well-conditioned problems in the limit of perfect sampling ($B\to\infty$) become ill-conditioned as soon as sampling is imperfect (finite $B$). The same statement holds for the sparsity-based criterion mentioned above: when data are generated by a sparse interaction network, the solution to the inverse Ising model is not sparse as a consequence of imperfect sampling. Only the presence of an explicit and additional regularization forces the solution to be sparse. In much the same way, the procedure we present hereafter builds a well-conditioned inverse Ising problem, which prevents overfitting of the noise. This procedure is based on the expansion of the entropy at fixed frequencies in clusters of spins, a notion closely related to the neighborhoods appearing in the localized responses.

The plan of the article is as follows. In Section \ref{secstatement} we give the notations and precise definitions of the inverse Ising problem, and briefly review some  of the resolution procedures in the literature. In Section \ref{secclusterprincipal}, we explain how the entropy can be expanded as a sum of contributions, one for each cluster (or sub-set) of spins, and review the properties of those entropic contributions. The procedure to truncate the expansion and keep only relevant clusters is discussed in Section \ref{sect99}. The pseudo-codes and details necessary for the implementation of the algorithm can be found in Section \ref{sect98}. Applications to artificial data are discussed at length in Section \ref{secappli}. Finally, Section \ref{sec-conclusion} presents some perspectives and conclusions. To improve the readability of the paper most technical details have been relegated to technical appendices.
 
\section{The Inverse Ising Problem: formulations and issues}
\label{secstatement}

\subsection{Maximum Entropy Principle formulation}

We consider a system of $N$ binary variables, $\sigma_i=0,1$, where $i=1,2,\ldots ,N$. The average values of the variables, $p_i$, and of their correlations, $p_{kl}$, are measured, for instance through the empirical average over $B$ sampled configurations of the system, see equations (\ref{mc}). As the correlations $p_{kl}$ are obtained from the empirical measure, the problem is realizable \cite{kuna07}. Let ${\bf p}=\{p_i, p_{kl}\}$ denote the data. The Maximum Entropy Principle (MEP) \cite{maxent} postulates that the probabilistic model $P ({\boldsymbol\sigma })$ should maximize the entropy $S$ of the distribution $P$ under the constraints
\begin{equation}
\sum _{\boldsymbol\sigma} P(\boldsymbol\sigma) =1\ , \quad
\sum _{\boldsymbol\sigma} P(\boldsymbol\sigma) \; \sigma_i=p_i\ , \quad
\sum _{\boldsymbol\sigma} P(\boldsymbol\sigma) \; \sigma_k\, \sigma_l=p_{kl}\ .
\end{equation}
In practice these constraints are enforced by the Lagrange multipliers $\lambda$ and ${\bf J}=\{ h_i,J_{kl}\}$. The maximal entropy is~\footnote{We have to minimize here rather than maximize since the true Lagrange multipliers take imaginary values, the couplings and fields being their imaginary part. }
\begin{eqnarray}
\label{maxentro}
S ({\bf p})= \min_{\lambda ,{\bf J}} \;  \max_{P({\boldsymbol\sigma}) }
\bigg[  & -&\sum_{\bf \boldsymbol\sigma} P({\boldsymbol \sigma}) \log { P({\boldsymbol\sigma })} + \lambda\left(\sum_{\boldsymbol\sigma} P({\boldsymbol\sigma})-1\right)\
\nonumber \\ &+& \left. \sum_i  h_i
\left ( \sum_{\boldsymbol \sigma} P({\boldsymbol \sigma})\,\sigma_i -p_i\right )+ \sum_{k<l} J_{kl} \left (\sum_{\boldsymbol\sigma} P({\boldsymbol\sigma})\,\sigma_k\,\sigma_l -  p_{kl}
\right)  \right] \ .
\end{eqnarray}
The maximization condition over $P$ shows that the MEP probability corresponds to the Gibbs measure $P_{\bf J}$ of the celebrated Ising model,
\begin{equation}
P_{\bf J}[{\boldsymbol\sigma}]=\frac {e^{-H_{Ising}[{\boldsymbol\sigma}|{\bf J}]}}{Z[{\bf J}]} \label{pisingvrai}
 \end{equation}
where the energy function is
\begin{equation}
H_{Ising}[{\boldsymbol\sigma}|{\bf J}] = -\sum_i\; h_i\,\sigma_i -\sum_{k<l}\; J_{kl}\, \sigma_k\,\sigma_l
\label{h2}
\end{equation}
and $Z[{\bf J}]= \sum_{\boldsymbol\sigma} \exp( -H_{Ising}[{\boldsymbol\sigma} | {\bf J}])$ denotes the partition function. The values of the couplings and fields \footnote{The vocable 'field', should strictly speaking, be used when the variables $\sigma_i$ are spins taking $\pm 1$ values. For $0,1$ variable, the use of the denomination 'chemical potential' would be more appropriate. We keep to the simpler  denomination 'field' hereafter.}  are then found through the minimization of
\begin{equation} \label{s2}
S _{Ising}[{\bf J}| {\bf p}] = \log Z[{\bf J}]-\sum_i h_i\; p_i-\sum_{k<l} J_{kl} \; p_{kl} \ .
\end{equation}
over $\bf J$. The minimal value of $S_{Ising}$ coincides with $S$ defined in (\ref{maxentro}).

The cross-entropy $S_{Ising}$ has a simple interpretation in terms of the Kullback-Leibler divergence between the Ising distribution $P_{\bf J}[{\boldsymbol\sigma}]$ and the empirical measure over the observed configurations, $P_{\text{obs}} [\boldsymbol\sigma ]$. Assume $B$ configurations of the $N$ variables, $\boldsymbol\sigma ^\tau$, with $\tau =1,2,\ldots , B$, are sampled. We define the empirical distribution through
\begin{equation}
P_{\text{obs}} [\boldsymbol \sigma] = \frac 1B \sum _{\tau =1} ^B\delta _{\boldsymbol\sigma , \boldsymbol\sigma^\tau} \ ,
\end{equation}
where $\delta$ denotes the $N$-dimensional Kronecker delta function. It is easy to check from (\ref{s2}) that
\begin{equation}\label{s2bis}
S_{Ising}[{\bf J}| {\bf p}] = - \sum _{\boldsymbol\sigma}  P_{\text{obs}} [\boldsymbol\sigma] \log P_{\bf J}[{\boldsymbol\sigma}]= - \sum _{\boldsymbol\sigma} P_{\text{obs}} [ \boldsymbol\sigma] \log P_{\text{obs}} [ \boldsymbol \sigma ]+ D\big( P_{\text{obs}}    || P _ {\bf J} \big) \ ,
\end{equation}
where $D$ denotes the KL-divergence. Hence, the minimization procedure over $\bf J$ ensures that the 'best' Ising measure (as close as possible to the empirical measure) is found.

\subsection{Regularization and Bayesian formulation}
\label{bayessec}
We consider the Hessian of  the cross-entropy $S_{Ising}$, also called Fisher information matrix, which is  a matrix of dimension $\frac 12 N(N+1)$, defined through
\begin{equation}\label{defchi}
\boldsymbol\chi = \frac{\partial ^2 S_{Ising}}{\partial {\bf J} \partial {\bf J}} =\left( \begin{array} {c c}
\chi_{i,i'} & \chi_{i,k'l'} \\ \chi_{kl,i'} & \chi_{kl,k'l'} \end{array} \right) \ .
\end{equation}
The entries of $\boldsymbol \chi$ are obtained upon repeated differentiations of the partition function $Z[{\bf J}]$, and can be expressed in terms of averages over the Ising Gibbs measure $\langle \cdot \rangle _{\bf J}$,
\begin{eqnarray}
\chi _{i,i'} &=& \langle \sigma_i \sigma_{i'} \rangle_{\bf J} -  \langle \sigma_i \rangle_{\bf J} \langle \sigma_{i'} \rangle_{\bf J} \ ,  \\
\chi _{i,k'l'} &=& \langle \sigma_i \sigma_{k'}\sigma_{l'} \rangle_{\bf J} -  \langle \sigma_i \rangle_{\bf J} \langle \sigma_{k'} \sigma_{l'} \rangle_{\bf J} \ , \nonumber \\
\chi _{kl,k'l'} &=& \langle \sigma_k\sigma_l \sigma_{k'}\sigma_{l'} \rangle_{\bf J} -  \langle \sigma_k\sigma_l \rangle_{\bf J} \langle \sigma_{k'} \sigma_{l'} \rangle_{\bf J} \ . \nonumber
\end{eqnarray}
Consider now an arbitrary $\frac 12 N(N+1)$-dimensional vector ${\bf x}=\{x_i,x_{kl}\}$. The quadratic form
\begin{equation}
{\bf x}^\dagger \cdot \boldsymbol \chi\cdot  {\bf x} = \left\langle \left( \sum _i x_i \big( \sigma _i -
\langle \sigma_i\rangle_{\bf J}\big) + \sum _{k<l} x_{kl} \big( \sigma _k \sigma_l -
\langle \sigma_k \sigma_l\rangle_{\bf J}\big) \right)^2 \right\rangle_{\bf J}
\end{equation}
 is semi-definite positive. Hence, $S_{Ising}$ is a convex function.
  
 However the minimum is not guaranteed to be unique if $\boldsymbol\chi$ has zero modes, nor to be finite. To circumvent those difficulties, one can 'regularize' the cross-entropy $S_{Ising}$ by adding a quadratic term in the interaction parameters, which forces $\boldsymbol\chi$ to become definite positive, and ensures the uniqueness and finiteness of the minimum of $S_{Ising}$. In many applications, no regularization is needed for the fields $h_i$. The reason can be understood intuitively as follows. Consider a data set where all variables are independent, with small but strictly positive means $p_i$. Then, the empirical average products, $p_{kl}$, may vanish if the number $B$ of sampled configurations is not much larger than $(p_kp_l)^{-1}$. This condition is often violated in practical applications, {\em e.g.} the analysis of neurobiological or protein data \cite{bialek, noi,Weigt}. Hence, poor sampling may produce infinite negative couplings. We therefore add the following regularization term to $S_{Ising}$, 
\begin{equation}\label{regul2}
\gamma\sum_{k<l} J_{kl}^2\, p_k(1-p_k) p_l(1-p_l)Ê\ . 
\end{equation} 
The precise expression of the regularization term is somewhat arbitrary, and is a matter of convenience. The dependence on the $p_i$'s in (\ref{regul2}) will be explained in Section \ref{secs0}. Other regularization schemes, based on the $L_1$ norm rather than on the $L_2$ norm are possible, such as
\begin{equation}\label{regul1}
\gamma \sum_{k<l} |J_{kl}|\, \sqrt{p_k(1-p_k) p_l(1-p_l)}Ê\ . 
\end{equation} 
The above regularization is especially popular among the graphical model selection community \cite{wain}, and favors sparse coupling networks, {\em i.e.} with many zero interactions.

 The introduction of a regularization is natural in the context of Bayesian inference. The Gibbs probability $P_{\bf J}[\boldsymbol\sigma]$ defines the likelihood of a configuration $\boldsymbol\sigma$. The likelihood of a set of $B$ independently drawn configurations $\boldsymbol\sigma ^\tau$ is given by the product of the likelihoods of each configuration. The posterior probability of the parameters (fields and couplings) $\bf J$ given the configurations $\boldsymbol\sigma ^\tau$, $\tau =1,2,\ldots , B$, is, according to Bayes' rule,
\begin{equation}\label{ppost}
P_{post}[{\bf J} | \{\boldsymbol\sigma ^\tau\}] \propto \prod_{\tau =1}^B P_{\bf J} [\boldsymbol\sigma ^\tau ] \; P_0[{\bf J}] \ ,
\end{equation}
up to an irrelevant ${\bf J}$-independent multiplicative factor. In the equation above, $P_0$ is a prior probability over the couplings and fields, encoding the knowledge about their values in the absence of any data. Taking the logarithm of (\ref{ppost}), we obtain, up to an additive $\bf J$-independent constant, 
\begin{equation}\label{logppost}
\log P_{post}[{\bf J} | \{\boldsymbol\sigma ^\tau\}] = -B\; 
 S_{Ising}[{\bf J}| {\bf p }] + \log P_0[{\bf J}] \ .
\end{equation}
Hence, the most likely value for the parameters $\bf J$ is the one minimizing  $S_{Ising}[{\bf J}| {\bf p }] -\frac 1B  \log P_0[{\bf J}]$. The regularization terms (\ref{regul2}) and (\ref{regul1}) then correspond to, respectively, Gaussian and exponential priors over the parameters. In addition, as the prior is independent of the number $B$ of configurations, we expect the strength $\gamma$ to scale as $\frac 1B$.  The optimal value of $\gamma$ can be  also  determined based on Bayesian criteria \cite{noi,McKay} (Appendix \ref{appchoice}).

We emphasize that the Bayesian framework changes the scope of the inference. While the MEP aims to reproduce the data, the presence of a regularization term leads to a compromise between two different objectives: finding an Ising model whose observables (one- and two-point functions) are close to the empirical values and ensuring that the interaction parameters $\bf J$ have a large prior probability $P_0$. In other words, a compromise is sought between the faithfulness to the data and the prior knowledge about the solution. The latter is especially important in the case of poor sampling (small value of $B$ or data corrupted by noise). For instance, the regularization term based on the $L_1$--norm (\ref{regul1}) generally produces more couplings equal to zero than its $L_2$--norm counterpart (\ref{regul2}). This property is desirable if one a priori knows that the interaction graph is sparse. Hence, the introduction of a regularization term can be interpreted as an attempt to approximately solve the inverse Ising problem while fulfilling an important constraint about the structure of the solution. We will discuss the nature of the structural constraints corresponding to our adaptive cluster algorithm in Section \ref{sec-chi}.

Knowledge of the inverse of the Fisher information matrix, ${\boldsymbol\chi}^{-1}$, allows for the computation of the statistical fluctuations of the inferred fields and couplings due to a finite number $B$ of sampled configurations. According to the asymptotic theory of inference, the posterior probability $P_{post}[{\bf J}|\{\boldsymbol\sigma ^\tau\}]$ over the fields and couplings becomes, as $B$ gets very large, a normal law centered in the minimum of $S_{Ising}[{\bf J}|{\bf p}]$. The covariance matrix of this normal law is simply given by $\frac 1B \, {\boldsymbol\chi}^{-1}$. Consequently the standard deviations of the fields $h_i$ and of the couplings $J_{kl}$ are, respectively,
\begin{equation}
\delta h_{i}=\sqrt{\frac 1B \; (\chi^{-1})_{i,i}}\ , \quad 
\delta J_{kl}=\sqrt{\frac 1B \; (\chi^{-1})_{kl,kl}} \ .
\label{deltaj}
\end{equation}
In order to remove the zero modes of $\boldsymbol\chi$ and have a well-defined inverse matrix ${\boldsymbol\chi}^{-1}$, the Ising model entropy  $S_{Ising}$ (\ref{s2}) can be added a regularization term, {\em e.g.} (\ref{regul2}), which guarantees that ${\boldsymbol\chi}$ is positively defined. 

The Fisher information matrix, $\boldsymbol\chi$, can also be used to estimate the statistical deviations of the observables coming from the finite sampling. If the data were generated by an Ising model with parameters $\bf J$, we would expect, again in the large $B$ setting, that the frequencies $p_i,p_{kl}$ would be normally distributed with a covariance matrix equal to $\frac 1B \, {\boldsymbol\chi}$. Hence, the typical uncertainties over the 1- and 2-point frequencies are given by
\begin{equation}
\delta p_{i}=\sqrt{\frac 1B \, {\chi}_{i,i} }=\sqrt{\frac{\langle \sigma_i\rangle _{\bf J}(1-\langle \sigma_i\rangle _{\bf J})}{B}}\ , \qquad 
\delta p_{kl}=\sqrt{\frac 1B \, {\chi }_{kl,kl} }=\sqrt{\frac{\langle \sigma_k\sigma_l\rangle _{\bf J}(1-\langle \sigma_k\sigma_l\rangle _{\bf J})}{B}}\ .
\label{def-deltac}
\end{equation}
In practice, we can replace the Gibbs averages above with the empirical averages $p_i$ and $p_{kl}$ to obtain estimates for the expected deviations. These estimates will be used to decide whether the inference procedure is reliable, or leads to an overfitting of the data in Section \ref{secappli}.

\subsection{Methods}\label{secmethods}

The inverse Ising problem has been studied in statistics, under the name of graphical model selection, in the machine learning community under the name of (inverse) Boltzmann machine learning, and in the statistical physics literature. Different methods have been developed, with various applications. Some of the methods are briefly discussed below. 

A direct calculation of the partition function $Z[{\bf J}]$ generally requires a time growing exponentially with the number $N$ of variables, and is not feasible when $N$ exceeds a few tens. Inference procedures therefore tend to avoid the computation of $Z[{\bf J}]$:
\begin{itemize} 
\item A popular algorithm is the {\em Boltzmann learning} procedure, where the fields and couplings are iteratively updated until the averages $\langle \sigma_i\rangle_{\bf J}$'s and $\langle \sigma_k \sigma_l\rangle_{\bf J}$'s, calculated from Monte Carlo simulations, match the imposed values \cite{ackley}. The number of updatings can be very large in the absence of a good initial guess for the parameters $\bf J$. Furthermore, for each set of parameters, thermalization may require prohibitive computational efforts for large system sizes $N$, and problems with more than a few tens of spins can hardly be tackled. Finally, learning data exactly leads to overfitting in the case of poor sampling.
\item the {\em Pseudo-Likelihood}-based algorithm by Ravikumar {\em et al.} \cite{wain, bento} is an extension to the binary variable case of Meinshausen and B\"uhlmann's algorithm \cite{mein06} and is related to a renormalisation approach introduced by Swendsen \cite{swendsen}. The procedure requires the complete knowledge of the configurations $\{\boldsymbol\sigma ^\tau\}$ (and not only of the one- and two-point functions $\bf p$). The starting point is given by well-known Callen's identities for the Ising model,
\begin{equation}
\langle \sigma_i \rangle_{\bf J} = \left\langle \frac 1{1+\exp \left( -\sum _j J_{ij} \sigma _j - h_i \right)} \right\rangle_{\bf J}
\simeq \frac 1B \sum _{\tau =1}^B  \frac 1{1+\exp \left( -\sum _j J_{ij} \sigma _j ^\tau- h_i \right)} 
\end{equation}
where the last approximation consists in replacing the Gibbs average with the empirical average over the sampled configurations. Imposing that the Gibbs average $\langle\sigma_i\rangle_{\bf J}$ coincides with $p_i$ is equivalent to minimizing the following pseudo-likelihood over the field $h_i$,
\begin{equation}
 S_{i, PL} [h_i, \{J_{ij}, j\ne i\}] =  \frac 1B\sum _{\tau =1}^B \log \bigg[ 1+ \exp\bigg( \sum _j J_{ij} \sigma _j^\tau + h_i \bigg)\bigg] - h_i\, p_i - \sum _{j (\ne i)} J_{ij} p_{ij} \ .
 \end{equation}
The minimization equations over the couplings $J_{ij}$, with $j\ne i$ (and fixed $i$), correspond to Callen identities for two-point functions. Informally speaking, the pseudo-likelihood approach simplifies the original $N$-body problem into $N$ independent 1-body problem, each one in a bath of $N-1$ quenched variables.  Note that the couplings $J_{ij}$ and $J_{ji}$ (found by minimizing $S_{j,PL}$) will generally not be equal. However, as far as graphical model selection is concerned, what matters is whether $J_{ij}$ and $J_{ji}$ are both different from zero. 

The pseudo-entropy $S_{i,PL}$ is convex, and can be minimized after addition of a $L_1$--norm regularization term \cite{tibshirani,wain,Aurell2}. The procedure is guaranteed to find strong enough couplings \footnote{The minimal strength of the couplings which can be 'detected' depend on the quality of the sampling, and scales as $\sqrt{\log N/B}$, see Section \ref{secutile}.} in a polynomial time in $N$, provided that  the data were generated by an Ising model (which is usually not the case in practical applications) and that a quantity closely related to the susceptibility $\boldsymbol\chi$ (\ref{defchi}) is small enough. The latter condition holds for weak couplings and may break down for strong couplings \cite{bento}. For a review of the literature in the statistics community,  see \cite{tibshirani}.
\end{itemize}

In specific cases, however, the partition function can be obtained in polynomial time. Two tractable examples are:
\begin{itemize}
\item {\em Mean-field models}, which are characterized by dense but weak interactions. An example is the Sherrington-Kirkpatrick model where every pair of spins interact through couplings of the order of $N^{-1/2}$ \cite{sher}. The entropy $S[{\bf p}]$ coincides asymptotically with
\begin{equation}\label{loopentro}
S_{MF} ({\bf p})= \frac 12\log \hbox{\rm det}\, M({\bf p}), \hbox{\rm where}\quad M_{ij}({\bf p})=\frac{p_{ij} -p_i p_j}{\sqrt{p_i(1-p_i)p_j(1-p_j)}}  ,
\end{equation}
which can be calculated in $O(N^3)$ time \cite{opper,diag}. Expression (\ref{loopentro}) has been obtained from the high temperature expansion  \cite{ple,geo91,geo04} of the Legendre transform of the free energy , and is consistent with the so-called TAP equations \cite{tap}. The derivative of $S_{MF}$ with respect to $\bf p$ gives the value of the couplings and the fields,
\begin{eqnarray}\label{hJMF}
(J_{MF})_{kl} &=& -\frac{\partial S_{MF}}{\partial p_{kl}} =- \frac{(M^{-1})_{kl}}{\sqrt{p_k(1-p_k)p_l(1-p_l)}} \ , \nonumber \\ 
(h_{MF})_i &=& -\frac{\partial S_{MF}}{\partial p_i} = \sum _{j (\ne i)} (J_{MF})_{ij} \left( c_{ij}\frac{p_i-\frac 12}{p_i(1-p_i)} - p_j\right)\ ,
\end{eqnarray}
where $c_{ij}=p_{ij}-p_ip_j$ is the connected correlation. From a practical point of view, expression (\ref{loopentro}) is a good approximation for solving the inverse Ising problem \cite{Tanaka,diag,aurell} on dense and weak interaction networks , but fails to reproduce dilute graphs with strong interactions. 
\item Ising models on tree-like structures, {\em i.e.} with no or few interaction loops. {\em Message passing methods} are guaranteed to solve the associated inverse Ising problems. For trees, the partition functions can be calculated in a time linear in $N$.  Sparse networks of strong interactions with long-range loops, such as Erd\"os-Renyi random graphs, can also  be successfully treated in polynomial time by message-passing procedures \cite{pel05,mora,Weigt}. However,  these methods generally break down in the presence of strongly interacting groups (clusters) of spins. 
\end{itemize}

When an exact calculation of the partition function is out-of-reach, accurate estimates can be obtained through cluster expansions. Expansions have a rich history in statistical mechanics, {\em e.g. } the virial expansion in the theory of liquids \cite{han,dedominicis}. However, cluster expansions  
suffer from several drawbacks. First, in cluster variational methods \cite{pel05,leb}, the calculation of the contributions coming from each cluster generally involves the resolution of non trivial and self-consistent equations for the local fields, which seriously limits the maximal size of clusters considered in the expansion. Secondly, the composition and the size of the clusters is usually fixed {\em a priori}, and does not adapt to the specificity of the data \cite{noi}. The combinatorial growth of the number of clusters with their size entails strong limits upon the maximal sizes of the network, $N$, and of the clusters, $K$. Last of all, cluster expansions generally ignore the issue of overfitting. 

Recently, we have proposed a new cluster expansion, where clusters are built recursively, and are selected or discarded, according to their contribution to the cross-entropy $S$ \cite{coc11}. This selection procedure allows us to fully account for the complex interaction patterns present in experimental systems, while preventing a blow-up of the computational time. The purpose of this paper is to illustrate this method and discuss its advantages and limitations.

\section{Cluster expansion of the cross-entropy}
\label{secclusterprincipal}

\subsection{Principle of the expansion} 
\label{secexpa}

\begin{figure}
\begin{center}
\epsfig{file=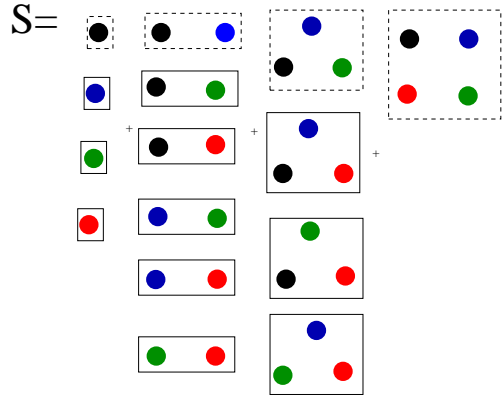,width=10cm}
\caption{Decomposition of the cross-entropy $S({\bf p})$ for a system of 4 spins, indicated with different colors, as the sum of cluster contributions. Each cluster-entropy  $\Delta S_\Gamma ({\bf p})$ depends only on the one- and two-point frequencies of the variables in the cluster: it can be calculated in a recursive way, see main text. Dotted clusters are decomposed into a diagrammatic expansion in Fig.~\ref{fig-diag}.}
\label{cluster}
\end{center}
\end{figure}

In this Section, we propose a cluster expansion for the entropy $S({\bf p})$. A cluster, $\Gamma$, is defined here as a non-empty subset of $(1,2,\ldots , N)$. To illustrate how the expansion is built we start with the simple cases of systems with a few variables ($N=1,2$), in the absence of the regularization term~(\ref{regul2}). 

Consider first the case of a single variable, $N=1$, with average value $p_1$. The entropy $S(p_1)$ can be easily computed according to the definitions given in Section \ref{secstatement}, with the result
\begin{eqnarray}\label{deltas1}
S_{(1)}(p_1) &=& \min _{h_1} S_{Ising}[h _1|p_1] =  \min _{h_1} \bigg[\log \big(1 + e^{h_1}\big) -h_1 p_1
\bigg] \nonumber \\
&=& -p_1 \log p_1 - (1-p_1) \log (1-p_1) \ .
\end{eqnarray}
We recognize the well-known expression for the entropy of a 0-1 variable with mean value $p_1$. For reasons which will be obvious in the next paragraph, we will hereafter use the notation $\Delta S_{(1)}(p_1)$ to denote the same quantity as $S_{(1)}(p_1)$. The subscript $(1)$ refers to the index of the (unique) variable in the system.

Consider next a system with two variables, with mean values $p_1,p_2$ and two-point average $p_{12}$. The entropy $S_{(1,2)}(p_1,p_2,p_{12})$ can be explicitly computed:
\begin{eqnarray}\label{s2tot}
S_{(1,2)}(p_1,p_2,p_{12}) &=& \min _{h_1,h_2,J_{12}} S_{Ising}[h_1,h_2,J_{12} |p_1,p_2,p_{12}] 
\nonumber \\
&=&  \min _{h_1,h_2,J_{12}} \bigg[\log \big(1 + e^{h_1}+ e^{h_2}+ e^{h_1+h_2+J_{12}}\big) -h_1p_1-h_2 p_2 - J_{12}p_{12} \bigg] \nonumber \\
&=& -(p_1-p_{12})\,\log{(p_1-p_{12})} -(p_2-p_{12})\log(p_2-p_{12})\nonumber \\
&-&p_{12}\log p_{12} -(1-p_1-p_2+p_{12})\,\log{(1-p_1-p_2+p_{12})} \ .
\end{eqnarray}
We now define the entropy $\Delta S_{(1,2)}$ of the cluster of the two variables $1,2$ as the difference between the entropy $S _{(1,2)}(p_1,p_2,p_{12})$ calculated above and the two single-variable contributions $\Delta S_{(1)}(p_1)$ and $\Delta S_{(1)} (p_2)$ coming from the variables $1$ and $2$ taken separately:
\begin{equation}\label{deltas2}
\Delta S_{(1,2)} (p_1,p_2,p_{12}) = S_{(1,2)}(p_1,p_2,p_{12})-\Delta S_{(1)}(p_1)-\Delta S_{(2)} (p_2) \ .
\end{equation}
In other words, $\Delta S_{(1,2)}$ measures the loss of entropy between the system of two isolated variables, constrained to have means equal to, respectively, $p_1$ and $p_2$, and the same system when, in addition, the average product of the variables is constrained to take value $p_{12}$. Using expressions (\ref{deltas1}) and (\ref{s2tot}), we find
\begin{eqnarray}\label{deltas2b}
\Delta S_{(1,2)} (p_1,p_2,p_{12}) &=& - (p_1-p_{12})  \log \left( \frac{ p_1-p_{12} }{p_1 - p_1p_2}\right) - (p_2-p_{12})  \log \left( \frac{ p_2-p_{12} }{p_2 - p_1p_2}\right)\nonumber \\
&-&p_{12}\log \left( \frac{p_{12}}{p_1p_2}\right) -(1-p_1-p_2+p_{12})\log \left( \frac{1-p_1-p_2+p_{12}}{1-p_1-p_2+p_1p_2}\right) \ .
\end{eqnarray}
The entropy of the cluster $(1,2)$ is therefore equal to the Kullback-Leibler divergence between the true distribution of probability over the two spins and the one corresponding to two independent spins with averages $p_1$ and $p_2$. It vanishes for $p_{12}=p_1p_2$.

Formula (\ref{deltas2}) can be generalized to define the entropies of clusters with larger sizes $N\ge 3$. Again ${\bf p}=\{p_i,p_{kl}\}$ denotes the data. For any non-empty subset $\Gamma$ including $1\le K\le N$ variables, we define two entropies:
\begin{itemize}
\item the subset-entropy $S_\Gamma({\bf p})$, which is the entropy of the subset of the $K$ variables for fixed data. It is defined as the right hand side of  (\ref{maxentro}), when the variable indices, $i,k,l$ are restricted to $\Gamma$. Note that, when the subset $\Gamma$ includes all $N$ variables, $S_\Gamma({\bf p})$ coincides with $S({\bf p})$.
\item the cluster-entropy $\Delta S_\Gamma ({\bf p})$, which is the remaining contribution to the subset-entropy $S_\Gamma  ({\bf p})$, once all other cluster-entropies of smaller clusters have been substracted.  The cluster entropies are then implicitly defined through the identity
\begin{equation}\label{recur-entro} 
S_{\Gamma}({\bf p})=\sum _{\Gamma' \subset \Gamma}  \Delta S_ {\Gamma'} ({\bf p})\ ,
\end{equation} 
where the sums runs over all $2^K-1$ non-empty clusters $\Gamma'$ of variables in $\Gamma$.
\end{itemize}
Identity (\ref{recur-entro}) states that the entropy of a system (for fixed data) is equal to the sum of the entropies of all its clusters. Figure~\ref{cluster} sketches the cluster decomposition of the entropy for a system of $N=4$ variables.

For $\Gamma=(1)$, equation (\ref{recur-entro}) simply expresses that $S_{(1)}(p_1)=\Delta S_{(1)}(p_1)$. For $\Gamma=(1,2)$, equation (\ref{recur-entro}) coincides with (\ref{deltas2}). For $\Gamma=(1,2,3)$, we obtain the definition of the entropy of a cluster made of a triplet of variables:
\begin{eqnarray} \label{deltas3}
\Delta S_{(1,2,3)} (p_1,p_2,p_3,p_{12},p_{13},p_{23})&=& S _{(1,2,3})( p_1,p_2,p_3,p_{12},p_{13},p_{23})- \Delta S_{(1)}(p_1) - \Delta S_{(2)}(p_2) - \Delta S_{(3)}(p_3)\nonumber \\
&-&  \Delta S_{(1,2)}(p_1,p_2,p_{12}) -  \Delta S_{(1,3)}(p_1,p_3,p_{13})- \Delta S_{(2,3)}(p_2,p_3,p_{23}) \ .
\end{eqnarray}
The analytical expression of the cluster-entropy $\Delta S_{(1,2,3)}$ is given in Appendix~\ref{entroK3}.
  

The examples above illustrate three general properties of cluster-entropies: 
\begin{itemize}
\item the entropy of the cluster $\Gamma$, $\Delta S_ \Gamma$, depends only on the frequencies $p_i,p_{ij}$ of the variables $i,j$ in the cluster $\Gamma$ (and not on all the data in ${\bf p}$).
\item  the entropy of a cluster with, say, $K$ variables, can be recursively calculated from the knowledge of the subset-entropies $S_{\Gamma'}({\bf p})$ of all the subsets $\Gamma'\in \Gamma$ with $K'\le K$ variables. According to M\"obius inversion formula,
\begin{equation}\label{mobius} 
\Delta S_{\Gamma}({\bf p})=\sum _{\Gamma'  \subset \Gamma}  (-1)^{K'-K} \; S_ {\Gamma'} ({\bf p}) \ .
\end{equation}
\item the  sum of the entropies of all $2^N-1$ clusters of  a system of $N$ spins is the exact entropy of the system, see (\ref{recur-entro}) with $\Gamma= (1,2,\ldots , N)$.
\end{itemize}

In practice,  to calculate $S({\bf p})$, one first computes the partition function $Z[{\bf J}]$ by summing over the $2^K$ configurations $\boldsymbol\sigma$ and, then, minimizes $S_{Ising}[{\bf J} | {\bf p}]$ (\ref{s2}) over the interaction parameters ${\bf J}$. The minimization of a convex function of $\frac 12 K(K+1)$ variables can be done in time growing polynomially with $K$. Moreover the
addition of the  regularization term (\ref{regul2}) can be easily handled. 
The limiting step is therefore the calculation of $Z$,  which can be done exactly for clusters with less than, say, $K=20$ spins.

Hence, only a small number of the $2^N-1$ terms in (\ref{recur-entro}) can be calculated. In the present work we claim that, in a wide set of circumstances, a good approximation to the entropy $S({\bf p})$ can be already obtained from the contributions of well-chosen clusters of small sizes,
 \begin{equation}\label{recur-entro-approx} 
S({\bf p})\simeq \sum _{\Gamma \in L}  \Delta S_ \Gamma ({\bf p})\ ,
\end{equation} 
We will explain in Section \ref{sect99}  how the list of selected  clusters, $L$, is established.

\subsection{The reference entropy $S_0$}
\label{secs0}

So far we have explained how the entropy $S({\bf p})$ can be expanded as a sum of contributions $\Delta S_{\Gamma} ({\bf p})$ attached to the clusters $\Gamma$. In this Section we present the expansion against a reference entropy, $S_0 ({\bf p})$, and two possible choices for the reference entropy.

The idea underlying the introduction of a reference entropy is the following. Assume one can calculate a (rough) approximation $S_0({\bf p})$ to the true entropy $S({\bf p})$. Then, the difference $S({\bf p})-S_0({\bf p})$ is smaller than $S({\bf p})$, and it makes sense to expand the former rather than the latter. We expect, indeed, the cluster-entropies to be smaller when the reference entropy $S_0({\bf p})$ is substracted from the true entropy. We substitute the original definition (\ref{recur-entro}) with the new definition
\begin{equation}\label{recur-entro0} 
S({\bf p}) = S_0({\bf p})+\sum _{\Gamma \subset (1,2,\ldots ,N)}  \Delta S_ \Gamma ({\bf p})\ .
\end{equation} 
With this new definition, the values of the cluster-entropies $\Delta S_\Gamma$ depend on the choice of $S_0$; the previous definition (\ref{recur-entro}) is found back when $S_0=0$.  The procedure for the calculation of the cluster-entropies $\Delta S_{\Gamma}({\bf p})$ is the same as in Section \ref{secexpa}, upon replacement of $S({\bf p})$ with $S({\bf p})-S_0({\bf p})$. The three properties of the cluster expansion listed above still hold.
 
Our final estimate for the entropy will be, compare to (\ref{recur-entro-approx}),
 \begin{equation}\label{recur-entro-approx2} 
S({\bf p})\simeq S_0({\bf p}) + \sum _{\Gamma \in L}  \Delta S_ \Gamma ({\bf p})\ .
\end{equation} 
Hence, the cluster expansion is a way to calculate a correction to the approximation $S_0$ to the true entropy $S$. Obviously, the introduction of a reference entropy is useful in practice only if  $S_0({\bf p})$ can be quickly calculated for the entire system of size $N$. In other words, the computational effort required to obtain $S_0$ should scale only polynomially with $N$. A natural choice for the reference entropy is $S_0=S_{MF}$ (\ref{loopentro}), the mean-field entropy discussed in Section \ref{secmethods}. As the calculation of $S_{MF}$ requires the one of the determinant of the matrix $M({\bf p})$, it can be performed in a time scaling as $N^3$ only. In addition, we expect $S_{MF}$ to be a sensible approximation to $S$ for systems with rather weak interactions. Corrections coming from the strongest interactions will be taken care of by the cluster expansion. 

Regularized versions of the Mean Field entropy can be derived as follows. First, we use the MF expression for the cross-entropy at fixed couplings $J_{kl}$ and frequencies $p_i$, see (\ref{s2}) and \cite{geo91}, to rewrite
\begin{equation}
S_{Ising}(\{ p_i\},\{J_{kl}\}) =   - \frac 12 \log \hbox{\rm det}   \big( \hbox{\rm Id} - J' \big)-\sum_{k<l} J_{kl}\,(p_{kl}-p_k\, p_l)\ , \quad \hbox{\rm where} \quad  J'_{kl}= J_{kl} \sqrt{ p_k(1-p_k) p_l(1-p_l)} \ ,
\end{equation}
and Id denotes the $N$-dimensional identity matrix. We consider the $L_2$-norm regularization (\ref{regul2}). The entropy at fixed data $\bf p$ is 
\begin{eqnarray}
S_0({\bf p}) &=& \min _{\{J_{kl}\}} \bigg[ S_{Ising}(\{ p_i\},\{J_{kl}\})  + \gamma\sum_{k<l} J_{kl}^2\, p_k(1-p_k) p_l(1-p_l) \bigg]Ê\nonumber \\
&=& \min _{\{J'_{kl}\}} \bigg[    - \frac 12 \log \hbox{\rm det} \big( \hbox{\rm Id} - J' \big) -\frac 12 \hbox{\rm Trace} \big( J' \cdot M({\bf p}) \big)+ \frac \gamma 2\hbox{\rm Trace} \big( J')^2 \bigg]Ê \ ,
\end{eqnarray}
where $M({\bf p})$ is defined in (\ref{loopentro}). The optimal interaction matrix $J'$ is the root of the equation
\begin{equation}
\big( \hbox{\rm Id} - J' \big) ^{-1} - M({\bf p}) + \gamma \, J' =0 \ .
\end{equation}
Hence, $J'$ has the same eigenvectors as $M({\bf p})$, a consequence of the dependence on $p_i$ we have chosen for the quadratic regularization term in (\ref{regul2}). Let $j_q$ denote its $q^{th}$ eigenvalue, and $\hat m _q= (1-j_q)^{-1}$. Then,
\begin{equation}\label{loopentrogamma}
S_0({\bf p}, \gamma)=\frac 12 \sum_{q=1}^N \big(\log \hat m_{q} +1-\hat m_{q}\big)\ ,
\end{equation}
where $\hat m_q$ is the largest root of $\hat m_{q}^2-\hat m_{q}(m_{q}-\gamma)=\gamma$, and $m_q$ is the $q^{th}$ eigenvalue of $M({\bf p})$. Note that $\hat m_q=m_q$ when $\gamma=0$, as expected.


\subsection{Properties of the cluster entropies $\Delta S_\Gamma$}

\subsubsection{Diagrammatic expansion in powers of the connected correlations}
\label{secexpdiag}

\begin{figure}
\begin{center}
\epsfig{file=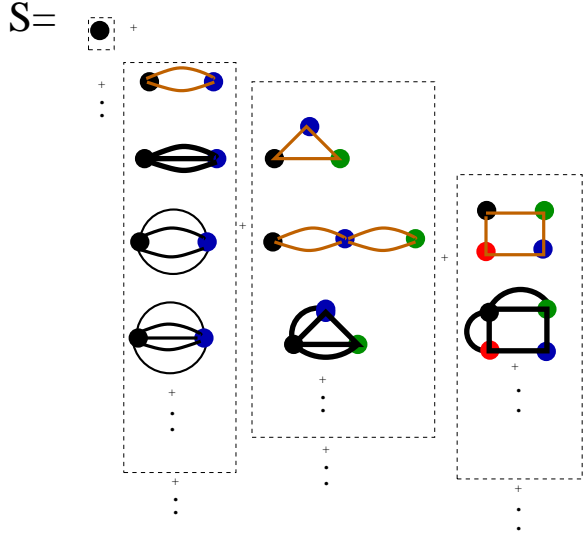,width=10cm}
\caption{Diagrammatic expansion of the cross-entropy $S({\bf p})$. A cluster-entropy (see Fig.~\ref{cluster}) is the infinite sum of all the diagrams in a box (dashed contour), linking the $K$ sites in the cluster. Each link in a diagram carries $M_{ij}$, and each site $p_i$; in addition, each diagram carries a multiplicative factor, which is a function of the $p_i$'s. In the Figure only one cluster among all ${N \choose K}$ clusters is represented. Only the first diagrams with non--zero coefficients are drawn. Loop diagrams are analytically summed up and removed from the expansion through the reference entropy $S_0=S_{MF}$; Eulerian circuit diagrams (brown/gray) are partly removed, see main text. Diagrams giving the largest contributions to the universal central peak of the cluster-entropy distribution (Appendix~\ref{app-diag}) are shown in bold.}
\label{fig-diag}
\end{center}
\end{figure}

A better understanding of the cluster expansion and of the role of the reference entropy $S_0$ can be gained through the diagrammatic expansion of the entropy $S({\bf p})$ in powers of the connected correlations (high-temperature expansion),
\begin{equation}
c_{ij} = p_{ij}-p_ip_j \ .
\end{equation}
Note that the entry $M_{ij}$ of the matrix $M$ defined in (\ref{loopentro}) vanishes linearly with $c_{ij}$. Thus, an expansion in powers of $c_{ij}$ is equivalent to an expansion in powers of $M_{ij}$. A procedure to derive in a systematic way the diagrammatic expansion of $S({\bf p})$ is proposed in \cite{diag}. The diagrammatic expansion provides a simple representation of the cluster-entropies, in which the entropy $S({\bf p})$ can be represented as a sum of connected diagrams (Fig.~\ref{fig-diag}). Each diagram is made of sites, connected or not by one or more edges. Each point symbolizes a variable, and carries a factor $p_i$. The presence of $n(\ge 0)$ edges between the sites $k$ and $l$ results in a multiplicative factor $(c_{kl})^n$. The contribution of a diagram to the entropy is the product of the previous factors, times a function of the $p_i$ specific to the topology of the diagram, see \cite{diag}.
 Diagrams of interest include (Fig.~\ref{fig-diag}):
\begin{itemize}
\item the $N$ single-point diagrams, whose contributions are $\Delta S_{(i)} (p_i)$;
\item the 'loop' diagrams, which consist of a circuit with $K$ edges going through $K$ sites $i_1 \to i_2\to \ldots  \to i_K\to i_1$, whose contributions to the entropy are  
\begin{equation} \label{entroploop2}
S_{loop} ({\bf p}|i_1,i_2,\ldots ,i_K) =  (-1)^{K-1} \,M_{i_1,i_2}M_{i_2,i_3}\ldots M_{i_{K-1},i_K}M_{i_K,i_1} \ ;
\end{equation}
\item the Eulerian circuit diagrams, for which there exists a closed path visiting each edge exactly once; 
\item the non-Eulerian diagrams, with the lowest number of links (smallest power in $M$).  
\end{itemize}
The entropy for two variables $i,j$, $S(p_i,p_j,p_{ij})$ (\ref{s2tot}), is the sum of the two single-point diagrams $i$ and $j$, plus the sum of all connected diagrams made of the two sites $i$ and $j$ with an arbirtrary large number of edges ($n\ge 2)$ in between (first two columns in Fig.~\ref{fig-diag}). According to (\ref{deltas2}), the cluster-entropy $\Delta S_{(i,j) }(p_i,p_j,p_{ij})$ is equal to the latter sum (second column in Fig.~\ref{fig-diag}). More generally, the entropy of a cluster $\Delta S_{\Gamma} ({\bf p})$ is the infinite sum of all diagrams whose sites are the indices in $\Gamma$.

We now interpret the Mean Field expression for the entropy, $S_{MF}$, in the diagrammatic framework. We start from identity (\ref{loopentro}), and rewrite, 
\begin{equation}
S_{MF}({\bf p}) =  \frac 12  \hbox{\rm Trace} \log M = \frac 12  \hbox{\rm Trace} \log \big[ \hbox{\rm Id}- \big(\hbox{\rm Id}-M\big)\big]= \sum _{K\ge 1} \frac{-\hbox{\rm Trace} \big[\big(\hbox{\rm Id}-M\big)^K\big]}{2K} \ .
\label{sloopa}
\end{equation}
Using the fact that the diagonal elements of $M$ are equal to unity, the term corresponding to $K=1$ above vanishes. For $K\ge 2$, we have
\begin{eqnarray}
-\hbox{\rm Trace} \big[\big(M-\hbox{\rm Id}\big)^K\big] &=& -
\sum _{i_1,i_2,\ldots ,i_K} \big(\delta _{i_1,i_2}-M_{i_1,i_2}\big)  \big(\delta _{i_2,i_3}-M_{i_2,i_3}\big)\ldots  \big(\delta _{i_{K-1},i_K}-M_{i_{K-1},i_K}\big) \big(\delta _{i_K,i_1}-M_{i_K,i_1}\big) \nonumber \\ &=& \sum _{i_1,i_2,\ldots ,i_K}
(-1)^{K-1} \,\hat M_{i_1,i_2}\hat M_{i_2,i_3}\ldots \hat M_{i_{K-1},i_K}\hat M_{i_K,i_1} \ , \label{sloopb}
\end{eqnarray}
where the matrix $\hat M$ has the same off-diagonal elements as $M$, and has zero diagonal elements. Each term in the above sum corresponds to an Eulerian circuit over $K'\le K$ sites, where $K'$ is the number of distinct indices in $(i_1,i_2,\ldots, i_K)$. Note that the same circuit can be obtained from different $K$-uplets of indices. Consider for instance the longest circuits, obtained for $K'=K$, {\em i.e.} all distinct indices. $2K$ different $K$--uplets $(i_1,i_2,\ldots, i_K)$ correspond to the same circuit, as neither the starting site nor the orientation of the loop matter. For instance, $i_1\to i_2\to i_3\to i_1$, $i_2\to i_3\to i_1\to i_2$, $i_1\to i_3\to i_2\to i_1$, ... are all equivalent. This multiplicity factor $2K$ precisely cancels the $2K$ at the denominator in (\ref{sloopa}). The contribution corresponding to a circuit therefore coincides with expression (\ref{entroploop2}) for the loop entropy. We conclude that 
\begin{itemize}
\item $S_{MF}({\bf p})$ sums up all loop diagrams exactly;
\item $S_{MF}({\bf p})$, in addition, sums up Eulerian circuit diagrams, but with weights {\em a priori} different from their values in the cross-entropy $S({\bf p})$ \footnote{In mean-field spin-glasses, as the couplings scale as the inverse square root of the number $N$ of spins, only loop diagrams have non-zero weights in the thermodynamical limit.}. An exception is the three-variable Eulerian diagram shown in Fig.~\ref{fig-diag}, whose weights in $S_{MF}$ and $S$ coincide.
\item no non-Eulerian diagram is taken into account in $S_{MF}({\bf p})$.
\end{itemize}
As a conclusion, the diagrammatic expansion provides a natural justification for the choice of the reference entropy $S_0({\bf p})=S_{MF}({\bf p})$. In addition, it provides us with the dominant contribution to the cluster-entropies once the Mean-Field entropy is substracted, see Fig.~\ref{fig-diag}. A detailed study of those dominant contributions is presented in Appendix~\ref{app-diag}.

\subsubsection{Dependence on the cluster size and on the interaction path length}
\label{secpropdeltas}

\begin{figure}
\begin{center}
\epsfig{file=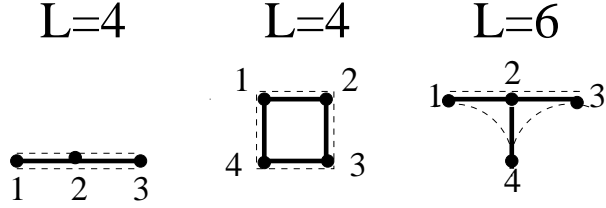,width=8cm}
\caption {Examples of contour paths for three different graphs. Spins are labelled by $1,2,3,4$ and first-neighbor interactions are represented by bold lines. The contour path is depicted with a dotted line. The contour length $L$, which can be calculated as the sum of distances along the contour path (dotted arcs have length 2) is indicated above each graph. Different clusters may have the same contour path and contour length. Left:  $(1,3)$ and  $(1,2,3)$ have contour length $L=4$, while $(1,2)$ has $L=2$. Middle: clusters $(1,3)$, $(1,2,3)$, $(1,2,3,4)$, $(1,3,4)$ have the same contour path. Right: $(1,2,3,4)$ has length $L=6$. }
\label{fig-contourpath}
\end{center}
\end{figure}

To reach a better understanding of what the cluster-entropy means, we consider the case of finite-dimensional Ising model, {\em e.g.} with coupling $J>0$ between nearest-neighbors on a $D$-dimensional lattice. We call $\xi$ the correlation length: the connected correlation $c$ between two sites at large distance $d$ decays as $\sim\exp (-d/\xi)$. We want to characterize the behavior of the cluster-entropy $\Delta S_{\Gamma}$  when the $K$ sites in the cluster $\Gamma$ are far apart on the lattice. We first choose no reference entropy ($S_0=0$). According to the above diagrammatic expansion, the lowest order diagram (in powers of $c$) with $K$ sites has the loop topology. We look for the shortest closed path joining all the sites in $\Gamma$; let $L(\Gamma)$ be this contour length, that is, the sum of the distances between neighboring sites along the path (Fig.~\ref{fig-contourpath}). Then, according to (\ref{entroploop2}), the largest contribution (in absolute value) to the cluster entropy is
\begin{equation}\label{asymptotics_deltas}
\Delta S_\Gamma \simeq A(p,K)\;(-1)^{K-1} \;\exp\big(\!- L(\Gamma)/\xi\big) \ ,
\end{equation}
where $A$ is a positive function $K$ and of $p$, the representative value of the frequencies $p_i$ of the variables in $\Gamma$. We conclude that the sign of the cluster-entropy depends on the parity of the number of sites. Furthermore, $\Delta S_{\Gamma}$ decreases exponentially fast (in absolute value) with the length of the shortest path joining the sites in the cluster. As soon as one site is very far away from the remaining $K-1$ ones, the cluster-entropy is small.

As a consequence, the sum (\ref{recur-entro}) is alternate, and we expect cancellation between contributions coming from clusters sharing the same shortest path, but with different sizes. This crucial point is perfectly illustrated by the one-dimensional Ising model. The correlation between two sites at distance $d_{ij}=j-i$ is, in one dimension, $c_{ij}= \sqrt{p_i(1-p_i)p_j(1-p_j)}\exp (-d_{ij}/\xi)$ (Appendix~\ref{app-ising1d}). The matrix $\boldsymbol M$ defined in (\ref{loopentro}) has  elements 
\begin{equation}
M_{ij}=e^{-d_{ij}/\xi}
\end{equation}
 Then, according to (\ref{entroploop2}), the largest contribution (in absolute value) to the cluster entropy of a cluster containing the $K$ spins $i_1<i_2< \ldots < i_K$  is given by (\ref{asymptotics_deltas}) with
\begin{equation}
 L\big(\Gamma=(i_1,i_2,\ldots ,i_k)\big)= 2(i_k-i_1) \ ,
\end{equation}
and $A(p,K)=\frac 12$. An exact calculation, reported in Appendix \ref{app-ising1d}, shows that
\begin{equation}\label{deltas_D1}
\Delta S_{(i_1,i_2,\ldots ,i_K)} = (-1)^{K} \, F \left( \exp\Big(-\frac{i_k-i_1}{\xi}\Big) \right) \ ,
\end{equation}
where $F $ is a smooth function given in (\ref{expF}), such that $F(0)=F'(0)=0,F''(0)=-1$. This identity is in agreement with (\ref{asymptotics_deltas}), since the shortest path encircling all sites has length $L(\Gamma) =2 (i_K-i_1)$. Hence, all clusters sharing the same 'extremities, {\em } i.e. the same values of $i_1$ and $i_K$, have the same entropies in absolute value. The sign is determined by the parity of $K$ as mentioned above. Let $i_K-i_1\equiv d$. $\Gamma=(i_1,i_K)$ is the unique cluster of size $K=2$ having its 'extremities' equal to $i_1$ and $i_K$; its entropy is $\Delta S^*_{(i_1,i_K)}= F(\exp(-d/\xi))$. There is $(d-1)$ clusters of size $K=3$ with the same extremities, each having an entropy equal to $-\Delta S^*_{(i_1,i_K)}$. More generally, there are ${d-1\choose K-2}$ clusters of size $K$ with the same extremities, each having an entropy equal to $(-1)^{K-2}\Delta S^*_{(i_1,i_K)}$. The total contribution to the entropy of all those clusters (at fixed extremities $i_1,i_k$) is
\begin{equation}\label{idcan}
\Delta S_{\hbox {\rm fixed}\ i_1,i_k} = \sum _{K=2}^{d+1} (-1)^{K-2}\, {d-1\choose K-2}\, \Delta S^* _{(i_1,i_K)}= (1-1)^{d-1}\;\Delta S^* _{(i_1,i_K)}=\left\{ \begin{array}{c c c}
\Delta S^* _{(i_1,i_K)}&\hbox{\rm if} & d = 1 \ , \\
0 &\hbox{\rm if} & d\ge 2  \ .
\end{array}\right.
\end{equation}
The above calculation nicely exemplifies the cancellation of cluster-entropies. The contributions of all clusters sharing the same extremities exactly compensate each other, unless those extremities are nearest-neighbors on the lattice. We show in Appendix \ref{app-ising1d} that this exact cancellation is a consequence of the existence of a unique interaction path along the unidimensional chain. As a result, in dimension $D=1$, the cross-entropy $S$ is simply the sum of the entropies of the clusters made of nearest neighbours.

In the presence of a reference entropy, $S_0=S_{MF}$, the asymptotic scaling of the cluster-entropy with its contour length $L$ changes, as the dominant contribution coming from loop diagrams is removed from the cluster expansion and absorbed into $S_0$. The subleading contribution to the cluster-entropies is depicted in bold in Fig.~\ref{fig-diag} and derived in Appendix \ref{app-diag}. In dimension $D=1$, formula (\ref{asymptotics_deltas}) is replaced with 
\begin{equation}\label{deltas_D1_bis}
\Delta S_{(i_1,i_2,\ldots ,i_K)} = A'(p,K)\; (-1)^{K-1} \; \exp\big( -3 (i_k-i_1)/\xi \big ) \ .
\end{equation}
Note the sharper asymptotics decay with the distance between the extremities of $\Gamma$ than in the absence of reference entropy. As expected, the terms in the expansion of $S-S_0$ are smaller than the one in the expansion of $S$ alone. Remarkably, the exact cancellation property studied above also holds when the reference entropy is non-zero, as proven in Appendix \ref{app-ising1d}. 
\begin{figure}
\begin{center}
\epsfig{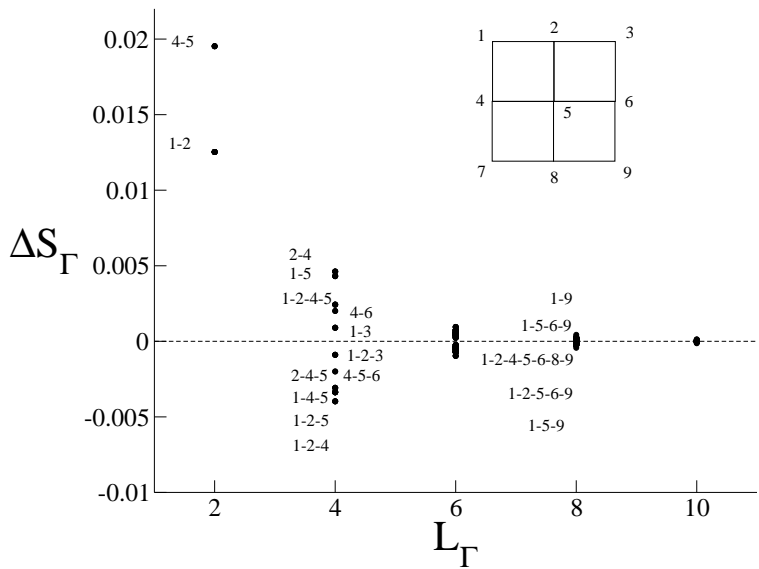}
\caption {Cluster entropy contribution $\Delta S_{\Gamma}$ for a $3\times 3$ grid (top-right) with nearest-neighbour couplings $J=1.777$ (in units of $k_BT$) as a function of  the contour length $L_{\Gamma}$ of the shortest closed path on the lattice joining the spins in $\Gamma$. To illustrate cancellation effects, some $\Delta S_{\Gamma}$  are  labelled with the indices in $\Gamma$, see main text. The values of $J$ and of the fields $h_i=-\frac 1 2 \sum_{j(\neq i)} J_{ij}$ \cite{spin01} are chosen to make the system critical in the infinite grid size limit, see Section \ref{secappligrid}.}
\label{fig-entrovsd}
\end{center}
\end{figure}

In dimension $D=2$ or higher, more than one interaction path connect any two spins, and cluster-entropies with the same contour path do not  cancel exactly as in the $D=1$ case. However, partial cancellations are present. Figure~\ref{fig-entrovsd} shows the values of the cluster-entropies  versus the length of the shortest path, $L(\Gamma)$, for a small bidimensional 3$\times$3 grid.  For such a small system all data ${\bf p}$ and cluster-entropies $\Delta S_\Gamma$ (with up to $K=9$ spins) can be calculated by exact enumeration methods. We observe that:
\begin{itemize}
\item $|\Delta S_{\Gamma}|$ is sensitive to the value of $L_\Gamma$ 
more than to the size $K$ of the cluster;
\item $|\Delta S_{\Gamma}|$ rapidly decreases with  $L_\Gamma$;
\item  the values of the cluster-entropies reflect the structural properties of the lattice, {\em e.g.}  clusters made of central  sites, such as 4-5, have a larger entropy than the clusters including pairs of edge spins, such as 1-2;
\item the sign of $\Delta S_{\Gamma}$ changes with the parity of the size of the cluster.
\end{itemize} 
As a result,  the  contributions to the entropies coming from the clusters sharing the same path, of length $L$,  partially cancel each other. Consider for example the path 1-2-4-5 of length $L=4$; all $7$ clusters that share this path have similar $|\Delta S|$, ranging between $ 0.0024$ and $0.0046$, and so does their sum, $\Delta S_{(2,4)}+\Delta S_{(1,5)} +\Delta S_{(2,4,5)}+\Delta S_{(1,4,5)} +\Delta S_{(1,2,5)}+ \Delta S_{(1,2,4)}+ \Delta S_{(1,2,4,5)}=-0.00242$ \footnote{A further theoretical argument supporting the existence of the cancellation property is, in the case of perfect sampling, the fact that the entropy $S$ must be extensive in $N$. As $S$ is the sum of $\sim 2^N$ cluster entropies, those contributions must compensate each other.}. The sum of the entropies of the clusters sharing the same path is generally of the same order of magnitude as, or even smaller than the single contributions. Figure~\ref{fig-truncation2D} shows that the sum of the $12$ clusters of contour length $L=2$ and of the 4 square--path contributions ($|\Delta S| \ge .0024)$ approximates the entropy within $10^{-6 }$. 

\section{Truncation of the cluster expansion}
\label{sect99}

In this Section we present a truncation scheme for the cluster expansion, which consists in discarding all clusters with entropies smaller than a threshold $\Theta$. We explain why this scheme is efficient, in particular in the presence of sampling noise, and robust against strong correlations in the data (large correlation length). The behavior of the expansion as a function of the threshold is discussed. 
 
\subsection{Schemes for truncating the expansion in the noiseless case}
\label{sectruncscheme}

Expansion (\ref{recur-entro}) for $S({\bf p})$ includes $2^N-1$ terms, and is useless unless an accurate truncation scheme is available. A naive truncation consists in keeping the contributions from the clusters with $\leq K$ spins, where $K$ is an arbitrary size. This procedure was  applied to neurobiological data (with $N \le 40$, $K=7$) in \cite{noi}, which are characterized by large negative fields. However it suffers from two drawbacks. First, the combinatorial growth of the number of clusters with $N$ and $K$ impedes its application to very large systems. Secondly, the truncation does not converge properly with increasing $K$ if the correlation length of the system is large.

As an illustration, consider again the 1D-ferromagnetic Ising model, with correlation length $\xi$. The sign of $\Delta S_{\Gamma}$ alternates with the parity of the size $K$ of $\Gamma$; its modulus decays asymptotically as $\exp(-\Omega d/\xi)$, where $d$ is the maximal distance between any two spins in $\Gamma$ (Section \ref{secpropdeltas}), and $\Omega=2$ if there is no reference entropy ($S_0=0$), $\Omega=3$ if $S_0=S_{MF}$. Let $\Delta S(K)$ be the sum of $\Delta S_\Gamma $ over all the clusters $\Gamma$ with $K$ spins. In the thermodynamic limit ($N\to\infty$), 
\begin{equation}
\frac 1N \; \Delta S (K) \sim  (-1)^{K-1} \sum _{d\ge K-1} {d-1\choose K-2} \exp\big(\! -\Omega\,d/\xi\big)= \frac {(-1)^{K-1}}{\big(\exp ( \Omega/\xi) -1\big)^{K-1}} \ .
\end{equation}
Consider then the series summing all $\frac 1N\Delta S(K)$ with $K\ge 2$. The series is convergent if $\xi <\xi_c=\frac \Omega{\log 2}$, and divergent when $\xi > \xi_c$. In the latter case, for a finite--$N$ system, the maximum of $|\Delta S(K)|$ is exponentially large in $N$, and is reached in $K \simeq \frac N2$. As a consequence, for $\xi> \xi_c$, the sum (\ref{recur-entro}) can not be truncated according to the size of the clusters.  This result is not specific to the dimension unity, and holds for other interaction networks. The expansion of $S({\bf p})$ defines an alternate series, and the order of its terms matters for its convergence in the $N\to\infty$ limit. For an Ising model on a generic lattice with fixed degree (number of neighbours) $v$, the largest value of $\xi$ such that the series (\ref{recur-entro}) (after division by $N$) is absolutely convergent in the $N\to\infty$ limit is $\xi_c = \frac \Omega{\log v}$ (Appendix~\ref{app-bound}).  

\begin{figure}
\begin{center}
\epsfig{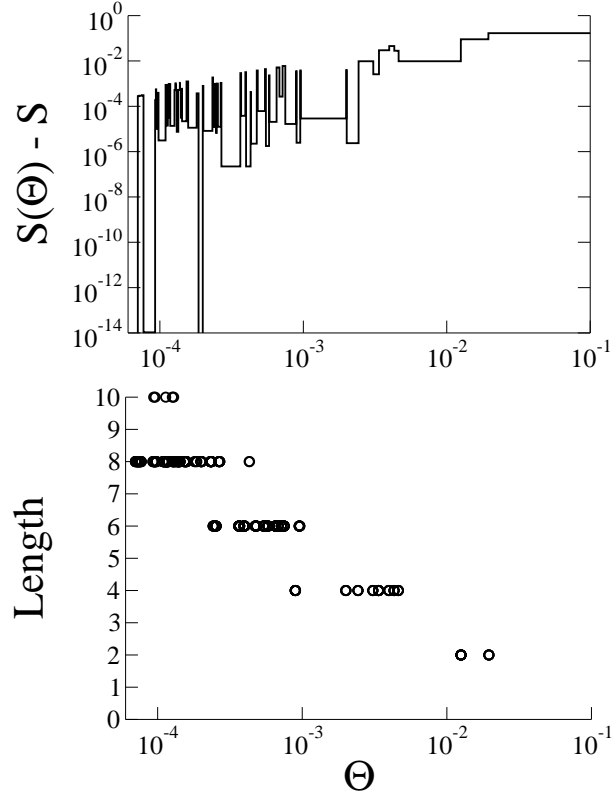}
\caption{Effect of truncation on the bidimensional 3$\times$3 grid with $J=1.777$ (units of $k_BT$) and fields $h_i=-\sum _{j(\ne i)}J_{ij}$ \cite{spin01}, for these parameters values the model has a phase transition between the paramagnetic and  ferromagnetic phase and therefore the correlation length is proportional to the linear size of the system. Top: difference between the truncated and the true cross-entropies as a function of the cut-off on the absolute cluster-entropies, $\Theta$. Bottom: contour lengths $L(\Gamma)$ vs. $\Theta$. The fluctuations of $S(\Theta)-S$ reflect the cancellation phenomenon. Summation of the 12 clusters of nearest-neighbours with $K=2$ and $L=2$ gives  $S(\Theta=0.1)-S\simeq 0.01$, of the 21 clusters contributions corresponding to squared paths, {\em e.g.} 1-2-4-5), with $K=2,3,4$ and $L=4$ gives $S(\Theta=0.002)-S\simeq 10^{-6}$.  Fluctuations arise if only a part of the clusters that share the same interaction path are summed up, and cancellation is incomplete. For instance, fixing $\Theta=0.0025$ discards $(1,2,4,5)$, which has the same interaction path as $(2,4)$.}
\label{fig-truncation2D}
\end{center}
\end{figure}

A better truncation scheme consists in keeping cluster-entropies larger than a threshold $\Theta$ only. Let us define
\begin{equation}\label{truncated1}
S({\bf p},\Theta ) = \sum _{\substack{\Gamma \subset (1,2,\ldots , N)\\ |\Delta S_\Gamma({\bf p}) | > \Theta}} \Delta S_\Gamma ({\bf p})  \ .
\end{equation}
The rationale is that, due to the properties of the cluster entropies and to the cancellation mechanism exposed in Section \ref{secpropdeltas}, summing large cluster-entropies may provide a good approximation to the true value of $S({\bf p})$. In the $D=1$ Ising model case, the exact value of $S({\bf p})$ is, indeed, obtained as soon as $\Theta <\Delta S_{(1,2)}$. We show in Fig.~\ref{fig-truncation2D} the residual error in the cross-entropy due to the truncation as a function of the threshold $\Theta$ for the same small $D=2$ grid as in Fig.~\ref{fig-entrovsd}. The error $S({\bf p},\Theta)-S({\bf p})$ is  very small, and equal to $10^{-6}$ when all clusters with contour length smaller than 4 are taken into account. As $\Theta$ is made smaller, clusters with larger contour lengths are summed up, and the error reaches the numerical accuracy $\sim 10^{-14}$. On top of this trend, positive fluctuations, corresponding to larger errors, arise when not all the clusters with the same interaction path (and length $L$) are summed up, and the cancellation of those contributions is not effective (Fig.~\ref{fig-truncation2D} and caption). We will study in more details this phenomenon in Section \ref{sec-convergence}.

We now explain why the presence of noise in the data provides a compelling argument supporting the introduction of the cut-off $\Theta$.

\subsection{Distribution of  small cluster-entropies in the presence of noisy data}

In this Section, we investigate how limited sampling affects the values of the cluster-entropies. We assume that $B$ configurations $\boldsymbol\sigma ^\tau$ are sampled from the Gibbs distribution of an Ising model with interaction parameters ${\bf J}$ using a Monte Carlo procedure to generate the data ${\bf p}$. 

\subsubsection{Universality at small $|\Delta S|$: numerical evidence}

\begin{figure}[t]
\begin{center}
\epsfig{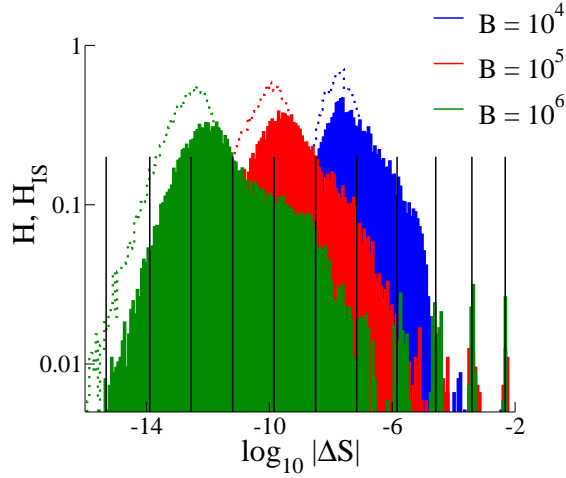}
\caption{Histograms of $\Delta S_{(i,j,k)}$ for the 1D-Ising (full distributions, $p=.02$, $\xi=1$) and Independent Spin (dotted distribution) models, with $N=50$ spins. Each histogram correspond to one random sample of $B$ configurations. Impulses show the histogram for perfect sampling ($B=\infty$), with Dirac peaks located at $\log| \Delta S_\Gamma| = - \frac{3 d_\Gamma}\xi+\hbox{\rm Cst}$. The cut-off at small entropies comes from the finite value of $N$.}
\label{fig-histo-a}
\end{center}
\end{figure}

The empirical correlations, $c_{ij}=p_{ij}-p_{i}p_{j}$, differ from the Gibbs correlations, $\langle \sigma_i\sigma_j\rangle _{\bf J}- \langle \sigma_{i}\rangle _{\bf J}\langle \sigma_{j}\rangle _{\bf J} $, by random fluctuations of amplitude 
\begin{equation}\label{valuecb}
c_B \simeq \frac{p(1-p)}{\sqrt B}\ , 
\end{equation}
where $p$ is the typical value of the $p_i$. For pairs $i,j$ with weak Gibbs correlations ($<c_B$ in absolute value), the experimental correlations are dominated by the noise. As a consequence, the distribution of the cluster-entropies is universal for small $\Delta S$. Its structure is a consequence of the noise in the data, and not of the interaction network of the model used to generate the data.

Figure~\ref{fig-histo-a} shows the histograms $H$ (full distributions) of the entropies $\Delta S_{(i,j,k)}$ for the $K=3$--clusters for a one-dimensional Ising model, for three values of the numbers $B$ of sampled configurations. The histograms are made of two components: a bell-shaped distribution at small $|\Delta S|$, and isolated peaks at larger $|\Delta S|$. The cluster-entropies corresponding to the isolated peaks have the same values as in the perfect sampling case ($B=\infty$, impulses). When $B$ increases, the bell shapes move towards smaller entropies (in the log-scale of Fig.~\ref{fig-histo-a}), and more peaks are unveiled in $H$.

We show also in Fig.~\ref{fig-histo-a} the histograms $H_{IS}$ for a system of Independent Spins (IS), with the same $p_i$'s as the original system, and the same number $B$ of sampled configurations. Contrary to $H$, $H_{IS}$ does not exhibit isolated peaks at well-defined, $B$--independent cluster-entropies. The histograms $H_{IS}$ concentrate around smaller $|\Delta S|$ as the number $B$ of configurations increases. Note that the histograms $H_{IS}$ roughly correspond to the bell-shape parts of the distributions $H$ for the same value of $B$. We have checked that these features are largely independent of the particular sample and of the cluster size, $K$. 

\begin{figure}[h]
\begin{center}
\epsfig{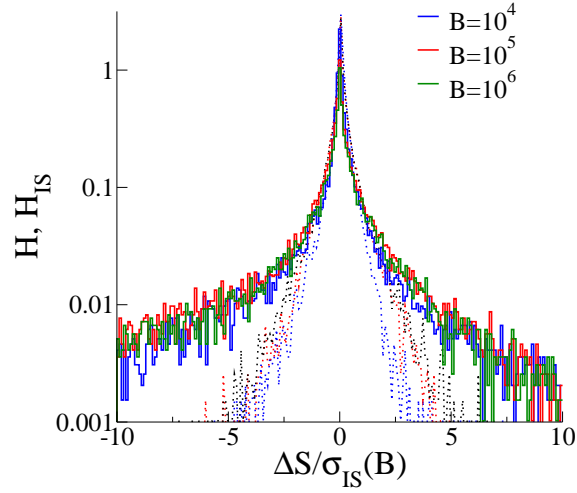}
\caption{Same as Fig.~\ref{fig-histo-a}, after rescaling by the standard deviation $\sigma_{IS}(B)$(\ref{deltasisb}) of the Independent Spin model. Note the linear scale of the $x$-axis. As a result of the presence of the interaction network, the Ising histograms $H$ are asymmetric in $\Delta S\to -\Delta S$ for large values of $\Delta S$, while the IS distributions $H_{IS}$ are obviously symmetric when averaged over the realizations of $B$ configurations (not shown).}
\label{fig-histo-b}
\end{center}
\end{figure}

The histograms $H_{IS}$ depend on $B$ through their standard deviation, $\sigma _{IS} (B)$. The calculation of  $\sigma _{IS} (B)$ from the dominant contribution (\ref{hatsk}) in the diagrammatic expansion of the cluster entropies (Section~\ref{secexpdiag}) is presented in Appendix~\ref{secisdominant}.
We obtain that, for clusters of size $K$ and in the case of uniform averages $p_i=p$ different from 0, $\frac 12$, and 1 \footnote{If $p=\frac 12$, $\Delta S_{IS}$ is of the order of $B^{-K}$}, 
\begin{equation}\label{deltasisb}
\sigma_{IS}(B)\simeq \sqrt{\frac{3^K K!}8}\,\frac{(2p-1)^2} {p(1-p)} \Big(\frac 1B\Big)^{K-\frac 12} \ .
\end{equation}
Figure~\ref{fig-histo-b} shows how the small-entropy regions of the histograms $H$ obtained for different $B$ collapse onto each other once rescaled by $\sigma_{IS}(B)$. As expected, the rescaled $H$ coincide with $H_{IS}$ in the small $|\Delta S| \leq \sigma_{IS}(B)$ region, which concentrates most of the distribution (Fig.~\ref{fig-histo-b}). The universality of the distribution at small  $\Delta S$  is not specific to the one-dimensional Ising model, but holds, in the thermodynamic limit, for all interacting spin systems when the measured connected correlations are corrupted by noise. For a finite system in dimension $D$ with correlation length $\xi$, we expect that the small-$\Delta S$ is universal when $N> {\ell}^D$, where $\ell =\xi \;\log (1/c_B)$. Indeed, the number of large $c_{ij}$ coming out of the noisy background is $\approx N\,{\ell}^D$, while, for most of the $N \choose 2$ pairs of spins $i,j$, the connected correlations have random values of amplitude $c_{B}$. 

The full  distribution $H_{IS}$ can be characterized analytically in the $N\to\infty$ limit. Details can be found in Appendix \ref{secisdominant}. We find the following scalings, depending on the value of the cluster size, $K$:
\begin{eqnarray}\label{his}
H_{IS}(\Delta S) &\sim &\frac 1{|\Delta S|^{2/3}} \ \text{for}\ K=2\;,\  \frac {(-\log \Delta S)^{K-2}}{\sqrt{|\Delta S|}} \ \text{for}\ K\ge 3 \qquad (|\Delta S| \to 0) \ , \nonumber \\
&\sim &\exp\left( - C_2(B)\; |\Delta S|^{2/(2K-1)} (1+o(1))\right) \ \text{for every}\ K\ge 2 \qquad (\text{large}\ |\Delta S| ) \ ,
\end{eqnarray}
where $C_2(B)=2 \times 3^{(K-1)/(2K-1)}\;K^{(2K-3)/(2K-1)}/(2K-1)^{2} \;(\sigma _{\Delta S})^{-2/(2K-1)}$ is proportional to $B$, see Appendix~\ref{secisdominant} and equation (\ref{e22}). The distribution is therefore characterized by a divergence at the origin, and stretched exponential tails. The scalings above were derived with the choice $S_0=S_{MF}$; in the absence of the reference entropy, the stretched exponential has exponent $\frac 2K$ instead of $\frac 2{2K-1}$.

\subsubsection{Finite--$N$ effects and lower bound to the threshold $\Theta$}
\label{secutile}

The discussion about the localized peaks and the bell-shape distribution in $H_{IS}$ in the previous Section is an oversimplification. In reality, for finite systems, large fluctuations of the sampled correlations take place, and no clear-cut boundary exist between cluster-entropies due to the noise and the ones deriving from the interaction network. From extreme value theory \cite{EVT}, the largest value of the correlations are of the order of $c_{ij}^{MAX}=c_B \sqrt{4\log N}$. Therefore, the largest cluster-entropy is, according to (\ref{hatsk}), of the order of
\begin{equation}
\Delta S ^{max}\approx (4 \log N)^{(2K -1)/2} \sigma_{IS} \ .
\label{deltasmax}
\end{equation}

A more detailed calculation to estimate where this fuzzy boundary between the  signal and the noise in the  entropy distribution takes place is presented below. Let $M_K(\Theta)$ be the average number of clusters of size $K$ with entropies $|\Delta S|>\Theta$. According to (\ref{his}), 
\begin{equation} 
M(\Theta) = {N \choose K} \;  \int_\Theta d(\Delta S) \, H_{IS}(\Delta S) \simeq \exp \left( - C_2(B)\; \Theta^{2/(2K-1)} \right)\, \frac{(2K-1)\,N^K\, \Theta ^{(2K-3)/(2K-1)}}{2\, K!\, C_2(B)} \ ,
\end{equation}
for large $\Theta$ and $N$ (compared to $K$). The value of the threshold $\Theta$ such that $M(\Theta) = N^{\alpha}$, with $\alpha<K$, is, to the leading order in $N$,
\begin{equation}\label{thetaseuil}
\Theta (\alpha) \simeq  \left( \frac {K-\alpha}{C_2(B)} \, \log N \right) ^{K - \frac 12} \ .
\end {equation}
In particular, using the formula above for $\alpha=0$, it is likely that no cluster have entropy larger than $\Theta (0)$, in agreement with (\ref{deltasmax}).

\begin{figure}
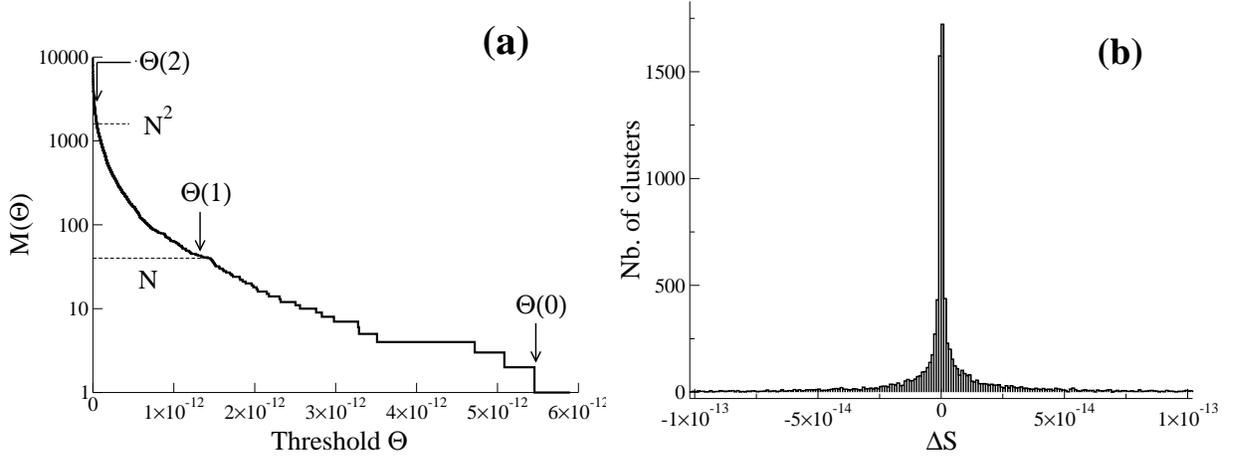

\begin{center}
\epsfig{file=fig9a.eps,width=8cm}
\epsfig{file=fig9b.eps,width=8cm}
\caption{{\bf (a)} Number $M$ of clusters $(i,j,k)$ with $|\Delta S_{(i,j,k}|>\Theta$ as a function of the threshold  $\Theta$, for one realization of $B=10^6$ configurations ($N=40$ independent spins with $p=.0248$). The theoretical values for the threshold,  $\Theta (2)= 5\,10^{-14},\Theta (1)=1.3\,10^{-12}, \Theta(0)=5.5\,10^{-12}$, corresponding to, respectively, $M=N^2,N,1$, are shown.
{\bf (b)} Number of clusters as a function of their entropy $\Delta S$. Same data as in {\bf (a)}, on a smaller entropy scale.}
\label{fig-diag3vssoln40-alea}
\end{center}
\end{figure}

We have tested formula (\ref{thetaseuil}) through a computation based of a system of $N=40$ Independent Spins, with uniform mean $p=.0248$; these parameters were chosen to mimick real data described in~\cite{bialek}. Figure \ref{fig-diag3vssoln40-alea}(a) shows the number of clusters with entropies larger than $\Theta$ in absolute value, for a random set of $B=10^6$ configurations ($K=3$). The theoretical predictions based on (\ref{thetaseuil}) are in very good agreement with the simulations. The vast majority of clusters have entropies smaller than, say, $\Theta (2)$. On a smaller entropy scale, the histogram $H_{IS}$ of the small cluster entropies is strongly concentratred around zero as predicted in \ref{his} (Fig.~\ref{fig-diag3vssoln40-alea}(b)).

As a conclusion, due to the sampling noise, most small cluster-entropies are random quantities, and provide no information about the underlying interactions parameters.  Imposing a threshold $\Theta$ allows one to remove these artifact contributions. A lower bound to the value of $\Theta$ is given by (\ref{thetaseuil}), with, say $\alpha=1$ or $2$. In practice, we will see that higher values of $\Theta$ may be sufficient for an accurate solution of the inverse Ising problem. 

\subsection{Properties of the susceptibility matrix and of its inverse}
\label{sec-chi}

We now present a theoretical argument suggesting that the truncation scheme we have introduced is robust against an increase of the correlation length of the system. More precisely,  the maximal size of the clusters to be summed up to reach an accurate solution of the inverse problem is not directly related to the correlation length, but rather depends on the structure of the interaction graph.

The susceptibility matrix $\boldsymbol\chi$ (\ref{defchi}) characterizes how the observables of the Ising model, such as the averages and correlations in $\bf p$, are modified in response to an infinitesimal change in one or more interaction parameters in $\bf J$.  As far as the inverse Ising problem is concerned, it is more natural to ask the following question. Assume the inverse problem has been solved for a set of data $\bf p$ and the corresponding interations $\bf J$ have been found. Now imagine that the data are slightly changed, ${\bf p}\to {\bf p}+{\boldsymbol\delta\bf p}$. How large will be the resulting change $\boldsymbol\delta \bf J$ in the interactions? The response function characterizing the inverse problem,
\begin{equation}\label{inverseresponse}
\frac{\boldsymbol\delta \bf J}{\boldsymbol\delta \bf p} = - \frac{\partial ^2 S ({\bf p})}{\partial {\bf p} \partial {\bf p}} = \boldsymbol\chi ^{-1} \ ,
\end{equation}
is simply the inverse of the susceptibility matrix $\boldsymbol\chi$. Whether the inverse problem is well-behaved or not will therefore depend on the properties of $\boldsymbol\chi ^{-1}$. In particular, it will depend on the largest eigenvalues of $\boldsymbol\chi ^{-1}$ and on the structure of the corresponding eigenvectors.

A quantity which is  closely related to (\ref{inverseresponse}) in liquid theory is the  Ornstein-Zernike direct correlation function. The direct correlation is widely believed to be short--ranged, as the interaction potential \cite{fisher}. This property is used in closure schemes such as the Percus-Yevick scheme to obtain the equation of state \cite{percusyevick}. We discuss below in details the property of the inverse susceptibility matrix in the case of the spherical model and of the unidimensional Ising model.

\subsubsection{Case of perfect sampling: properties of $\boldsymbol\chi$ and $\boldsymbol\chi^{-1}$}

Consider first the $O(m)$ model, where each site $i=1,2,\ldots , N$ carries a $m$-dimensional real-valued spin vector $\boldsymbol \sigma_i =(\sigma_i^1,\sigma_i^2,\ldots , \sigma_i^m)$, of norm $\sqrt m$.  As usual, two spins, say, $i$ and $j$, are coupled through the dot product of their spin vectors, $-J_{ij}\, \boldsymbol \sigma_i\cdot \boldsymbol \sigma_i$ (units of $k_BT$). Hence the interaction $J_{ij}$ couples the same component ($\alpha$) of the spins in the pair $i,j$. The fields $h_i^\alpha$, with $\alpha=1,2,\ldots ,m$,  are chosen to vanish for simplicity. In the large--$m$ limit the model can be exactly solved \cite{brayandmoore}. The cross-entropy is equal to
\begin{equation}
S({\bf p}) = \frac m2 \; \log \text{det}\ {\bf \hat p} + O\big( \log m \big) \ ,
\end{equation}
where $\bf \hat p$ is the $N\times N$ matrix with diagonal elements $\hat p_{ii}=1$ and off-diagonal elements $\hat p_{ij}$, equal to the average of the product of the components $\alpha$ of spins $i$ and $j$. The elements of the inverse susceptibility matrix are obtained by differentiating $S({\bf p})$ twice with respect to $\bf \hat p$,
\begin{equation}
\big( \chi ^{-1}\big)_{kl,k'l'} =\frac 12 \big ( J_{k,k'}J_{l,l'}+ J_{k,l'}J_{l,k'}\big) \ .
\end{equation}
Hence, the inverse susceptibility has the same structure as the interaction graph. In particular, if the coupling matrix $\bf J$ is sparse (has many zero elements), so is $\boldsymbol\chi ^{-1}$. On the contrary, the susceptibility matrix $\boldsymbol\chi$ is generally not sparse.

The observation above is not specific to spherical spins. Consider now the $D$--dimensional Ising model with $\sigma_i=0,1$ spins on a hypercubic lattice, with nearest neighbour interactions $J_{ij}$. In the $D=1$ case the susceptibility matrix (top right corner in matrix (\ref{defchi})) is non zero for all $i,i'$: $\chi_{i,i}\propto x^{|i-i'|}$, where the proportionality constant does not depend on $i,i'$, and $x=\exp(-1/\xi)$. The inverse susceptibility matrix is a tridiagonal matrix \cite{bor86}: the only non-zero elements are 
\begin{equation}
(\chi^{-1})_{ii}=\frac{1+x^2}{1-x^2}\ \text{and}\ (\chi^{-1})_{i,i\pm1}=-\frac x{1-x^2}\ .
\end{equation}
As in the spherical model, the structure of the inverse susceptibility matrix is the same as the one of the interaction matrix.  In dimension $D\ge 2$ the inverse susceptibility matrix is not, strictly speaking, sparse. However it exhibits a much faster decay with the distance $r=|i-i'|$ than the susceptibility itself \footnote{This statement is widely believed to be true in the theory of liquids literature. The fast decay of the inverse susceptibility, $\chi^{-1}(r)$, or, equivalently, of the direct pair correlation, $g(r)$, is used to approximately close the hierarchy of correlation functions. The Percus-Yevik closure scheme, which gives an accurate equation of state for liquids of hard spheres, assumes that the inverse susceptibility vanishes above the interaction range of the potential (diameter of a particle).}. At the critical point, the latter decays as $\chi(r) \sim r^{-(D-2+\eta)}$, where the critical exponent attached to the decay of the spin-spin correlation, $\eta$, vanishes in dimension $D\ge 4$, and is positive and small in dimension $D\le 3$, {\em i.e.} $\eta =\frac 14$ for $D=2$. The inverse susceptibility is the Laplacian in dimension $D\ge 4$, a purely local operator, and decays as $\chi^{-1}(r) \sim r^{-(D+2-\eta)}$ for $D\le 3$. While both quantities decrease as power laws in $r$, the inverse susceptibility has a much sharper decay than the susceptibility itself. In particular, the integrated contribution to the susceptibility coming from distances larger than $R$, 
\begin{equation}
\int_{R}^{\infty} d^Dr\; \chi (r)=R^{1-\eta}  \ , 
\end{equation}
diverges when $R\to \infty$, while the same quantity calculated for the inverse susceptibility,
\begin{equation}
\int_{R}^{\infty} d^Dr\; \chi ^{-1}(r)=\frac 1{R^{3-\eta}}  \ ,
\end{equation}
tends to zero as $R\to \infty$. This fact is a good news for the inverse problem. According to (\ref{inverseresponse}) the error on the field $h_i$ done when discarding all the spins at distance $R > \epsilon ^{-1/(3-\eta)}$ is of the order of $\epsilon$ only. In this regard, the inverse Ising problem remains local even at the critical point.

While the discussion above is related to the response of a field $h_i$ to a change in the average $p_{i'}$ of spin $i'$, the response of a coupling $J_{kl}$ following a modification of the 2-point average $p_{k'l'}$, see (\ref{defchi}), is also of interest. Unfortunately, to our best knowledge, this quantity has not been studied in the case of the Ising model so far. As a first step, we focus here on the $D=1$-Ising model with uniform nearest-neighbour interactions, and in the thermodynamical limit ($N\to\infty$). The four-spin connected correlation function is, up to a $p$-dependent multiplicative constant, equal to
\begin{equation}\label{chi4}
\chi_{ij,kl}= x^{i_4-i_3+i_2-i_1}-x^{j-i+k-l}\ ,
\end{equation}
where $i_1\le i_2\le i_3\le i_4$ are the same indices as $i,j,k,l$ but sorted in increasing order, and $x=\exp(-1/\xi)<1$. We show in Appendix \ref{app-chiquatre} that the inverse susceptibility matrix is given by
\begin{equation}
\big(\chi^{-1}\big)_{ij,kl}= \left\{\begin{array}{c c c} 
\frac{(1+x^2)^2}{(1-x^2)^2} &\text{if} & i=k, j=l \ \text{and}\ j\ge i+2\ ,\\
\frac{1+x^2+x^4}{(1-x^2)^2} &\text{if} & i=k, j=l \ \text{and}\ j= i+1\ , \\
-\frac{x(1+x^2)}{(1-x^2)^2} &\text{if} & i=k\pm 1, j=l \ \text{or}\ i=k, j=l\pm 1 \ ,\\
\frac{x^2}{(1-x^2)^2} &\text{if} & i=k\pm 1, j=l\pm 1\ , \\
0 & \text{otherwise} & \ .
\end{array} \right. \label{invchi4}
\end{equation}
Hence, the inverse susceptibility matrix is sparse, with at most $9$ non-zero elements per line, while the dimension of the matrix is $\frac 12N(N-1)\to\infty$.
In dimension $D\ge 2$, we do not expect $\boldsymbol\chi^{-1}$ to be sparse. However we conjecture that $\big(\chi^{-1}\big)_{ij,kl}$ decays quickly with the minimal distance between the four points $i,j,k,l$ (each index, {\em e.g.} $j$, is now a $D$-dimensional vector). 

\subsubsection{Influence of the sampling noise on the norms of $\boldsymbol\chi$ and $\boldsymbol\chi^{-1}$}

\begin{figure}
\begin{center}
\epsfig{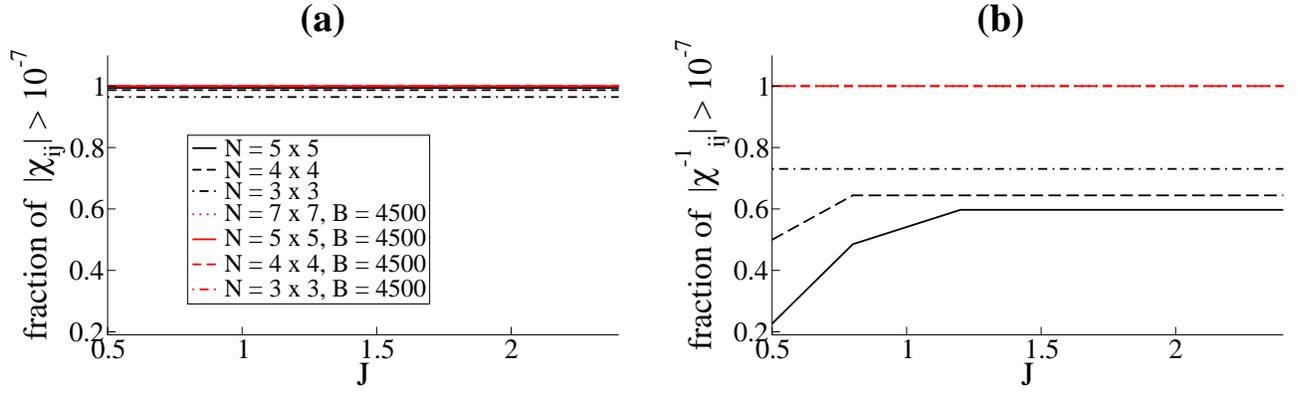}
\caption{Fraction of elements larger than $10^{-7}$ (in absolute value) for the susceptibility $\boldsymbol\chi$ {\bf (a)} and the inverse susceptibility $\boldsymbol\chi ^{-1}$ {\bf (b)} matrices vs. strength $J$ of the nearest-neighbour coupling. The sizes $N$ of the grids are indicated. Data were obtained from exact numerations for sizes $3\times 3$, $4\times 4$, $5\times 5$ (perfect sampling, black) and from Monte Carlo  simulations for all sizes (one realization of $B=4500$ configurations, red/gray). Periodic boundary conditions were used.}
\label{fig-smallgrid1}
\end{center}
\end{figure}

To corroborate this statement we have carried out exact numerical analysis of small bidimensional grids (Section \ref{secpropdeltas}). We show in Fig.~\ref{fig-smallgrid1}(a) the fraction of elements $\chi_{ij,kl}$ of the susceptibility matrix larger than $\epsilon = 10^{-7}$ in absolute value (the largest elements have magnitude $\sim 1$). This fraction is closed to 1 for all the values $J$ of the coupling we have studied. As expected, the inverse susceptibility matrix has many more small elements (Fig.~\ref{fig-smallgrid1}(b)). In addition, the fraction  of entries in $\boldsymbol\chi^{-1}$ smaller than $\epsilon$ seem to increase with the size $N$ of the grid.

In the presence of noise in the sampling process the inverse matrix $\boldsymbol\chi^{-1}$ loses its quasi-sparse structure. More precisely, for the number $B$ of sampled configurations chosen in Fig.~\ref{fig-smallgrid1}(b), all the elements $(\chi^{-1})_{ij,kl}$ are larger than $\epsilon$ in absolute value. Indeed, the quasi-sparsity $\boldsymbol\chi^{-1}$ in the perfect sampling case reflects the sparse structure of the underlying interaction matrix. When data are corrupted by noise, the Ising model (over)fitting the data has no reason to be sparse anymore, and neither has the inverse susceptibility. 

\begin{figure}
\begin{center} 
\epsfig{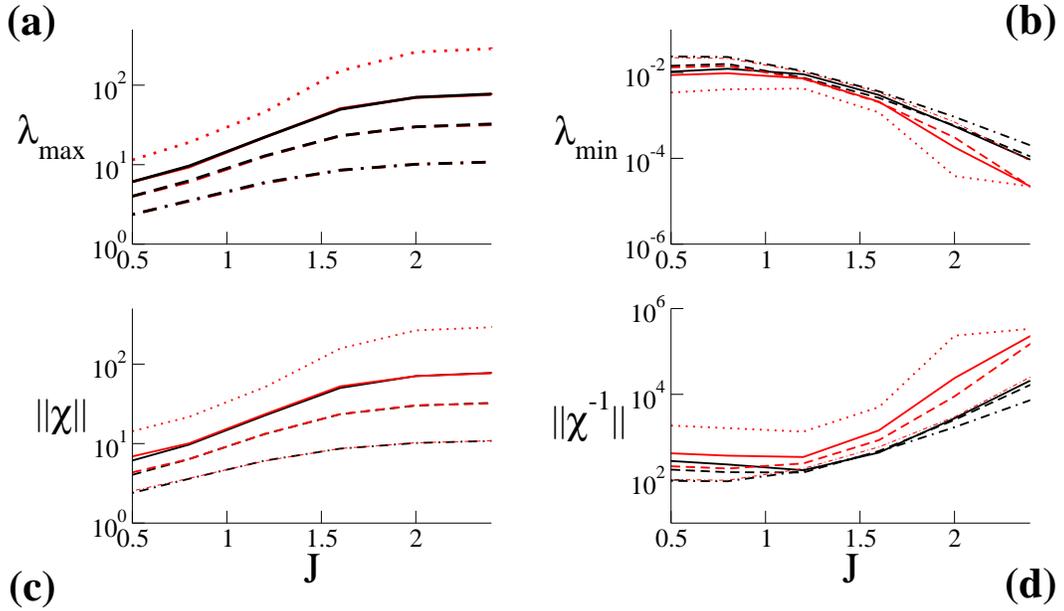}
\caption {Largest {\bf (a)} and smallest {\bf (b)} eigenvalues of the susceptibility matrix, and norms of $\boldsymbol\chi$ {\bf (c)} and of its inverse $\boldsymbol\chi^{-1}$ {\bf (d)} as functions of the coupling strength $J$ for the $3\times 3$ grid. See Fig.~\ref{fig-smallgrid1} for explanations regarding the color code and the line styles. }
\label{fig-smallgrid2}
\end{center}
\end{figure}

The influence of the sampling noise on the susceptibility matrix and on its inverse can be measured through the largest and smallest eigenvalue of $\boldsymbol\chi$, denoted by, respectively, $\lambda_{max}$ and $\lambda_{min}$. According to Figs.~\ref{fig-smallgrid2}(a,b), we have that:
\begin{itemize}
\item $\lambda_{max}$ increases with the size of the system (we expect $\lambda_{max}$ to diverge at the critical coupling $J\simeq 1.778$ in the thermodynamical limit), but is not affected by the sampling noise (the black and red/gray curves associated to the same size are nearly indistinguishable in Fig.~\ref{fig-smallgrid2}(a)).
\item $\lambda_{min}$ is not strongly affected by the system size in the case of perfect sampling. In case of noisy sampling, $\lambda_{min}$ acquires a smaller value. The effect of the noise increases with the system size (Fig.~\ref{fig-smallgrid2}(b)).
\end{itemize}
Those facts are observed from the study of the norms of the two matrices $\boldsymbol\chi$ and $\boldsymbol\chi ^{-1}$. Here, we define the norm of the matrix $A$ through
\begin{equation}\label{defnorm}
||A|| = \max_i \sum_{j} |A_{i,j}| \ .
\end{equation}
Figures~\ref{fig-smallgrid2}(c,d) show that the behaviours of the norms $||\boldsymbol\chi ||$ and $||\boldsymbol\chi ^{-1}||$ are, from a qualitative point of view, similar to the ones of, respectively, $\lambda_{max}$ and $1/\lambda_{min}$.
However the norms are directly related to the magnitudes of the elements of the matrices, according to (\ref{defnorm}). The independence of $||\boldsymbol\chi ^{-1}||$ from the size $N$, contrary to the strong increase of $||\boldsymbol\chi ||$, supports the notion that most elements of $\boldsymbol\chi ^{-1}$ are very small (or even zero) in the case of perfect sampling. This property is lost when the sampling is not perfect: the presence of noise in the correlation makes the norm $||\boldsymbol\chi ^{-1}||$ increases with $N$ (Fig.~\ref{fig-smallgrid2}(d)). 

\subsection{Dependence of the truncated entropy on the threshold}
\label{sec-convergence}

Hereafter, we study how the error on the entropy $S(\Theta)$ resulting from the truncation varies with the threshold $\Theta$ and we discuss the fluctuations of $S(\Theta)- S$ observed in Fig.~\ref{fig-truncation2D}. We start by sorting the absolute values of the cluster-entropies $|\Delta S_\Gamma|$  in decreasing order:
\begin{equation}
\Delta S_1 \ge \Delta S_2\ge \Delta S_3 \ge \ldots \ge \Delta S_n \ge \ldots \ge 0 \ .
\end{equation}
We call $\eta _n =\pm 1$ the sign of the cluster-entropy $\Delta S_\Gamma$ attached (equal in absolute value) to $\Delta S_n$. Given the threshold $\Theta$, we define $n^*(\Theta)$ as the index of the smallest cluster-entropy larger than  $\Theta$: $\Delta S_{n^*(\Theta)} \ge \Theta > \Delta S_{n^*(\Theta)+1}$. The truncated entropy (\ref{truncated1}) can be rewritten as $S({\bf p},\Theta )=\Sigma(\Theta)$, where
\begin{equation} \label{truncated2}
\Sigma (\Theta ) = \sum _{n =1}^{n^*(\Theta)} \eta _n \; \Delta S_n  \ .
\end{equation}
We want to study how $\Sigma (\Theta)$ behaves when $\Theta$ is made small. In particular, how does the difference $\epsilon_s(\Theta) = \Sigma (\Theta) - \Sigma(0)$  behave as a function of $\Theta$? Is it a smooth function, or does it exhibit large and irregular fluctuations? From a mathematical point of view, it is convenient to imagine that $N\to\infty$. The above question can be formalized as whether $\frac 1N\Sigma (\Theta)$ converges to some limit value; the normalization factor comes from the fact that we expect the cross-entropy to be extensive in the system size $N$. Depending on the system under consideration, different situations can be encountered.

The most favorable case is when 
\begin{equation}\label{condconv}
\lim _{N\to\infty} \frac 1N \sum _{n\ge 1} \Delta S_n < \infty \ .
\end{equation}
If this condition holds, the difference $\epsilon_s(\Theta )$ can be made arbitrarily small if $\Theta$ is small enough. An illustration is provided by the one-dimensional Ising model with small correlation length $\xi$ and perfect sampling ($B=\infty$). For this model, the sequence of $\Delta S_n$ is highly degenerate, and its distinct values are in one-to-one correspondence with the integer distances $d\ge 1$ between the extremities of the clusters 
(Fig.~\ref{fig-histo-a}). The cluster-entropy $\Delta S(d)$ asymptotically decays as $\exp(-3d/\xi)$, and has multiplity $2^{d-1}$, since each point between the extremities may or may not belong to the cluster. We find
\begin{equation}
\frac 1N \sum _{n\ge 1}\Delta S_n \simeq \sum _{d\ge 1} 2^{d-1}\;  \exp(-3d/\xi) \ ,
\end{equation}
which converges if $\xi < \xi_c=\frac 3{\log 2}$. The calculation above is very similar to the one of Section \ref{sectruncscheme}. Indeed, when the series with general term $\Delta S(K)$  is absolutely convergent, any ordering of the cluster-entropies is possible. In particular, one is allowed to sum all the clusters of a given size $K$ as proposed at the beginning of Section \ref{sectruncscheme}.

\begin{figure}
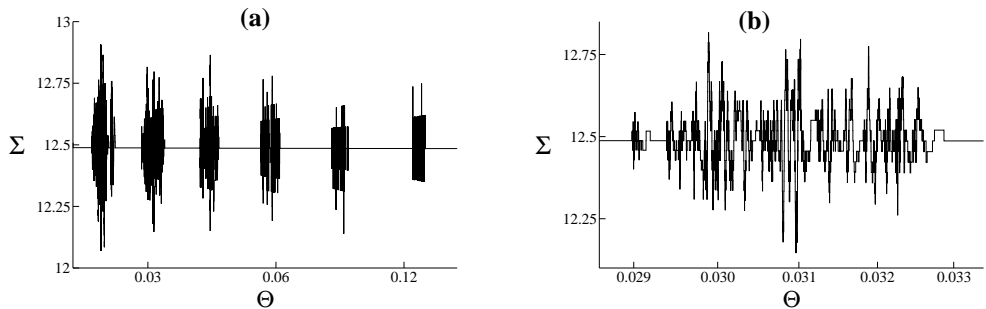

\begin{center}
\epsfig{file=fig12a.eps,width=6cm}\hskip 1cm
\epsfig{file=fig12b,width=6cm}
\caption{{\bf (a)} Sum $\Sigma$ of the cluster-entropies larger than $\Theta$ (in absolute value) for the nearest neighbour one-dimensional Ising model with $\xi \simeq 8.97 \gg \xi_c\simeq 4.33$, and $B=10^5$ configurations. The initial increase from zero, taking place at small $\Theta \simeq \Delta S(1)\simeq 0.41$, is not shown. {\bf (b)}~magnification of {\bf (a)} in the range $.029<\Theta<.033$. Within the random sign model, the behavior of $\Sigma$ within a packet is similar to a Brownian bridge, see main text.}
\label{fig-packets}
\end{center}
\end{figure}

What happens when condition (\ref{condconv}) is violated? Again consider the one-dimensional Ising model. For perfect sampling, the cancellation property discussed in Section \ref{secpropdeltas} ensures that $\frac 1N\Sigma (\Theta)$ has reached its limit $\Delta S(d=1)$ as soon as $\Theta < \Delta S(1)$. In the case of noisy sampling (finite $B$), the situation is more complex. In the presence of noise in the correlations $c_{kl}$ the cluster-entropies with the same distance $d$ between extremities are not degenerate any longer. We show in Fig.~\ref{fig-packets}(a) the value of $\Sigma$ as function of $\Theta$ for a large correlation length $\xi$ compared to $\xi_c$, and $B=10^5$ sampled configurations. We observe the appearance of 'packets' of cluster-entropies, located around the noiseless values $\Delta S(d\ge 2)$. The width of a packet depends on the amount of noise due to the sampling, {\em i.e.} on the number $B$ of sampled configurations. The values of $\Sigma$ at the two edges of the packet are very close to one another due to the cancellation property. As $\Theta$ spans the range of cluster-entropies in the packet, $\Sigma$ fluctuates. The maximal amplitude of the fluctuations seems to weakly increase as we look at packets with smaller and smaller entropies (Fig.~\ref{fig-packets}(a)).

\begin{figure}
\begin{center}
\epsfig{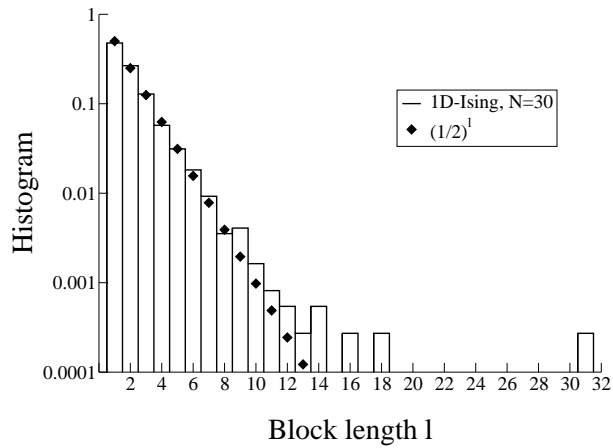}
\caption{Frequences of the block length $\ell$ for the 1D-Ising model, with $N=30$ spins, $\xi=8.96$, and one set of $B=10^5$ sampled configurations. The statistics takes into account the cluster-entropies used to draw Fig.~\ref{fig-packets}(a) only.} 
\label{fig-blocklength}
\end{center}
\end{figure}

We have analyzed the statistics of the signs $\epsilon_n$ of the cluters -entropies in (\ref{truncated2}). Writing the sequence of signs $\boldsymbol\eta=(\eta_1,\eta _2,\eta _3 , \ldots )$, we consider the blocks $j$ of contiguous and equal signs, and defines their lengths $\ell_j$. For instance, the block lengths corresponding to $\boldsymbol\eta=(+,+,-,+,+,+,+,-,-,-,-,-,+,-, ....)$ are $\ell_1=2,\ell_2=1,\ell_3=4,\ell_4=5,\ell_5=1, ...$. The histogram of the block-lengths is shown in Fig.~\ref{fig-blocklength}. The two main features are:
\begin{itemize}
\item The frequence of $\ell$ decreases exponentially when $\ell \ll N $, and is in very good agreement with the exponential law $\left( \frac 12\right)^\ell$. 
\item  A large 'structural' block of length $\ell\simeq N$ is present. This block corresponds to the $N$ clusters of size $K=2$ (having all sign $+$), and the cluster of size $K=3$ with largest entropy (which has the same sign $+$). 
\end{itemize}
We have calculated the correlation between successive block lengths, normalized by the variance of the block length,
\begin{equation}
\rho = \frac{\overline{l_{j} l_{j+1} } - \overline{l_{j} }^2}{\overline{l_{j} ^2} - \overline{l_{j} }^2} \ ,
\end{equation}
where $\overline{(\cdot) }$ denotes the average over the blocks $j$. For the model and the data shown in Figs.~\ref{fig-packets} and \ref{fig-blocklength}, we find $\rho \simeq .012$. Changing the set of sampled configurations does not affect the amplitude of the ratio $\rho$, which is always found to be about $1\%$. This ratio coincides with the inverse of the square root of the number of blocks, equal to a few thousands. Hence, the analysis is compatible with the absence of any correlation between the lengths of successive blocks. The same conclusion is reached with experimental data, {\em e.g.} multi-electrode recordings of the activity of a neural population \cite{bialek,schni} (not shown).

The simple statistics sets above suggests the following 'random sign' model, allowing us to deepen our theoretical understanding of the behavior of $\Sigma(\Theta )$. In the random sign model, the signs $\eta_n$ are replaced with random variables, equal to $\pm$ with probabilities $\frac 12$, and independent from each other. We emphasize that, in $\Sigma (\Theta)$ defined in (\ref{truncated2}), the signs are deterministic (for given data {\bf p}). The random sign model is therefore an approximation motivated by the statistical analysis above. Assume now that the value chosen for the threshold $\Theta$ falls within a packet $p$ including ${\cal N}_p$ clusters. Fluctuations of the order of  
\begin{equation}
\Delta \Sigma\sim \pm \Theta \sqrt{{\cal N}_p} 
\end{equation} 
are expected on the entropy. As $\Theta$ decreases, the size of the packets, ${\cal N}_p$, tends to be bigger. Loosely speaking, smaller entropies correspond to longer interaction paths, shared by many more clusters. In the case of the one-dimensional Ising model, as $\Theta$ decreases, the distance between the extremities of the clusters involved in a packet, $d$, increases. We have ${\cal N}_p =2^{d-1}$; hence, $ \Delta \Sigma\sim \exp(-3d/\xi)\sqrt{2^d}$. We conclude that the error on the entropy tends to zero if $\xi < 2\,\xi_c$.

From the above discussion, it appears that a general, sufficient condition for the amplitude of the fluctuations to vanish as $\Theta\to 0$ is
\begin{equation}\label{ds2cv}
\lim_{N\to\infty} \frac 1N  \sum _n (\Delta S_n)^2  < \infty \ .
\end{equation}
Indeed, if condition (\ref{ds2cv}) is fulfilled, the sum of the fluctuations due to {\em all} packets corresponding to cluster-entropies smaller than $\Theta$ is guaranteed to vanish with $\Theta$. Hence condition (\ref{ds2cv}) not only ensures that the fluctuations $\Delta \Sigma$ attached to the packet 'cut' by $\Theta$ vanishes, but also that the error on the entropy, $\epsilon_s(\Theta)$, tends to zero when $\Theta\to 0$. It is important to realize that the  guarantee is of probabilistic nature. Arbitrary large fluctuations are possible (in the $N\to\infty$ limit), but are very unlikely. More precisely, within the random sign model, the error is a normal variable, 
\begin{equation} \label{rserror}
\epsilon _s (\Theta) = {\cal N}\bigg( 0, \frac 1N \sum_{n > n^*(\Theta) }  (\Delta S_n)^2 \bigg) \ ,
\end{equation}
with a variance vanishing with $\Theta$ according to
(\ref{ds2cv}). The true error is expected to be even smaller than the random sign estimate (\ref{rserror}). Indeed, packets need not be isolated from each other as in Fig.~\ref{fig-packets}. In the presence
of a strong sampling noise, or in higher dimension than $D=1$, packets
will overlap. As a consequence, the number of packets 'cut' by the
threshold and their size will determine the amplitude of $\Delta
\Sigma$. Further investigations of those points are needed.

\section{Adaptive algorithm for the inverse Ising problem}
\label{sect98}

\subsection{Procedure to construct and select clusters}

As explained above discarding the cluster-entropies smaller than a threshold $\Theta$ is an efficient step against overfitting of the sampling noise. In addition, for systems with dilute and strong interactions, we expect that only the clusters of neighboring sites on the interaction network will have substantial entropies. These arguments provide a heuristic basis for the threshold-based truncation of the expansion (\ref{recur-entro}). 

How can we implement the truncation scheme in practice? The combinatorial explosion of the number of clusters of size $K$ among $N$ sites impedes any brute force computation approach, as soon as $N$ is larger than a few tens. Even for small--$N$ system for which it is feasible, computing $\sim 2^N$ cluster-entropies and, then, discarding most of them does not sound like an efficient procedure.

We propose below an alternative approach, based on a recursive and selective construction of relevant clusters. The approach is based on the principle that clusters with large entropies should be compatible with the interaction network to be inferred. Suppose that two clusters $\Gamma$ and $\Gamma '$ have both large entropies, and share most of their spins. Then, the union $\Gamma\cup \Gamma'$ is a good candidate for a bigger cluster. If the entropy of the union cluster is large, a new part of the interaction network will be unveiled. Conversely, if it is small, no new interaction path with respect to the one discovered from $\Gamma$ and $\Gamma'$ separately exists. Hence, combining strongly overlapping clusters should allow us to progressively deepen our knowledge of the local structure of the interaction graph.

The above heuristics is formalized as follows:

\begin{enumerate}
\item[{\bf A1.}] Initial step: build the list of all clusters of size one: $L_1=\{ (i): i=1,2,\ldots , N\}$. All the other lists $L_K$ for $K\ge 2$ are empty.
\item[{\bf A2.}] Iteration: assume the current size of clusters is $K\ge 1$, {\em i.e.} $L_K$ is not empty while $L_{K+1}$ is empty. For every pair $\Gamma_1,\Gamma_2$ in $L_K$:
\subitem {\bf A21.} Construction: build $\Gamma = \Gamma _1\cup \Gamma_2$ 
\subitem {\bf A22.} Selection: if $\Gamma$ is of size $K+1$ and if $|\Delta S_\Gamma({\bf p})|\ge\Theta$, then select $\Gamma$ and add it to $L_{K+1}$.  
\item[{\bf A3.}] Recursion: if at least one cluster has been selected, then add 1 to $K$, and go to step 2 to pursue the construction process. If no cluster has been selected, the construction process is over.
\end{enumerate}

The first condition in {\bf A22} is about the size of $\Gamma$. The union of two clusters of size $K$ has size $K+1$ if and only if they have exactly $K-1$ common spins. $\Gamma_1=(i_1,i_2\ldots i_{K-1},x)$ and $\Gamma_2=(i_1,i_2,\ldots,i_{K-1},y)$ can be merged into $\Gamma = (i_1,i_2,\ldots, i_{K-1},x,y)$; the ordering of  $x$, $y$, and of the $i_l$'s is irrelevant here.

\subsection{Calculation of $\Delta S_\Gamma({\bf p})$}
 
Step {\bf A22} requires the calculation of the cluster-entropy $\Delta S_\Gamma({\bf p})$ for each selected cluster $\Gamma $ (of size $K$). In order to do so we make use of the formula
\begin{equation}\label{recur-entro2}
\Delta S_\Gamma({\bf p}) = S_\Gamma ({\bf p})- (S_0)_{\Gamma}({\bf p}) - \sum _{\substack{\Gamma ' \subset \Gamma \\(\Gamma' \ne \Gamma)}}  \Delta S_{\Gamma '}({\bf p})\ ,
\end{equation}
which can be easily deduced from (\ref{recur-entro}).
The procedure is as follows:
\begin{enumerate}
\item[{\bf B1.}] calculate the subset-entropy $S_\Gamma ({\bf p})$ through the minimization of $S_{Ising}({\bf J}|{\bf p})$ (\ref{s2}) with respect to the fields and couplings. The partition function $Z[{\bf J}]$ is computed as the sum over the $2^K$ configurations of the spins in $\Gamma$. 
\item[{\bf B2.}]  substract the reference entropy $(S_0)_{\Gamma}({\bf p})$. For the mean-field reference entropy, $(S_0)_\Gamma ({\bf p})= \frac 12 \log \text{det} M_{\Gamma}({\bf p})$, according to formula (\ref{loopentro}); $M_\Gamma({\bf p})$ is the $K\times K$ restriction of matrix $M({\bf p})$ to the indices $i_1,i_2,\ldots , i_K$ in $\Gamma$. In presence of a regularization term (\ref{regul2}) equation (\ref{loopentrogamma}) has to be  used instead of  (\ref{loopentro})  to calculate $(S_0)_{\Gamma}({\bf p})$.
\item[{\bf B3.}]  Substract the entropies $\Delta S_{\Gamma '}({\bf p})$ of all the sub-clusters $\Gamma '$ of size $K'<K$, included in $\Gamma$. 
\end{enumerate}
The last step ({\bf B3}) assumes that the entropies of all the sub-clusters of $\Gamma$ are known, {\em i.e.} have been computed  at a previous step in the algorithm. This is true for $K'=2$, but not necessarily so for $K'\ge 3$. To circumvent this difficulty we maintain at all times during the execution of the algorithm the list $L_{all}$ of all the clusters and of their entropies calculated so far; $L_{all}$ is a larger list than the one of the selected clusters (union of all $L_K$). The procedure to compute $\Delta S_\Gamma({\bf p})$ is then: 
\begin{enumerate}
\item[{\bf B0.}] build the list $\hat L$ of all the sub-clusters $\Gamma '$ in $\Gamma$ {\em not} already present in $L_{all}$. For each $\Gamma'\in \hat L$,  starting from the smallest sub-cluster and ending up with the largest one, run steps {\bf B1}, {\bf B2}, {\bf B3} to obtain $\Delta S_{\Gamma '}({\bf p})$, and add $\Gamma'$ and its entropy to the list $L_{all}$. 
\end{enumerate}
The ordering of $\hat L$ ensures that all the sub-clusters of $\Gamma'$ required to calculated its entropy are in $L_{all}$ when step {\bf B3} is executed.

\subsection{Calculation of the cross-entropy, couplings and fields}

Once the construction process is finished, the list $L_{sel}=L_1\cup L_2\cup L_3\cup \ldots \cup L_{ K_{max}}$ of all selected clusters is available. Here, $K_{max}$ is the size of the largest cluster selected by the construction procedure. We then
\begin{enumerate}
\item[{\bf C1.}] 
estimate the cross-entropy through
\begin{equation}
S({\bf p}) = S_0 ({\bf p}) +  \sum _{\Gamma \in L_{sel}} \Delta S_{\Gamma}({\bf p}) \ .
\end{equation}
\end{enumerate}
Next we need to estimate the values of the fields and of the couplings, solution to the inverse Ising problem. One possibility would be to use recursion relations similar to (\ref{recur-entro2}) for $\Delta h_{i,\Gamma}({\bf p})$ and $\Delta J_{ij,\Gamma}   ({\bf p})$, that is, the contributions to, respectively, the field $h_i$ and the coupling $J_{ij}$ coming from the cluster $\Gamma$. Next we could sum up those contributions over the clusters included in $L_{sel}$. However, to save memory space, it is possible to resort to the following, alternative procedure:
\begin{enumerate}
\item[{\bf C2.}] define the 'multiplicities' $m_\Gamma$ of the subsets $\Gamma$ through: 
\subitem {\bf C21.} let $L_{sub}$ be the list of all clusters in $L_{sel}$ {\em and} of all their subsets. Initialize $m_\Gamma=0$ for every $\Gamma \in L_{sub}$. 
\subitem {\bf C22.} For each $\Gamma \in L_{sel}$, and for each $\Gamma'\subset \Gamma$, add $ (-1)^{K-K'}$ (see (\ref{mobius})) to $m_{\Gamma'}$, where $K,K'$ are the sizes of, respectively, $\Gamma,\Gamma'$. The sub-clusters $\Gamma'=\Gamma$ must be taken into account in the addition process. 
\item[{\bf C3.}]  estimate the fields and the couplings through
\begin{eqnarray}\label{final-hj}
h_i ({\bf p}) &=& (h_0)_{i} ({\bf p})+ \sum _{\Gamma \in L_{sub} :  (i)\subset \Gamma}  m_\Gamma\, \Big( h_{i,\Gamma}({\bf p})-(h_0)_{i,\Gamma}({\bf p})\Big) \ , \nonumber \\
J_{ij} ({\bf p})&=&(J_0)_{ij} ({\bf p})+ \sum _{\Gamma \in L_{sub} : ( i,j)\subset \Gamma}   m_\Gamma\, \Big(J_{ij,\Gamma}   ({\bf p})- (J_0)_{ij,\Gamma}   ({\bf p}) \Big) 
\ .
\end{eqnarray}
\end{enumerate}
The fields $h_{i,\Gamma}$ and the couplings $J_{ij,\Gamma}$ in step {\bf C3} above are the ones obtained through the minimization of $S_{Ising}({\bf J}|{\bf p})$ over ${\bf J}=\{h_{i,\Gamma},J_{ij,\Gamma}\}$ in step {\bf B1}. The fields $(h_0)_{i}$ and the couplings $(J_0)_{ij}$ are (minus) the derivatives of the reference entropy $S_0({\bf p})$ with respect to $p_i$ and $p_{ij}$, see formulas (\ref{hJMF}). The fields $(h_0)_{i,\Gamma}$ and the couplings $(J_0)_{ij,\Gamma}$ are their counterparts for the subset $\Gamma$ only, {\em i.e.} the derivatives of $(S_0)_\Gamma({\bf p})$; their expressions are given by (\ref{hJMF}) again, upon substitution of the $N\times N$ matrix $M({\bf p})$ with the $K\times K$ matrix $M_\Gamma({\bf p})$ restricted to the $K$ elements of $\Gamma$ only. 

\subsection{Pseudo-code of the algorithm}

We now give the pseudo-code useful for the implementation of the procedures above. To improve the readability the code is broken into several parts.

We start with Algorithm 1, which computes the cross-entropy and the reference entropy for a subset $\Gamma$. The energy function $H_{Ising}$ is defined in (\ref{h2}). The minimization over $\bf J$ can be done using standard numerical algorithms for convex functions. A speed-up is generally obtained when we start with ${\bf J}_{MF}$, the value of the interaction parameters obtained from the MF approximation (\ref{hJMF}), as an initial guess for the value of $\bf J$ \cite{john}. In the absence of regularization, the parameter $\gamma$ is set to 0. It is straightforward to change the pseudo-code to introduce the $L_1$-regularization instead of the $L_2$-norm, see formulas (\ref{regul1}).

\begin{algorithm}[h]
\caption{Computation of entropy $S_\Gamma({\bf p})-(S_0)_{\Gamma} ({\bf p})$}         
\label{alg1}                          
\begin{algorithmic}                    
\REQUIRE $\Gamma$ (of size $K$), data {\bf p}, regularization parameter $\gamma$
\STATE
\STATE {\em Computation of $S_\Gamma$:}
\STATE \hskip .5cm Define $\displaystyle{\big(S_{Ising}\big)_\Gamma  [{\bf J}|{\bf p}]\leftarrow \log \left( \sum _{\boldsymbol\sigma\in \{0,1\}^K}\exp \big ( -H_{Ising}[{\boldsymbol\sigma}|{\bf J}]\big)
\right)-\sum _{i\in \Gamma} h_i \,p_i -\sum _{i<j \in \Gamma} J_{ij}\, p_{ij}} + \gamma \sum _{i<j} J_{ij}^2 \;p_i\,(1-p_i)\,p_j\,(1-p_j) $\ , where
\STATE \hskip .5cm ${\bf J}=\{h_i,J_{ij}\}$ is of dimension $\frac 12 K(K+1)$.
\STATE \hskip .5cm $S_\Gamma({\bf p}) \leftarrow\displaystyle{ \min _{\bf J} \ \big(S_{Ising}\big)_\Gamma [{\bf J}|{\bf p}]} $
\STATE
\STATE {\em Computation of $(S_0)_\Gamma$:}
\STATE \hskip .5cm${\bf M}_\Gamma \leftarrow K\times K$ matrix with elements $(M_\Gamma)_{ij} = \frac{p_{ij}-p_ip_j}{\sqrt{p_i(1-p_i)p_j(1-p_j)}}$ with $i,j\in \Gamma$
\STATE \hskip .5cm$( S_0)_\Gamma ({\bf p})\leftarrow \frac 12 \log \text{det}\, {\bf M}_\Gamma $ if $\gamma=0$, or use formula (\ref{loopentrogamma}) if $\gamma >0$.
\STATE
\STATE \hskip -.3cm {\bf Output:} $S_\Gamma ({\bf p})- (S_0)_{\Gamma} ({\bf p})$
\end{algorithmic}
\end{algorithm}

Algorithm 2 calculates the entropy $\Delta S_\Gamma$ of the cluster $\Gamma$ and maintains the list $L_{all}$ of all cluster-entropies computed so far.  It calls Algorithm 1 as a subroutine.

\begin{algorithm}[h]
\caption{Computation of cluster-entropy $\Delta S_\Gamma({\bf p})$}         
\label{alg2}                          
\begin{algorithmic}                    
\REQUIRE $\Gamma$ (of size $K$), data {\bf p}, list $L_{all}=\{\Gamma ',\Delta S_{\Gamma'}({\bf p})\}$ of known cluster-entropies.
\STATE 
\STATE $\hat L \leftarrow \{\Gamma ' : \Gamma' \subset \Gamma \ \text{and} \ \Gamma' \notin L_{all}\}$ \quad (ordered in increasing sizes)
\FOR {$\Gamma' \in \hat L$}
\STATE \hskip 1cm $\Delta S_{\Gamma '}({\bf p}) \leftarrow S_\Gamma ({\bf p})- (S_0)_{\Gamma} ({\bf p})- \displaystyle{\sum _{\substack{\Gamma '' \subset \Gamma' \\ (\Gamma '' \ne \Gamma ')}} \Delta S_{\Gamma ''}}({\bf p})$ using list $L_{all}$ of cluster-entropies calculated so far
\STATE \hskip 1cm update $L_{all} \leftarrow L_{all} \ \text{U}\ \{\Gamma ' , \Delta S_{\Gamma '}({\bf p})\}$
\ENDFOR
\STATE
\STATE \hskip -.3cm {\bf Output:} $\Delta S_\Gamma({\bf p})$ and $L_{all}$
\end{algorithmic}
\end{algorithm}

We can now give the core part of the procedure, which produces the list of selected clusters:

\begin{algorithm}[h]
\caption{Adaptive cluster algorithm for the inverse Ising problem} 
\label{alg3}                          
\begin{algorithmic}                    
\REQUIRE data {\bf p}, threshold $\Theta$
\STATE
\STATE $L_1\leftarrow \{ (i): i=1,2,\ldots, N\}$
\STATE $L_{sel}\leftarrow \emptyset$
\STATE  $K\leftarrow 1$
\STATE 
\STATE {\bf while} $L_K$ is not empty
\STATE  \hskip 1cm$L_{sel}\leftarrow L_{sel}\ \text{U}\ L_K$
\STATE \hskip 1cm $K\leftarrow K+1$
\STATE \hskip 1cm $L_{K} \leftarrow \emptyset$
\STATE \hskip 1cm {\bf for} $\Gamma_1,\Gamma_2 \in L_{K-1}$ {\bf do}
\STATE \hskip 2cm $\Gamma \leftarrow \Gamma_1 \cup \Gamma_2$
\STATE \hskip 2cm if $\Gamma$ is of size $K-2$ and if $|\Delta S_{\Gamma} ({\bf p})|< \Theta$, then 
$L_{K} \leftarrow L_{K} \ \text{U} \ \Gamma$  
\STATE \hskip 1cm{\bf end for}
\STATE {\bf end while}
\STATE 
\STATE \hskip -.3cm {\bf Output:} list $L_{sel}$ of selected clusters
\end{algorithmic}
\end{algorithm}

Algorithm 4 calculates the estimates for the total cross-entropy, and for the interaction parameters once the list of selected clusters $L_{sel}$ has been obtained. It requires Algorithms 1 and 2; function $(S_{Ising})_\Gamma$ and matrix $M_\Gamma$ are defined in the pseudo-code of Algorithm 1. 

\begin{algorithm}[h]
\caption{Estimates for the cross-entropy and for the interaction parameters} 
\label{alg4}                          
\begin{algorithmic}                    
\REQUIRE data {\bf p}, list $L_{sel}$ of selected clusters
\STATE
\STATE {\em Computation of cross-entropy $S$:}
\STATE \hskip 1cm compute $S_0({\bf p})$ using formula (\ref{loopentro}) or
(\ref{regul2})
\STATE \hskip 1cm$S_{\Gamma} ({\bf p}) \leftarrow S_0({\bf p}) +  \displaystyle{\sum _{\Gamma \in L_{sel}} \Delta S_\Gamma ({\bf p}) }$
\STATE
\STATE {\em Computation of fields and couplings ${\bf J}=\{h_i,J_{ij}\}$:}
\STATE \hskip 1cm compute $\{(h_0)_{i}({\bf p}),(J_0)_{ij}({\bf p})\}$ using formula (\ref{hJMF})
\STATE \hskip 1cm $L_{sub} \leftarrow \{ \Gamma ' \subset \Gamma : \Gamma \in L_{sel}\}$
\STATE \hskip 1cm {\bf for} $\Gamma'\in L_{sub}$ {\bf do}
\STATE \hskip 2cm $\displaystyle{m_{\Gamma '}} = \sum _{\Gamma \in L_{sel}: \Gamma' \subset \Gamma} (-1)^{|\Gamma|-|\Gamma'|}$, where $|\Gamma|,|\Gamma '|$ are the sizes of $\Gamma,\Gamma'$.
\STATE \hskip 2cm $\{h_i({\bf p}), J_{ij}({\bf p})\} \leftarrow  \displaystyle{\text{arg}\min _{\bf J} \ \big(S_{Ising}\big)_\Gamma[{\bf J}|{\bf p}]}$
\STATE \hskip 2cm compute $\{(h_0)_{i,\Gamma}({\bf p}),(J_0)_{ij,\Gamma}({\bf p})\}$ using formula (\ref{hJMF}), with $M_{\Gamma}$ replacing $M$.
\STATE \hskip 1cm {\bf end for}
\STATE \hskip 1cm compute $\{h_{i}({\bf p}),J_{ij}({\bf p})\}$ using formula (\ref{final-hj})
\STATE 
\STATE \hskip -.3cm {\bf Output:} $S({\bf p}), \{h_{i}({\bf p}),J_{ij}({\bf p})\}$ 
\end{algorithmic}
\end{algorithm}

\section{Applications}
\label{secappli}

In this Section we report the results of our algorithm when applied to data generated from Ising models with diverse interaction structures and various numbers $B$ of sampled configurations. We define: 
\label {definitions}
\begin{itemize}
\item the number $N_{clu}$ of clusters generated by the algorithm and the size $K_{max}$ of the largest clusters.
\item the average error on the inferred couplings and  fields:
\begin{equation}
\epsilon _h= \left( \frac 1 {N} \sum_{i} (h^{inf}_{i}-h_{i})^2 \right)^{\frac 12}\ , \quad
\epsilon _J= \left( \frac 2 {N\,(N-1)} \sum_{i<j} (J^{inf}_{ij}-J_{ij})^2 \right)^{\frac 12}
\ .\label{epsilonJ}
\end{equation} 
Here, $J^{inf}_{ij}$ and $h^{inf}_i$ denote the values of, respectively, the inferred couplings and fields, while $J_{ij}$ and $h_i$ are the values of the couplings and fields in the model used to generate the data.
\item
The error bars $\delta{h_{i}}$ and $\delta{J_{kl}}$ on the inferred couplings and fields, resulting from the finite sampling. Those statistical fluctuations are asymptotically given by the inverse of the susceptibility matrix of the cross-entropy $S_{Ising}$, see equation (\ref{deltaj}). The entries of $\boldsymbol\chi$ can be calculated from a Monte Carlo simulation, to estimate the multi-spins correlations. In practice, a  good approximation of $\boldsymbol\chi$ can already be obtained from the empirical average over the $B$ configurations in the sampling set. This procedure avoids the use of the Monte Carlo. In the presence of a $L_2$-regularization (\ref{regul2}), $\gamma \, p_k(1-p_k)p_l(1-p_l)$ is added to the diagonal element $\chi_{kl,kl}$ of the susceptibility matrix, before the inversion is performed. Hence, all the eigenvalues are strictly positive and the inverse is well defined. The inversion of $\chi$ can be done with standard linear algebra routines. 

Inferred couplings are called 'reliable' when their absolute value is larger than three times their statistical error-bar:  $|J_{kl}| > 3\; \delta J_{kl}$. 
\item 
 The reconstructed  observables, $p^{rec}_i$ and $c^{rec}_{ij}$, which we compare to the data, $p_i$ and $c_{ij}$. Those reconstructed averages are obtained using Monte Carlo simulations of the Ising model with the inferred fields, $h^{inf}_i$, and couplings, $J^{inf}_{ij}$. For those simulations the number of sampled configurations is chosen to be much larger than $B$, {\em e.g.} $100\,B$, to minimize the uncertainty on the reconstructed averages. 

\item the relative errors on the reconstructed averages $p_i$ and connected correlations $c_{ij}$, with respect to their statistical fluctuations due to 
finite sampling:
\begin{equation}\label{epsilonC}
\epsilon_p = \left( \frac 1 {N} \sum_{i} \frac{(p_i^{rec}-p_i)^2}{(\delta p_i)^2} \right)^{\frac 12} \ , \quad 
\epsilon _c = \left( \frac 2 {N\,(N-1)} \sum_{k<l} \frac{(c_{kl}^{rec}-c_{kl})^2}{(\delta c_{kl} )^2 } \right)^{\frac 12}
\ .
\end{equation} 
where the denominators in (\ref{epsilonC}) measure the typical fluctuations of the data expected at thermal equilibrium, see (\ref{def-deltac}), and 
\begin{equation}\label{fluctuationseq}
\delta c_{kl} = \delta p_{kl}+p_k \delta p_l +p_l\delta p_k \ .
\end{equation}
\end{itemize}
If not explicitly stated otherwise we start from the value $\Theta=1$ for the threshold and run the algorithm several times, dividing the threshold by 1.01 after each execution. The algorithm is stopped when both errors $\epsilon _p$ and $\epsilon_c$ are close to 1. We call $\Theta ^*$ the final value of the threshold corresponding to this criterion. Unless explicitly stated otherwise a $L_2$--regularization term (\ref{regul2}) is present, with $\gamma=1/(10\,B \,p^2\,(1-p)^{2})$, where $p$ is the average value of the $p_i$'s (Appendix \ref{appchoice}). As explained in Section \ref{bayessec} the regularization term is important in case of undersampling and guarantees the convergence of the numerical minimization of $S_{Ising}$.

\subsection{Independent Spin Model}

\begin{figure}
\begin{center}
\epsfig{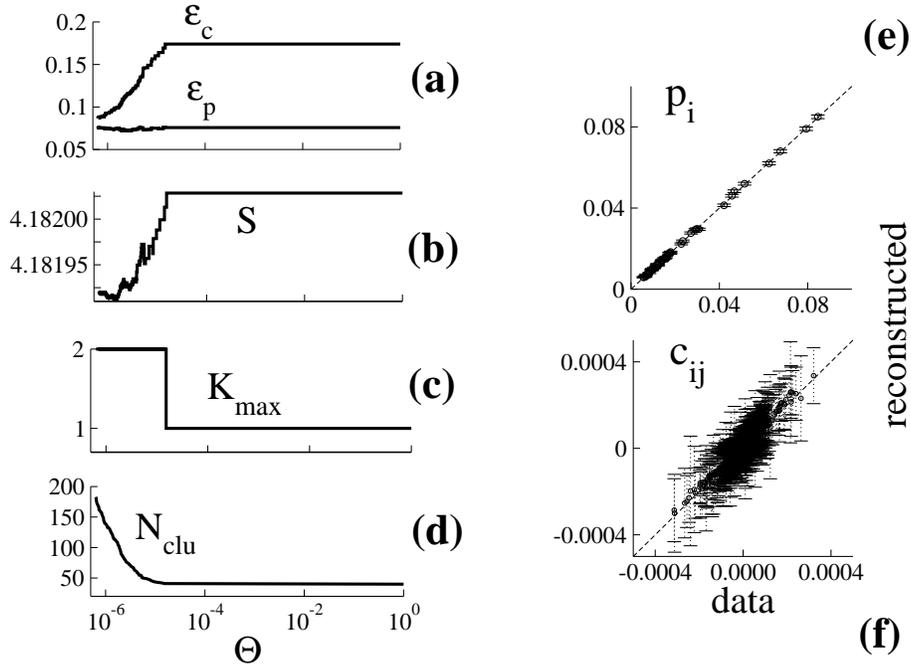}
\caption{Performance of the algorithm as a function of $\Theta$ for the Independent Spin model: {\bf (a)} errors $\epsilon _p$ and $\epsilon _c$; {\bf (b)} cross-entropy $S$ 
The entropy of the model in absence of sampling noise is $\simeq4.182071$;  {\bf (c)} size $K_{max}$ of largest clusters; {\bf (d)} number $N_{clu}$  of clusters. Panels {\bf (e)} and {\bf (f)} show the reconstructed $p_i$ and $c_{ij}$ vs. their values in the data. Error bars on $p_i$ and $c_{ij}$   are calculated through (\ref{fluctuationseq}). Data were obtained by sampling $B=10^5$ configurations of the $N=40$ spins, with spin dependent means $p_i$ equal to the average activity of the neurons in \cite{bialek}. The optimal threshold $\Theta^*$ is already obtained with clusters of size $K=1$.}
\label{indberry}
\end{center}
\end{figure}

\begin{figure}
\begin{center}
\epsfig{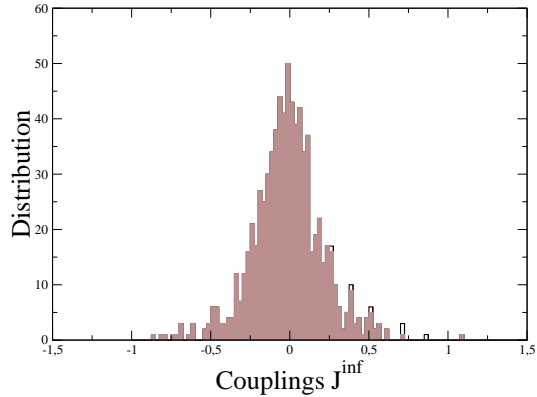}
\caption {Histogram of the inferred couplings for the Independent Spin model of Fig.~\ref{indberry}, and $\Theta^*=1$.815 of the 820 inferred couplings unreliable (in gray), because compatible with zero within
 three standard deviations}
\label{fig-istoj}
\end{center}
\end{figure}

It is instructive to run first the algorithm on the Independent Spins model, where each spin has a probability $p_i$ to be 1, $1-p_i$ to be 0, independently of the other variables. Due to the noise in the sampling (finite value of $B$), the connected correlations $c_{ij}$ are not equal to zero. Figure~\ref{indberry} shows the outcome for a system of size $N=40$, as a function of the threshold $\Theta$. The errors of reconstruction, $\epsilon_p$ and $\epsilon_c$, are already smaller than one for the initial threshold value $\Theta^*=1$. For this value of the threshold, cluster of size one only are selected. In other words, the interaction network $J_0$, calculated from the reference entropy $S_0=S_{MF}$ alone, is already overfitting the data as it attempts to reproduce the correlations due to statistical fluctuations. For smaller thresholds $\Theta$ contributions from clusters of size $K \ge 2$ allow for an even more precise reproduction of the data . 

The histogram of the inferred couplings, $J^{inf}_{ij}$, is shown in Fig.~\ref{fig-istoj}. It is centered around zero, and is approximately Gaussian.  The standard deviation of the distribution is compatible with the statistical error bar on couplings $ \delta J_{ij}$ (\ref{deltaj}) averaged on all the couplings. For the particular case of Fig.~\ref{indberry},  815 of the 820 inferred couplings are away from zero by less than three error bars, and are, therefore, classified as unreliable. This result is compatible with the fact that the non-zero couplings are the consequence of overfitting and do not reflect any real interactions.

Another possibility to avoid overfitting in this case is to apply the cluster expansion to the entropy $S$ in the absence of reference entropy ($S_0=0$). We  find that, for $\Theta^*=1$, the reconstruction errors are $\epsilon_p=0.07$ and $\epsilon_c=0.8$. Therefore only one--spin clusters are taken into account, and all couplings are equal to zero exactly (since $J_0=0$). 


\subsection{Unidimensional Ising model}
\label{secapplising}

We now test the algorithm on the unidimensional Ising model with first--neighbor interactions, $J_{i,i+1}=J$, and uniform fields, $h_i=h$. The model is placed on a ring with $N$ sites (periodic boundary conditions). Data ${\bf p}$ can be computed exactly (Appendix \ref{app-ising1d}) or through an average over $B$ configurations, sampled by Monte Carlo simulations. We compare the performance of the inference procedure for various values of $B$ and two values of $J,h$, corresponding to the correlation lengths $\xi \simeq 1$ ($h=-5,J=4$) and $\xi \simeq 9$ ($h=-5.95,J=6$). These values are, respectively, smaller than $\xi_c\simeq 4.3$,  the correlation length below which the cross-entropy expansion (\ref{truncated2}) is absolutely convergent, and larger than $2\,\xi_c\simeq 8.6$, above which condition (\ref{ds2cv}) is violated, see Section \ref{sec-convergence}. 

\subsubsection{Accuracy of the cluster expansion as a function of the threshold: errors on the entropy, couplings and fields }
 
\begin{figure}
\begin{center}
\epsfig{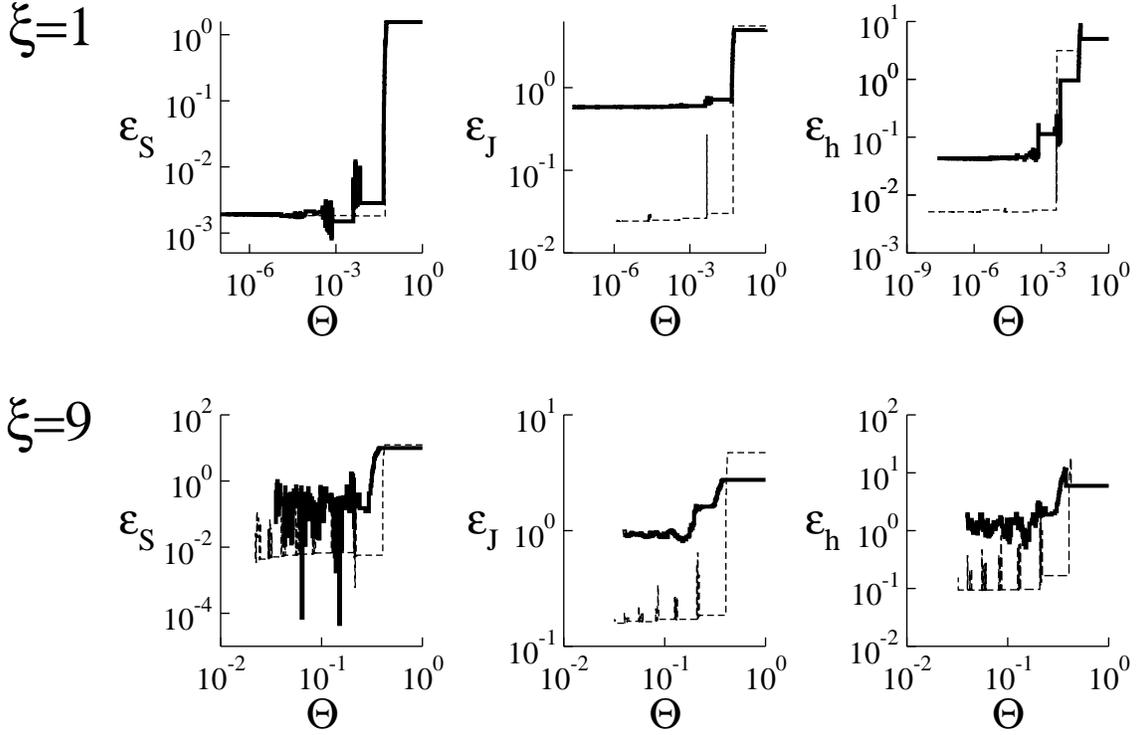}
\vskip 1cm
\caption{Errors on the entropy ($\epsilon_S$), the couplings ($\epsilon_J$), and the fields ($\epsilon_h$) vs. threshold $\Theta$ for the unidimensional Ising model with $\xi=1$ (top) and $\xi=9$ (bottom). Full lines correspond to large sampling noise ($B=10^5$ for $\xi=1$, $B=10^3$ for $\xi=9$), dashed line to weak sampling noise ($B=10^7$ for $\xi=1$, $B=10^5$ for $\xi=9$). The size is $N=30$.} 
\label{dsvsdTxi}
\end{center}
\end{figure}

We start with the small--$\xi$ case. Figure~\ref{dsvsdTxi}(top) shows $\epsilon_S$, the absolute value of the difference between the cross-entropy $S({\bf p},\Theta)$ (\ref{truncated1}) and the entropy of the model for perfect sampling, and the errors $\epsilon_J$ and $\epsilon_h$, for various values of $B$. We observe that $\epsilon_S$ sharply decreases around $\Theta_1=0.05$, that is, the entropy of nearest-neighbor clusters $\Delta S_{(i,i+1)}$; discarding all entropies smaller than this value would be exact in the perfect sampling case (Section \ref{secpropdeltas}). As $\Theta$ is decreased, $\epsilon_S$ exhibits fluctuations centered around a discrete sequence of threshold values, $\Theta _d$, with $d\ge 2$. As explained in Section \ref{secpropdeltas}, these values correspond to the cluster-entropies $\Delta S_{(i,i+d)}$ in the absence of noise, see identity (\ref{idcan}). Fluctuations spread over a small window around $\Theta_d$. They correspond to imperfect cancellations of 'packets' of entropies whose associated clusters share the same interaction path with length $L=2d$ (Section \ref{sec-convergence}). Since the correlation length $\xi$ is small, and, therefore, the cross entropy expansion is absolutely convergent, the magnitude of the fluctuations quickly decreases with $d$. For $B=10^5$ two bursts of fluctuations are visible (corresponding to $d=2,3$). For $B=10^7$, fluctuations are smaller and spread over narrower windows; only the $d=2$ burst can be observed. In between two bursts of fluctuations, $\epsilon _S$ reaches a plateau. Note that the value of $\epsilon_S$ on the plateau is not zero due to the sampling noise (finite $B$).    

The errors on the inferred couplings and fields have the same behaviour as $\epsilon_S$ (Fig.~\ref{dsvsdTxi}(top)). Their magnitude are comparable to the expected statistical fluctuations calculated from the 4-spin correlations (inverse susceptibility in (\ref{deltaj})), which decrease as $B^{-1/2}$.

We now test our algorithm on the unidimensional Ising model with a large correlation length, $\xi=9$. The errors $\epsilon_S,\epsilon_J,\epsilon_h$ are shown in Fig.~\ref{dsvsdTxi}(bottom) as functions of $\Theta$. The correlation length of the model is larger than $2\xi_c$,  at which the series of the squared cluster-entropies is divergent. Therefore, we expect large fluctuations of $\epsilon_S$, corresponding to packets of clusters with the same interaction path (Section \ref{sec-convergence}). Figure \ref{dsvsdTxi} shows, indeed, that fluctuations are much larger for $\xi=9$ (bottom) than for $\xi=1$ (top). Furthermore, fluctuations do not decrease much in amplitude as $\Theta$ is decreased. In the case of severe undersampling ($B=1000$) the notion of bursts of fluctuations separated by plateaus is blurred out. The distributions of cluster-entropies in a packet is so wide that it overlaps with the distributions of entropies associated to the neighbouring packets.
As for the previous case the magnitude of $\epsilon_J,\epsilon_h$ are comparable to
 the expected statistical fluctuations
 calculated from (\ref{deltaj}).

\subsubsection{Quality of reconstruction and choice of the threshold $\Theta^*$  }

\begin{figure}
\begin{center}
\epsfig{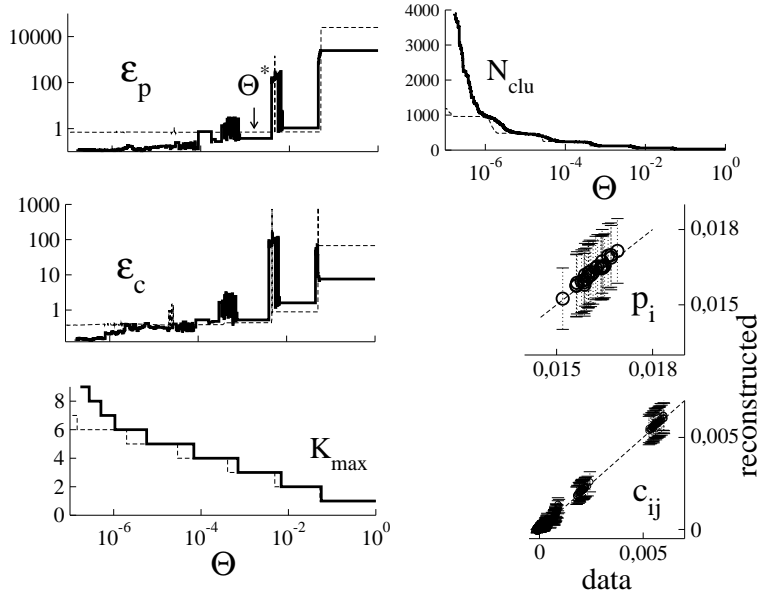}
\caption {Performance of the algorithm on the unidimensional Ising model with $\xi=1$, $B=10^5$ (full-bold line)  $B=10^7$ (dashed line). For both values of $B$ the optimal  threshold $0.0007<\Theta^*<0.003$ is reached with clusters of size $K=3$ and length $L=6$. The reconstruction of data $p_i$ and $c_{ij}$ is shown only for the $B=10^5$ case (large sampling noise). Error bars on $p_i$ and $c_{ij}$ are computed from (\ref{fluctuationseq}).}
\label{fig-xi1}
\end{center}
\end{figure}

\begin{figure}
\begin{center}
\epsfig{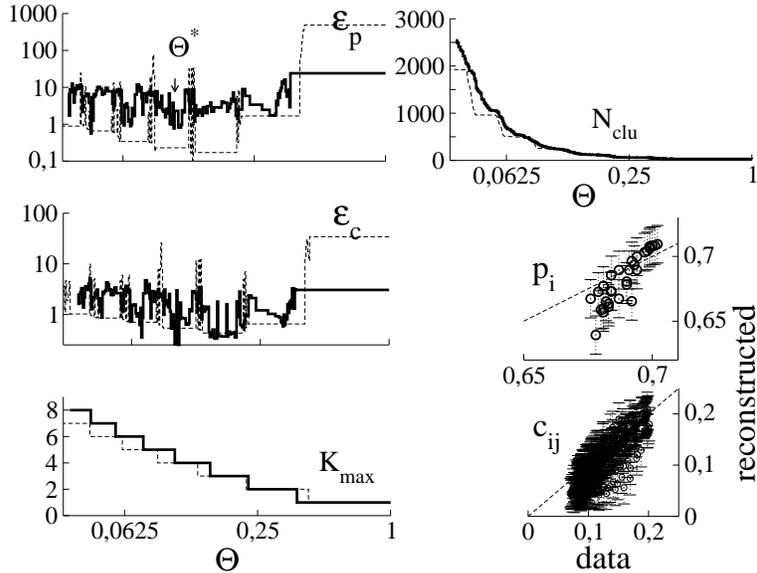}
\caption {Performance of the algorithm on the unidimensional Ising model with $\xi=9$, $B=10^3$ (full-bold line) and  $B=10^5$ (dashed line). For the
largest sampling noise case ($B=10^3$) at the threshold $\Theta^*\simeq .104$, 89 clusters of size $K=2$, 92 clusters of size $K=3$, 35 clusters of size $K=4$, and 2 clusters of size $K=5$ are selected. The reconstruction of data
$p_i$ and $c_{ij}$ is shown for the $B=10^3$ case.
 Error bars on $p_i$ and $c_{ij}$   are computed from (\ref{fluctuationseq}).}
\label{fig-xi9}
\end{center}
\end{figure}

We now study the reconstruction errors $\epsilon_c$ and $\epsilon_p$ as functions of $\Theta$, for the small and large correlation lengths and for the weak and strong sampling noises. We show in Fig.~\ref{fig-xi1} the errors $\epsilon_c$ and $\epsilon_p$, as well as the maximal size, $K_{max}$, and the number, $N_{clu}$, of clusters vs. the threshold $\Theta$ in the case of a small correlation length, $\xi=1$ . The threshold at which both $\epsilon_c$ and  $\epsilon_p$ are close to 1 can be chosen in the range $0.0007<\Theta^*<0.003$, for which all the clusters of lengths $L\le 4$ are processed. 
It is possible to check in Fig.~\ref{dsvsdTxi}(top) that  this  threshold  value gives 
 the minimum  values  of  $\epsilon_S$, $\epsilon_J$, $\epsilon_h $ .
Contrary to the case of perfect sampling, it is  not sufficient to take into account clusters with contour length $L=2$ only. The selected clusters correspond to three groups of $N=30$ clusters each; the first group gathers the clusters $(i,i+1)$ ($L=2$), the second one, the clusters $(i,i+2)$ ($L=4$) and the third one, the clusters $(i,i+1,i+2)$ ($L=4$). In Fig.~\ref{fig-xi1} we show the reconstructed averages $p_i$ and correlations $c_{ij}$, at $\Theta^*$ and for the largest sampling noise case $B=10^5$,  vs. their values in the data . The agreement is very good, and falls within the statistical fluctuations expected at equilibrium for the Ising model, given in (\ref{fluctuationseq}). We stress that the optimal value of the threshold and the maximal size of selected clusters depend on the particular realization of the data $\bf p$, and can vary from sample to sample. By further decreasing the threshold $\Theta$ below  $\Theta^*$, the reconstruction errors $\epsilon_p$ and $\epsilon_c$ decrease to values smaller than one (Fig.~\ref{fig-xi1}). This regime corresponds to an overffiting of the data, as the errors on the couplings, $\epsilon_J$, and on the fields, $\epsilon_h$, cease to decrease  (Fig.~\ref{dsvsdTxi}(top)). 

Results for the case of a larger correlation length ($\xi=9$) with $B=10^5$ and $B=10^3$ sampling configurations are reported in Fig.~\ref{fig-xi9}.
For the poor sampling case ($B=10^3$) plateaus are not present any longer, but a good inference is still obtained for a value $\Theta ^*$ of the threshold, which, as in the $\xi=1$ case, corresponds to the summation of all the clusters with contour length $L=4$. This finding supports the discussion of Section \ref{sec-chi}: the contour length required for a good inference is largely independent of the correlation length. However, in the poor sampling case, finding the right value for $\Theta^*$ is harder for larger $\xi$ due to the mixing of packets.

\subsubsection{Quality of the inference: histograms of couplings}
\begin{figure}
\begin{center}
\epsfig{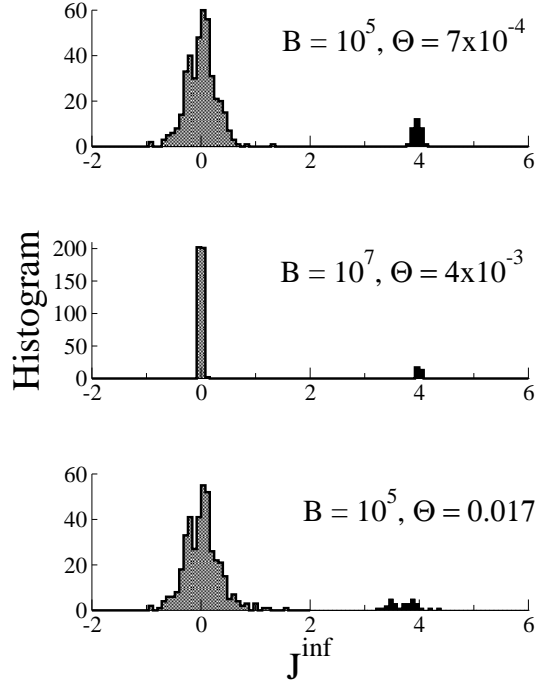}
\caption{Histograms of the inferred couplings $J_{ij}^{inf
}$ for the unidimensional Ising model with $\xi=1$. Couplings equal to $J=4$ and $J=0$ in the model are shown in, respectively, black and gray. Gray couplings are unreliable, as they differ from zero by less than three standard deviations.
 Values of $B$ and $\Theta$ are indicated on the figure. The bin width is $\Delta J=.08$.} 
\label{istojxi1}
\end{center}
\end{figure}

To better understand the quality of the inference  we plot in Fig.~\ref{istojxi1} (up and middle panels) the histogram of the inferred couplings $J_{ij}^{inf}$ at the threshold $\Theta^*$. The distribution is bimodal: a Gaussian-like peak centered in $J^{inf}=0$ and a smaller distribution around $J^{inf}=4$. The two sub-distributions are separated by a wide gap. The inference algorithm makes no classification error: the sub-distribution centered around $J^{inf}=0$ contains all the pairs $(i,j)$ such that $J_{ij}=0$, and the one around $J^{inf}=4$ includes all the pairs of nearest neighbours $(i,i+1)$.  All the couplings centered around zero are unreliable. Moreover the standard deviation of the distribution of the couplings around the zero value (equal to the minimal value of $\epsilon_J$ reached on the plateau in Fig.~\ref{dsvsdTxi} ) agrees with the statistical fluctuations (\ref{deltaj});  all the couplings around zero are therefore unreliable. The structure of the interaction network is perfectly recovered. 

\begin{figure}
\begin{center}
\epsfig{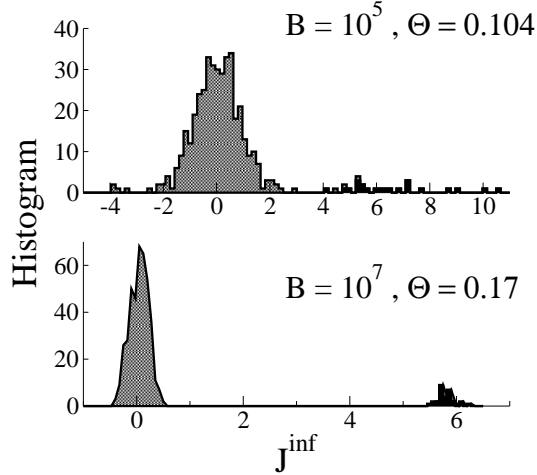}
\caption{Same as Fig.~\ref{istojxi1}, but with $\xi=9$ instead of $\xi=1$. Couplings equal to $J=6$ and $J=0$ in the model are shown in, respectively, black and gray. Gray couplings are unreliable, as they differ from zero by less than three standard deviations.}
\label{istojxi9}
\end{center}
\end{figure}

We show  the histogram of couplings for $\Theta>\Theta^*$, {\em i.e.} $\epsilon_c>1$, in  Fig.~\ref{istojxi1} (bottom);  the structure of the interaction network is still perfectly recovered but the values  of the  positive inferred couplings is less accurate.  The histogram of inferred couplings for the large correlation length , $\xi=9$, is shown in Fig.~\ref{istojxi9}. Even when the sampling noise is large, a good separation of the two sub-distributions corresponding to interacting and non-interacting pairs of spins is achieved for large values of the threshold $\Theta^*$, and the couplings are correctly inferred (up to the expected statistical fluctuations). 

\subsection{Regular bidimensional grid}\label{secappligrid}

We now analyze the performance of the algorithm on bidimensional grids of different sizes, $N=N'\times N'$. Nearest neighbors on the grid interact through the coupling $J$. The value of $J\simeq 1.778$ is chosen to make the grid critical in the thermodynamical limit, $N'\to\infty$ \cite{fisher67,zobin78}. Hence, the system is at the paramagnetic/ferromagnetic critical point, and the correlation length $\xi'$ diverges with $N'$. 

We have described in Section \ref{sectruncscheme} and Fig.~\ref{fig-truncation2D} the partial cancellation of the cluster-entropies for a small bidimensional grid ($N'=3$), and no sampling noise. Due to this cancellation property, taking into account clusters of contour length $L \le 4$ was sufficient to obtain a very accurate approximation to the cross-entropy. Hereafter, we show that this result is not affected by the presence of sampling noise. Furthermore, we will see that the size of the clusters necessary for a good inference of the interactions remains rather constant when the grid size is increased from $N'=7$ to $N'=20$, and is thus, as in Section \ref{secapplising}, largely independent of the correlation length $\xi$.

\subsubsection{The small $3\times 3$ grid revisited: influence of the sampling noise}

\begin{figure}
\begin{center}
\epsfig{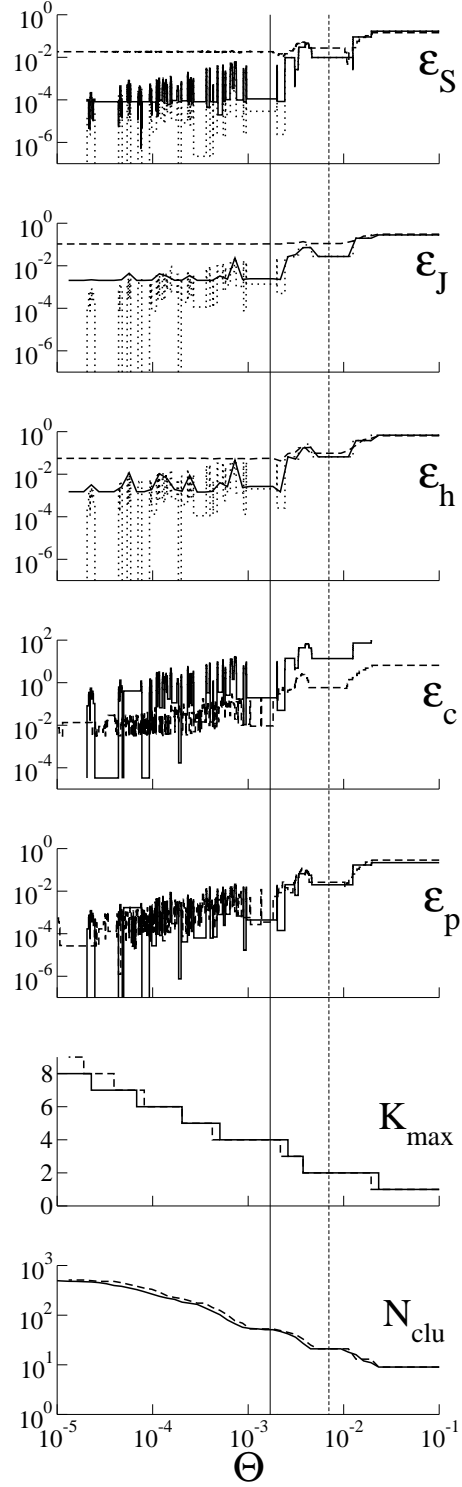}
\caption {Errors $\epsilon_s$, $\epsilon_J$, $\epsilon_h$, $\epsilon_c$, $\epsilon_p$, size $K_{max}$, and number $K_{clu}$ of clusters vs. $\Theta$ for a $3\times3$ grid with $J=1.778$.
The dashed line corresponds to $B=4500$ sampled configurations, the full line to $B=10^7$; the perfect sampling curves are shown with dotted lines in the top three panels (same data as Fig.~\ref{fig-truncation2D}). The values of $\Theta^*$ are shown with vertical lines: $\Theta^*=.0017$  (full, $B=10^7$), and $.007$ (dashed, $B=4500$). At $\Theta^*$ 21 clusters ($B=4500$) and 49 clusters ($B=10^7$) are selected}.
\label{score3x3}
\end{center}
\end{figure}

We start with the $3\times 3$ grid of Section \ref{sectruncscheme}, for which the summations of all clusters up to size $K=9$ gives the exact solution of the inverse problem, and all 1- and 2-point averages can be calculated exactly in the perfect sampling case. The reader is kindly referred to Fig.~\ref{fig-truncation2D} and to the related discussion. Figure~\ref{score3x3} shows the errors on the entropy, the couplings, the fields, the reconstructed 2- and 1-point averages, and the size and the number of clusters as functions of $\Theta$. For a given amount of sampling noise (set by the value of $B$), the errors $\epsilon_S,\epsilon_J, \epsilon_h$ follow their perfect-sampling counterparts, until a threshold value $\Theta_{sat}$, and they saturate for $\Theta<\Theta_{sat}$. The saturations are interrupted by fluctuations due to the imperfect cancellations of clusters within a packet (Section \ref{sec-convergence}). The saturation values of $\epsilon_J$ and $\epsilon_h$ decrease with increasing $B$, and are compatible with the expected statistical fluctuations $\delta J_{ij}$ given by (\ref{deltaj}). The value of $\Theta _{sat}$ approximately coincides with the threshold $\Theta^*$, at which both $\epsilon_c$ and $\epsilon_p$ are $\simeq 1$ (Fig.~\ref{score3x3}). For $B=4500$, only clusters of size $K=2$ are taken into account at $\Theta^*$. For $B=10^7$, clusters made of spins on the elementary squares of Fig.~\ref{fig-entrovsd} and of size up to $K=4$ are selected, {\em e.g.}  $1,5$ or $1,2,4,5$  with $L=4$ in Fig.~\ref{fig-entrovsd}, while clusters such as $(1,2,3)$ or $(1,3)$, which have the same contour length but are not on elementary squares, are discarded. At $\Theta^*$ the histogram of the inferred couplings is made of two far apart sub-distributions (not shown): the first one corresponds to the $12$ non zero couplings and is centered around $J^{inf}\simeq 1.8$, and the second one, peaked around $J^{inf}=0$,  contains the remaining 24 pairs of sites. Hence, the structure of the interaction graph is correctly found back.

\subsubsection{Study of larger critical grids}

\begin{figure}
\begin{center}
\epsfig{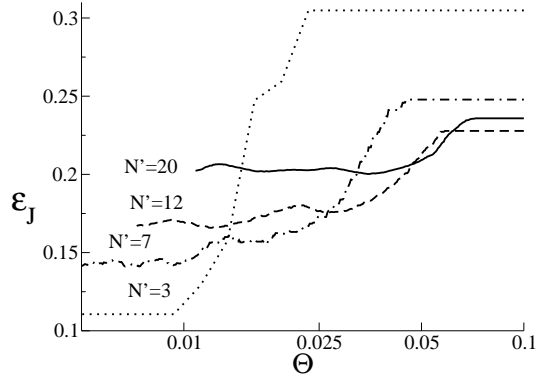}
\caption {Error $\epsilon_J$ for a bidimensional grid $N'\times N'$, with $N'=3,7,12,20$, and for $B=4500$ sampled configurations.}
\label{scorej-grid}
\end{center}
\end{figure}

\begin{figure}
\begin{center}
\epsfig{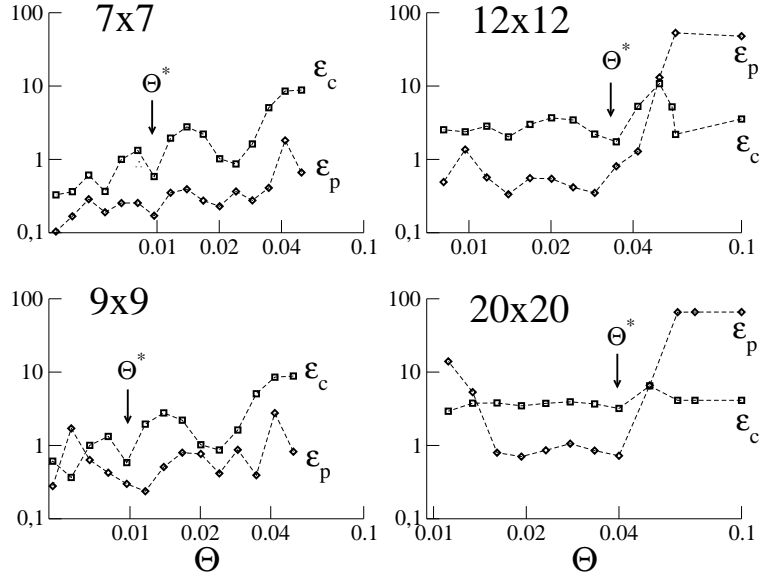}
\caption {Reconstruction errors $\epsilon_c$ (squares) and $\epsilon_p$ (diamonds) for grids of sizes $7\times 7$, $9\times 9$,$12\times 12$ and $20\times 20$. Optimal thresholds $\Theta^*$ are located by arrows. Other possible choices for $\Theta ^*$, for instance $\Theta^*=.024$ for the $7\times 7$ grid are equally possible.}
\label{score-grid}
\end{center}
\end{figure}

\begin{figure}
\begin{center}
\epsfig{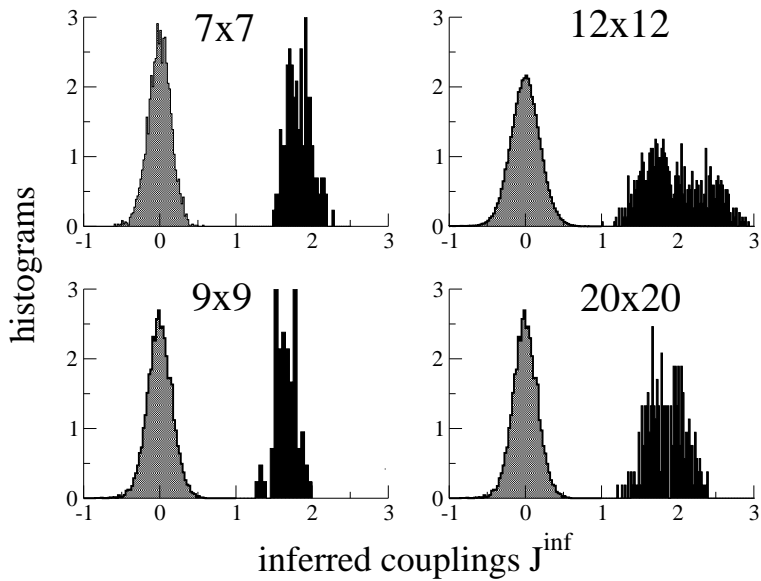}
\caption {Histograms of inferred couplings for bidimensional grids of different sizes. Couplings equal to $J=1.778$ and $0$ in the model are shown in, respectively, black and gray; the integral of each sub-distribution is normalized to one by hand. The values of $\Theta^*$ for each size are shown in Fig.~\ref{score-grid}. }
\label{istoj-grid}
\end{center}
\end{figure}

We now turn to larger grids $N'\times N$', where $N'$ ranges between 7 and 20. Data are calculated from $B=4500$ configurations sampled through a Monte Carlo simulation. The error $\epsilon _J$ on the couplings is shown in Fig.~\ref{scorej-grid}. As $\Theta$ decreases, $\epsilon_J$ saturates to a value close to the average of the expected statistical error, $\delta J_{ij}$, which lies between .1 and .2. Saturation begins at large values of the threshold, even when the linear size $N'$ of the grid is increased. The asymptotic values  depends strongly on $N$ due to the non periodic boundary conditions \footnote{We have verified that the  dependence on $N$ is weaker in the case of periodic boundary conditions.}.

The reconstructions errors $\epsilon_p$ and $\epsilon_c$ are shown in Fig.~\ref{score-grid}. The maximal size of the clusters at the optimal threshold  $\Theta^*$  is bounded ($K_{max}\le 4$), as the correlation length $\xi$ diverges with $N'$. As a consequence, the running time of the algorithm increases linearly with $N$. For  threshold values smaller than $\Theta^*$, $\epsilon_p$ and $\epsilon_c$ decrease. The $7\times 7$ grid in Fig.~\ref{score-grid} provides a clear illustration of data overfitting. Keeping $B$ fixed while $N'$ and $N$ increase make the data effectively more and more noisy. This is the reason why the value of $\Theta ^*$ slightly increases with $N$.

The histograms of the inferred couplings at the threshold $\Theta^*$ are shown in Fig.~\ref{istoj-grid}. The structure of the grid is perfectly reconstructed for all sizes $N'$. We find that all the inferred couplings in the sub-distribution centered around zero are smaller (in absolute value) than three times their standard deviation (corresponding to the asymptotic value of Fig.~\ref{scorej-grid}) and are, therefore, unreliable.


\subsection{Randomly diluted bidimensional grid}

We now remove a fraction $1-\rho$ of the couplings on the grid, independently and at random. Our goal is to study how the algorithm performs on such disordered systems, at the phase transition and in the low temperature phase. We will compare the performance with another low complexity algorithm, the regularized logistic regression algorithm, guaranteed to perform well at high temperature and to fail at low temperature \cite{wain}. To compare with the numerical experiments of \cite{bento}, we have generated $7\times 7$ bidimensional grids, and keep each bond with probability $\rho=.7$. The remaining bonds are all equal to $J$  \footnote{The  numerical experiments of \cite{bento} were done with $\pm 1$ spins and with coupling parameter $\theta$ and field $\nu=0$; in the present work where spins are equal to 0,1, the corresponding couplings and fields are: $J=4\theta$ and $h_i=-\frac 1 2 \sum_{j\neq i} J_{ij}$.}. We generate, for each value of $J$ ranging from 0.4 to 4.4, eight randomly diluted grids. For each grid we calculate the data $\bf p$ by sampling over $B=4500$ configurations generated by a Monte Carlo dynamics. 

\subsubsection{Inference of the network structure from the mean field entropy $S_{MF}$}
\label{secgrid490}

\begin{figure}
\begin{center}
\epsfig{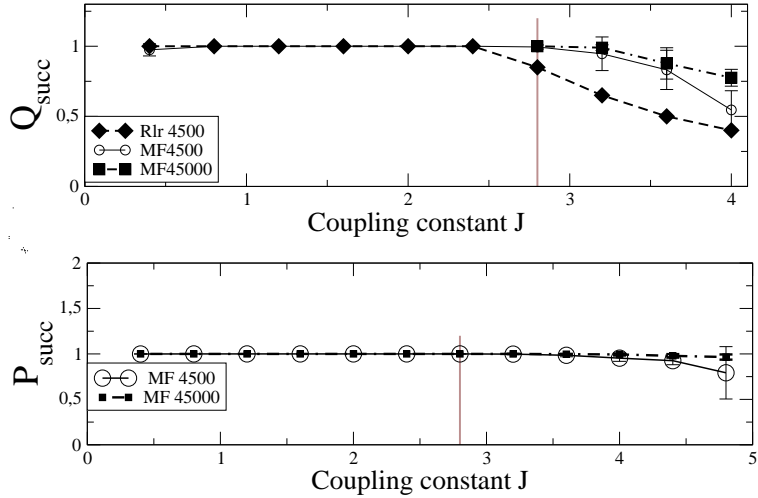}
\caption{Probabilities that a neighborhoods is reconstructed (top) and that a bond is inferred (bottom) as functions of the coupling $J$ for a bidimensional random $7\times 7$ grid density $\rho=.7$ and for various values of the numbers $B$ of sampled configurations. Error bars are calculated from the standard deviations over eight random grids. Probabilities were obtained from the simple Mean-Field algorithm ($S_0=S_{MF}$, $\Theta =1$). The value of $Q_{succ}$ (top) is compared to the performances of the pseudo-likelihood algorithm (Rlr) of \cite{wain,bento} in the top panel.}
\label{spfig}
\end{center}
\end{figure}

Our first task consists, as in \cite{bento}, in reconstructing the structure of the interaction graph only. We do not want to accurately determine the value of the coupling constants $J_{ij}$, but only if it is positive or null. This task is easier than the precise inference of the couplings, and we will first handle  by approximating the cross-entropy $S$ with the reference entropy $S_0=S_{MF}$ only. Equivalently, we choose $\Theta^*$ to be large enough that no cluster is selected by our algorithm. We compute the mean-field couplings, $(J_0)_{ij}$, and, for each pair $(i,j)$, and decide that a bond is present if $(J_0)_{ij}>J/2$, absent otherwise. The performances of this simple, Mean-Field algorithm are shown in Fig.~\ref{spfig}. We say that the neighborhood of a vertex $i$ is reconstructed if the sets of its neighbors $j$ ($J_{ij}\ne 0$) and of its non-neighbors ($J_{ij}=0$) are correctly inferred. In Fig. \ref{spfig} (top panel) we report the fraction $Q_{succ}$ of the neighborhoods which are reconstructed (straight line), as a function of the coupling strength $J$. We compare the performance of the mean-field algorithm with the pseudo-likelihood algorithm of \cite{wain} in Fig.~\ref{spfig}(right). Contrary to the pseudo-likelihood algorithm case, the neighborhoods are perfectly reconstructed at the phase transition. Furthermore, $Q_{succ}$ remains large in the ferromagnetic phase: for instance, for $J=3.2$, more than 80\% of neighborhoods are perfectly inferred. For very large $J$ (low temperatures) the average $p_i$ are too close to 0 in the down state and to 1 in the up state. Most of the sampled configurations coincide with one of the two ground states, and the inference is difficult. Fig.~\ref{spfig}(top panel) shows the increase of $Q_{succ}$ resulting from a ten-fold increase of $B$.

Another measure of the performances is shown in Fig.~\ref{spfig}(bottom panel). We plot $P_{succ}$, the fraction of bonds in the grid correctly predicted to exist, averaged over the data realizations, as a function of the coupling strength $J$. For $J=3.2$ more than 99.78\% (respectively 99.96\% ) of the bonds are correctly predicted with $B=4500$ (respectively $B=45000$) configurations. For even larger values of the coupling constant, $J=4$, more than 95 \% (respectively  99\%) of the bonds are correctly predicted with $B=4500$ (respectively $B=45000$)  with the Mean-Field algorithm.

In Section~\ref{gridcluster} we show that the probability of success increases in the ferromagnetic phase, when using our algorithm with a well-chosen threshold $\Theta$ rather than the simple Mean-Field procedure ($\Theta=1$), {\em e.g.} all neighborhoods are perfectly reconstructed  for $J=3.2$ ($Q_{succ}=1$). 

\subsubsection{Is thermalization relevant to the inference at low temperatures?}
 
\begin{figure}
\begin{center}
\epsfig{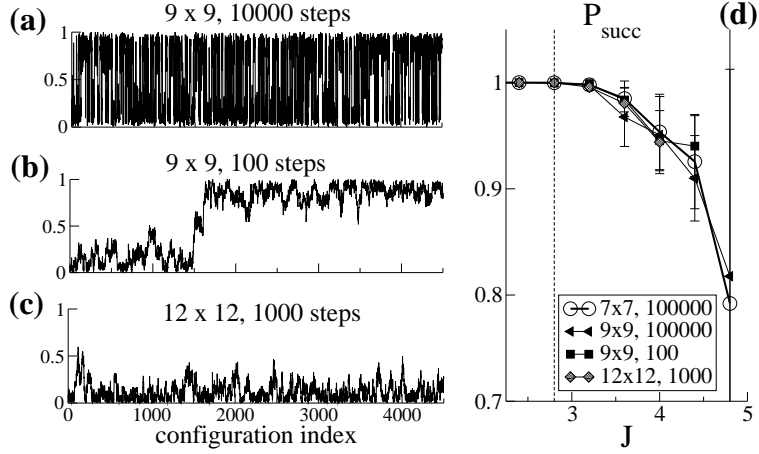} 
\caption{{\bf (a,b,c).} Average activity $\mu$ (\ref{defmu}) of the $B=4500$ sampled configurations for three grid sizes and numbers of spin-flip attempts in between two samplings. {\bf (d)} Probability of success for $J>2.8$, for the same data as in {\bf (a,b,c)} and for the $7\times 7$ grid of Fig.~\ref{spfig}.}
\label{mdt}
\end{center}
\end{figure}

The diluted bidimensional grid, with a fraction $\rho=.7$ of non-zero bonds, undergoes a transition from a paramagnetic to a ferromagnetic phase at the value $J_{crit}(\rho=.7)\simeq 2.8$ (vertical line in Fig.~\ref{spfig}) in the infinite size limit \cite{fisher67,zobin78}. In the ferromagnetic phase, $J>J_{crit}(\rho=0.7)$, and for small bidimensional grids, two competing 'states' coexist: the 'down' state, where most spins are 0, and the up state, where most spins are equal to 1. The system 'jumps' from one state to the other, as shown by the time-dependence of the average activity,
\begin{equation}\label{defmu}
\mu (t)=\frac{1}{N}\sum_{i=1}^N \sigma_i(t)\ ,
\end{equation}
where $t$ is the Monte Carlo time. Figure~\ref{mdt}(a) shows that the two states are equally sampled on a $9\times 9$ grid, with $N_A=10,000$ single spin-flip attempts with the Metropolis rule in between two sampled configurations (the results of Fig.~\ref{spfig} on the $7\times 7$ grid were obtained with the same value of $N_A$). To investigate the performance of the algorithm when the two states are not well sampled we have studied a $9\times 9$ grid, with $N_A=100$ and $N_A=1,000$. For $N_A=100$, fig.~\ref{mdt}(b) shows that few transitions occur, and that the two states will likely not be weighted equally. On larger $12\times 12$ grids no jump occurs, even with $N_A=1,000$ spin-flip attempts (Fig.~\ref{mdt}(c)). The values of the spin averages $p_i$ will strongly vary between the three cases above: $p_i\simeq .5$ in the mixed case (a), $p_i\simeq .7$ in the particular realization (b) of the partially mixed case, and $p_i$ close to zero in case (c). Remarkably the probability of success of the algorithm is not sensitive to the nature of the mixing. Fig.~\ref{mdt}(d) shows, indeed, that the reconstruction performances do not significantly decrease even in the partially and non mixed cases compared to the fully thermalized case.

\subsubsection{Inference of the couplings and reconstruction of 1- and 2-point averages with the cluster expansion}
\label{gridcluster}

\begin{figure}
\begin{center}
\epsfig{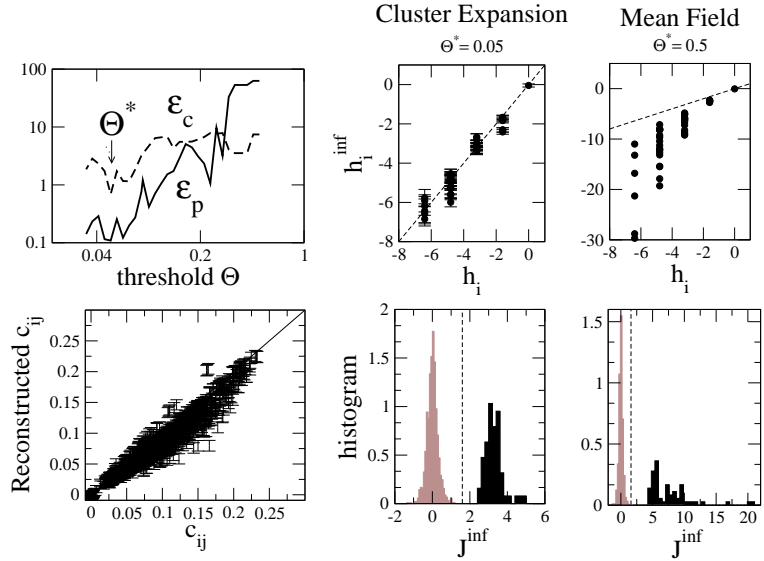}
\caption{Performances of the inference algorithm for the $7\times 7$ randomly diluted grid, with $J=3.2$, $B=4500$, $N_A=10^5$.  Left: Relative errors $\epsilon_p$, $\epsilon_c$ as functions of the threshold (top) and comparison of reconstructed and data correlations at $\Theta^*$ (bottom). 
Middle and Right: comparison of the inferred and true fields (top) and histograms of inferred couplings (bottom) for our cluster algorithm and the mean field procedure. Color code for the histograms: brown/gray: unreliable couplings (which also correspond to zero couplings in the true network), black : reliable couplings. The two sub-distributions are normalized separately.  }
\label{spfig3}
\end{center}
\end{figure}

\begin{figure}
\begin{center}
\epsfig{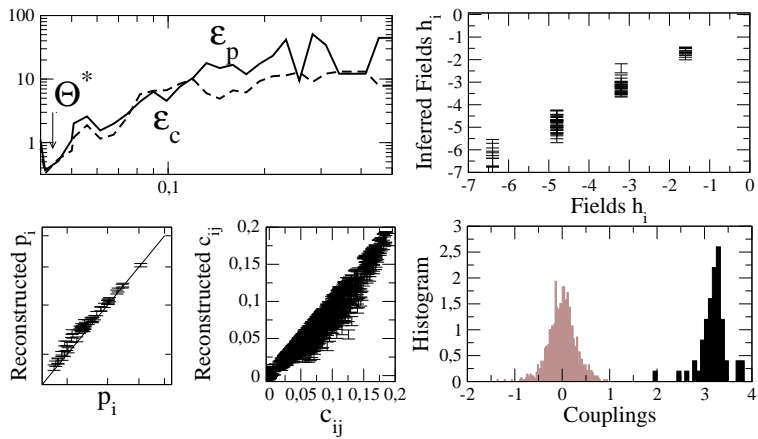}
\caption{Performance of the inference algorithm in the case of poor mixing, for a $7\times 7$ randomly diluted grid, with $J=3.2$, $B=4500$, $N_A=10^2$. Top \& left: relative errors $\epsilon_p$, $\epsilon_c$ as functions of the threshold $\Theta$. Bottom \& left: reconstructed $p_i$ and $c_{ij}$ versus their data values. Right: Inferred fields $h^{inf}_{i}$ vs. their true values for $\Theta^*=0.05$ (top) and histogram of  the inferred couplings $J^{inf}$ (bottom). As usual, unreliable couplings, which correspond to zero couplings in the true network, are depicted in gray, and reliable couplings in black. }
\label{spfig4}
\end{center}
\end{figure}

It is harder to determine the values of the fields and of the couplings and to reconstruct the frequencies than to infer the structure of the interaction graph alone. To this aim the minimization of the MF entropy $S^{MF}$ is generally not sufficient, and the cluster expansion of  $S-S^{MF}$ has to be carried out. In Fig.~\ref{spfig3} we show the relative errors $\epsilon_p$ and $\epsilon_c$ in the reconstruction of one- and two-site frequencies as a function of the threshold $\Theta$ for the same $7\times 7$ randomly diluted grid as in Section \ref{secgrid490}. We also compare the fields and couplings inferred with our cluster algorithm (middle panels in Fig.~\ref{spfig3}, $\Theta ^*=.05$) to the ones found with the simple MF procedure (right panels in Fig.~\ref{spfig3}, large value of $\Theta$). Note that the small error done in the graph learning for $J=3.2$ ($P_{succ}=0.997$) can be avoided when the threshold value is optimized for each data realization. 


We show in Fig.~\ref{spfig4} the performances of the algorithm in the case of poor mixing, when the two states are not equally sampled. For the particular realization corresponding to Fig.~\ref{spfig4}, the frequencies are $p_i\simeq 0.3$ instead of $p_i\simeq 0.5$. In spite of the poor mixing the inference of the fields and couplings is as accurate as in the case of well-mixed sampling. The difference between the fields corresponding to the apparent  frequencies $p_i\simeq 0.3$ and the true one ($p_i=.5$) are, indeed, smaller than the statistical uncertainty on the fields due to the limited sampling (finite value of $B$). The reason is that, near a critical point, a small variation in the field is sufficient to produce a large change in the average values of the spins.
 
\subsection{Erd\"os-Renyi random graphs}
 
In this Section we report the results of our inference algorithm when applied to disordered Ising models on random graphs. The random networks are generated from the Erd\"os-Renyi ensemble, where $M=\frac d2\,N$ edges are drawn, uniformly and at random, between $N$ points. Parameter $d$ is the average degree of a vertex on the network.

\begin{figure}[t]
\begin{center}
\epsfig{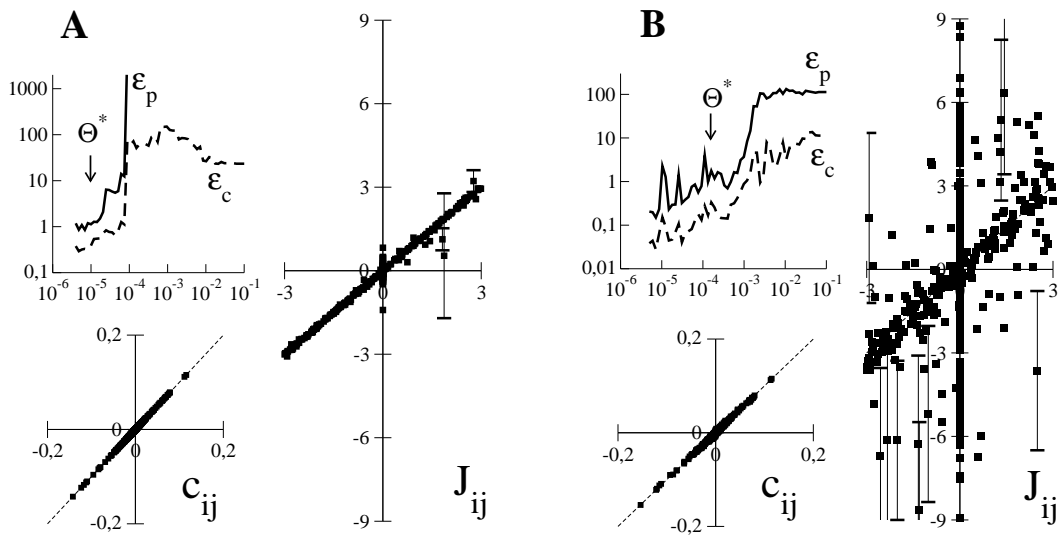}
\caption{Outcome of the inference algorithm for an Erd\"os-Renyi random 
graph with $N=50$ spins, connectivity $d=10$, and with $B=10^6$ {\bf (a)} and $B=10^3$ {\bf (b)} sampled configurations. For each value of $B$ we show the errors $\epsilon_p, \epsilon_c$ vs. $\theta$, the inferred vs. data values of the correlations $c_{ij}$, and of the couplings $J_{ij}$. A few large error bars over $J_{ij}$ (calculated from $\chi^{-1}$) are shown. Values of $J_{ij}$ were chosen uniformly at random in $[-3;3]$, and fields were set to $h_{i}=-1$.}
\label{fig:er}
\end{center}
\end{figure}

Figure~\ref{fig:er} shows the outcome of the algorithm when data are generated from an  Erd\"os-Renyi model of  connectivity $d=10$. On the selected bonds $(i,j)$ the couplings $J_{ij}$ were chosen uniformly at random in $[-3;3]$. All other couplings $J_{ij}$ were set to zero, and the fields were $h_{i}=-1$. Values of the parameters are such that the system is in the paramagnetic phase (in the thermodynamic limit). Panel $A$ shows the inference with good sampling (the data  are obtained by averaging over $B=10^6$ Monte Carlo configurations), while Panel $B$  shows the inference with poor sampling ($B=10^3$). At $\Theta^*$ the data are  reconstructed within the expected statistical fluctuations and, correspondingly, couplings are found back within the statistical error bars $\delta J_{ij}$. In the case of a large sampling noise case $B=10^3$, the statistical fluctuations $\delta J_{ij}$ are so large that most of the inferred couplings are unreliable. The inference of the complete network is thus not possible.  The maximal size of clusters at $\theta^*$ increases with the average degree (Section \ref{sectime}); we find $K_{max}=9$ and  $K_{max}=7$ for, respectively, $B=10^6$ and $B=10^3$.

\begin{figure}[t]
\begin{center}
\epsfig{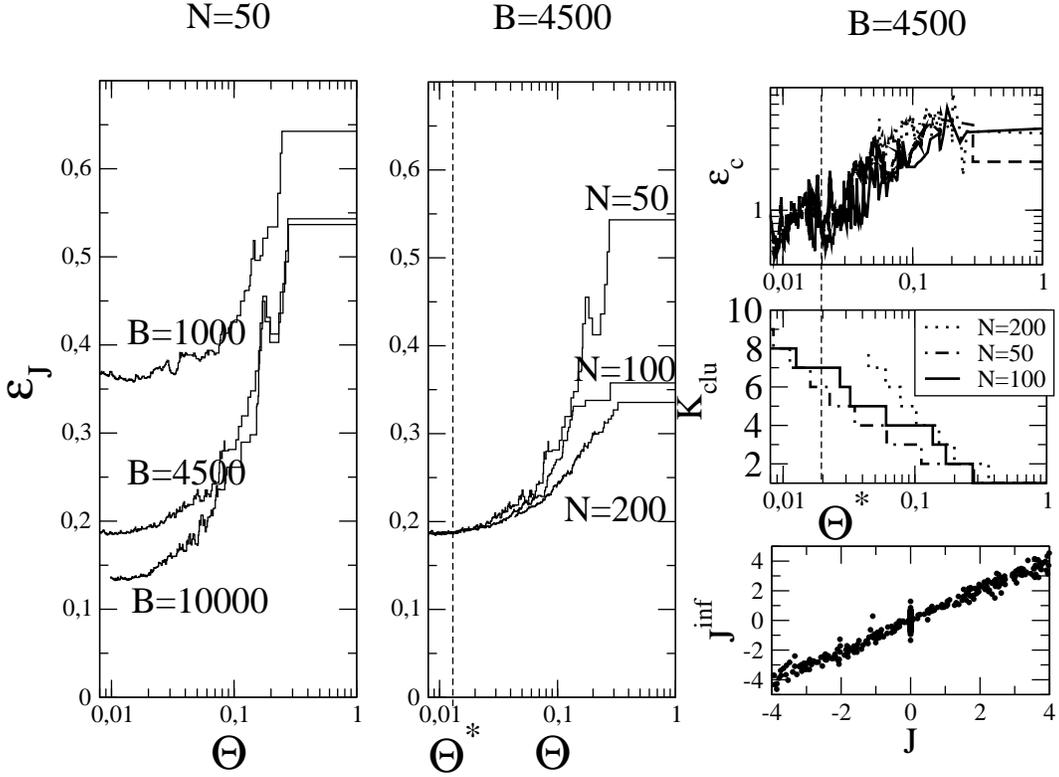}
\caption{Outcome of the inference algorithm for an Erd\"os-Renyi random graph  with $N$ spins, and connectivity $d=5$. Values of $J_{ij}$ were chosen uniformly at random in $[-4;4]$, and fields were set to $h_{i}=-\frac 12\sum_{j (\ne i)} J_{ij}$. Left: error $\epsilon _J$ on the inferred couplings as a function of  $\Theta$ for $N=50$ and $B=1000, 4500, 10000$ configurations. Middle:  error $\epsilon _J$  vs. $\Theta$ for $N=50, 100, 200$ and for $B=4500$. Right: $\epsilon_c$ (top) and $K_{clu}$ (middle) as functions of $\Theta$; the inferred couplings $\bf{J}^{inf}$ are compared to their true values in the bottom panel for $N=100$ and threshold $\Theta^*$. }
\label{erd5}  
\end{center}
\end{figure}

Fig.~\ref{erd5} shows the outcome of the algorithm on an Erd\"os-Renyi random graph with a smaller connectivity, $d=5$, and for values of the couplings $J_{ij}$  chosen uniformly at random in $[-4;4]$. The fields are set to  $h_{i}=-\frac 12\sum_{j (\ne i)} J_{ij}$, in such a way that the corresponding fields in spin variable $\pm 1$ vanish. This is an example of a  smaller connectivity system in the spin-glass phase. We have studied the performance of the algorithm as a function of the threshold $\Theta$, by varying the system size $N$ from 50 to 200 and the number of sampled configurations from $B=1000$ to $10000$. The algorithm is able to reach $\epsilon_c=1$ at large  thresholds, with a small number of selected clusters, {\em e.g.} $N_{clu} < 1000$ and $K_{max}=7$ for $N=100$. The threshold $\Theta^*$ for which $\epsilon_c=1$ corresponds to the beginning of the plateau for $\epsilon_J$. For smaller thresholds $\epsilon_c$ decreases and data are overfitted. The height of the plateau for $\epsilon_J$ coincides with the calculated statistical error $\delta J$; it scales as $1/\sqrt {B}$ and does not strongly depend on $N$.

\subsection{Computational Time}\label{sectime}

\begin{figure}[t]
\begin{center}
\epsfig{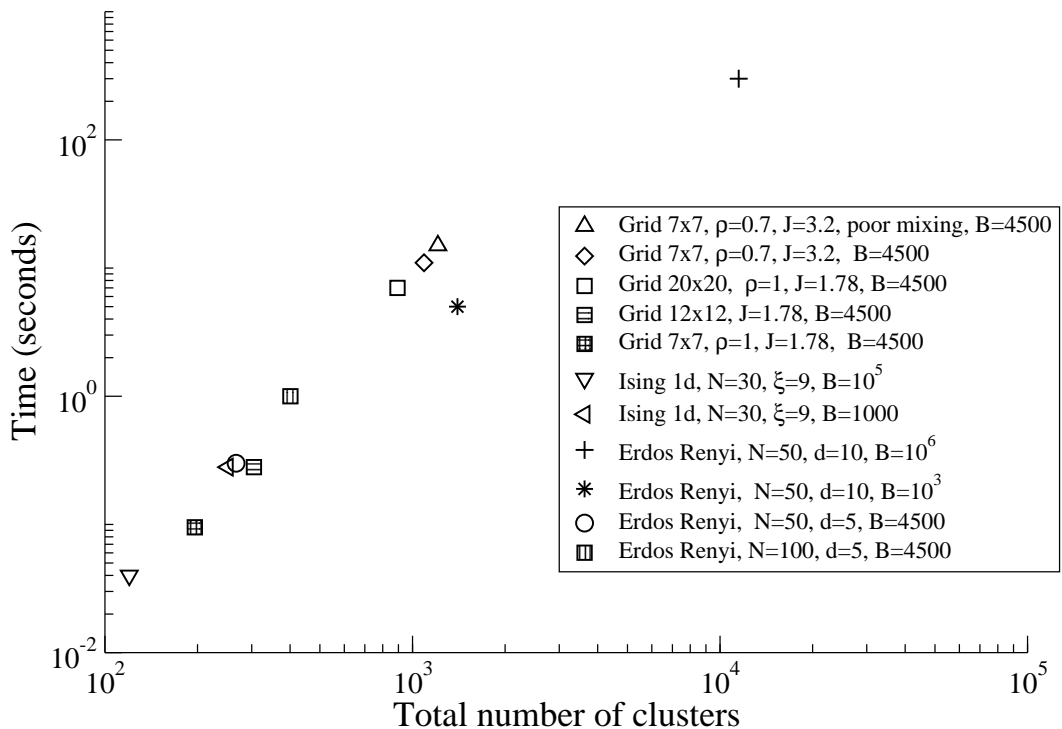}
\caption{Computational time of the cluster algorithm at threshold $\Theta^*$ for the different examples shown Section \ref{secappli}. The computational time grows with the number of processed clusters, which depends on  the structure of the interaction graph and on the number of sampled configurations more than the sole number of variables, $N$. For a fixed maximal size of clusters, $K_{max}$, the running time is roughly proportional to the number of clusters. Unless explicitly stated otherwise, the sampling is realized in good mixing conditions. Times were measured on one core of a 2.8 GHz Intel Core 2 Quad desktop computer.}
\label{tempi}
\end{center}
\end{figure}

For a given value $\Theta$ of the threshold the computational time can be estimated through
\begin{equation}
\text{time} \simeq \sum_{K=1}^{K_{max}} N_{clu}(K)\; 2^K \ ,
\end{equation}
where $N_{clu}(K)$ is the number of selected clusters of size $K$, and $2^K$ is the number of operations necessary to calculate exactly the partition function of a sub-system of size $K$. As the number of selected clusters depends on the interaction graph, the computational time is sensitive to the structure of the interaction graph, while it does not depend too much on the correlation length of the system. For instance, the number of processed clusters and the computational time for Erd\"os-Renyi graphs is larger for the connectivity $d=10$ than for $d=5$. 

Moreover, as the sampling noise increases (the value of $B$ is made smaller), so does  the threshold value $\Theta^*$. As less precision is needed in the reconstructed frequencies and correlations, the size of the selected clusters is reduced. As a consequence, the computational time is reduced. Figure~\ref{tempi} illustrates this statement for Erd\"os-Renyi random graphs: the running time increases with the quality of the sampling, {\em i.e.} increases with the value of $B$.  In some very noisy cases, however, large size clusters which are due only to the noise and do not reflect the interaction network can be processed. As an example,  the number of clusters for the unidimensional Ising model  with $\xi=9$, $B=1000$ is larger than the one for $B=4500$. Another illustration is given by the diluted $7\times7$ grid with $J=3.2$, which requires the processing of many clusters of large size ($K_{max}=8$). 

\section{Conclusion and perspectives}
\label{sec-conclusion}

In this paper, we have presented an adaptive cluster expansion to infer the interactions between a set of Ising variables from the measure of their equilibrium correlations. We have discussed the statistical mechanics of this expansion, and shown applications of the algorithm to artificial data generated using Ising models on unidimensional and bidimensional  lattices, as well as on Erdos-Renyi random graphs.

We have in particular underlined the important conditions on the inverse problem that should be fulfilled for our algorithm to be efficient. The essential condition is that the inverse susceptibility, which determines the change of a coupling (or a field) resulting from a change in the data (1- or 2-spin frequencies) should be well-conditioned. We stress that this property is not incompatible with the presence of a long-range susceptibility. Hence, the inverse problem can be easy to handle even in the presence of long-range correlations. As far as our algorithm is concerned, this condition entails that the maximal size $K_{clu}$ of the clusters which need to be taken into account remains small even if the correlation length of the system is large.

The origin of this condition is that our algorithm builds up, by definition, an interaction network defining a well-conditioned Ising model. Indeed, in the absence of reference entropy ($S_0=0$), the cross-entropy $S({\bf p})$ is approximated through a sum of a restricted number of cluster-entropies, see (\ref{recur-entro-approx}). For sufficiently large thresholds $\Theta$, most quadruplets of variables, say, $i,j,k,l$, do not appear in any selected cluster (of size $K\ge 4$); hence, most of the entries $(\chi^{-1})_{ij,kl}$ of the inverse susceptibility matrix entries vanish according to (\ref{inverseresponse}). In the presence of the reference entropy $S_0=S_{MF}$, $\boldsymbol\chi ^{-1}$ is not guaranteed to be sparse any more due to the contribution $\boldsymbol\chi ^{-1}_0= - \frac{\partial S_0}{\partial {\bf p}\partial {\bf p}}$. However, when a regularization is introduced, {\em e.g.} based on the norm $L_1$ (\ref{regul1}), the network of interactions $(J_0)_{ij}$ is highly diluted, and we expect $\boldsymbol\chi ^{-1}_0$ to be well-conditioned, too. Further investigations of this point would be very useful.

According to the discussion of Section \ref{sec-chi} inverse problems corresponding to Ising models on finite-dimensional lattices are well-conditioned in the perfect sampling limit. The introduction of a threshold over the minimal values of cluster-entropies allows us to force the inverse problem to be well-conditioned even in the presence of sampling noise. We have checked this statement on inverse problems corresponding to 'critical' Ising models. While the correlation length increases with the size of the system, the maximal size of the clusters, $K_{clu}$, remains roughly constant. Therefore, the computational complexity of the algorithm increases only linearly with the system size.
   
An essential feature of inverse problems is that data are generally obtained from a finite sampling and, therefore, frequencies and pairwise correlations are plagued by sampling noise. Avoiding overfitting is a primary goal for an inference algorithm. This goal is achieved, in our algorithm, by the introduction of the threshold $\Theta$. As a result most of the clusters are discarded, and in particular, those whose contributions would convey very little information about the true nature of the underlying interaction network. Fixing the threshold value such that the relative reconstruction errors $\epsilon_p$ and $\epsilon_c$ are of the order of one corresponds to the maximal accuracy allowed by the quality of the data. 
 
The cluster expansion introduced here differs from other classical cluster expansions, developed in the contexts of the theory of liquids and of computational physics. In particular we do not impose consistency equations for the marginal probabilities over the clusters. Our expansion scheme is simpler, and requires only the knowledge of the individual and pairwise frequencies of the variables in the cluster. Moreover, the cluster construction and selection rules prevents any combinatorial explosion of the computational time.

Several points would deserve further investigations. Among them the discussion of the convergence properties of the expansion, started in Section \ref{sec-convergence}, should be expanded and improved. A natural and interesting question is to ask how the series behaves when the packets of Fig.~\ref{fig-packets} start mixing, {\em i.e.} in the presence of a strong sampling noise. Another aspect which should be better understood is the influence of the construction rule. Our heuristic consists in merging two almost completely overlapping clusters of size $K$ to build a new cluster of size $K+1$ (provided its entropy is larger than $\Theta$). This rule has a simple intuitive interpretation, compatible with the notion of interaction path, and attempts with other rules have been less fruitful. However, a deeper theoretical understanding and justification is clearly needed. Last of all, the {\em a posteriori} validation of the method relies on the use of a Monte Carlo simulation to calculate $\epsilon_c$ and $\epsilon_p$. We have tested another procedure to avoid the use of a Monte Carlo calculation, based on a partial resummation of the cluster contributions corresponding to the free-energy (at fixed couplings and fields). This procedure, whose applicability goes beyond the inverse problem, will be detailed in a further publication.
 
\vskip .5cm
{\bf Acknowledgements:} We are grateful to J. Barton, J. Lebowitz, E. Speer for very useful and stimulating discussions, in particular regarding the correspondence between the inverse susceptibility and the direct correlation functions and the practical implementation of the inference algorithm.  We thank E. Aurell for pointing to us the difference between $P$ and $Q$, see Section \ref{secgrid490}.

\appendix

\section{Optimal choice for the regularization parameter $\gamma$}
\label{appchoice}

In this Appendix we discuss how the optimal value for the parameter $\gamma$ in the $L_2$-regularization (\ref{regul2}) can be determined. As explained in Section \ref{bayessec}, the regularization term can be interpreted as a Gaussian prior $P_0$ over the couplings. Let us call $\sigma^2$ the variance of this prior. Parameters $\gamma$ and $1/\sigma^2$ are related through
\begin{equation}
\gamma \,p^2\,(1-p)^2 = \frac{1}{2\,\sigma^2\, B}\ ,
\end{equation}
where we have assumed that the 1-site frequencies $p_i$ are uniformly equal to $p$. To calculate the optimal value for $\gamma$, or, equivalently, for $\sigma^2$, we start with the case of a single spin for the sake of simplicity, and then turn to the general case of more than one spin. 

\subsubsection{Case of $N=1$ spin}\label{appchoice1}

\begin{figure}[h]
\begin{center}
\epsfig{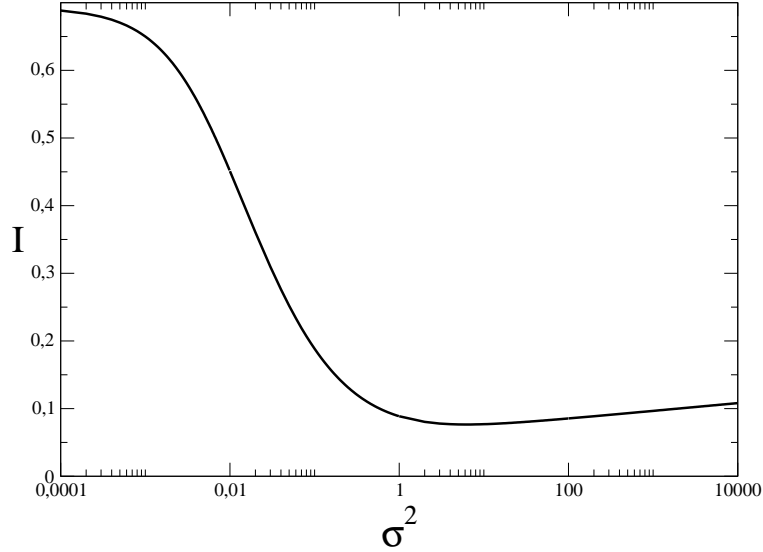}
\caption{Logarithm of the marginal likelihood (with a minus sign, and divided by the size $B$ of the data set) versus variance $\sigma^2$
of the prior distribution of the field. Parameters are $B=100$, $p=.02$.}
\label{fig-shapeI}
\end{center}
\end{figure}

For a unique spin subjected to a field $h$ the likelihood of the set of sampled spin values, $\{\sigma^{\tau}\}$, is $P_h[\sigma]=\exp(B\, p\, h)/(1+e^h)^B$. Here $p$ denotes the average value of the spin over the sampled configurations (\ref{mc}). We obtain the {\em a posteriori} probability (\ref{ppost}) for the field $h$ given the frequency $p$,
\begin{equation}
P_{post} [h|p] = \frac{\exp (-h^2/(2\sigma^2)+ B\;  p\, h -B\log (1+e^h))/\sqrt{2\pi\sigma^2}}{{\cal P}(p,B,\sigma^2)}
\end{equation}
where the denominator ${\cal P}(p,B,\sigma^2)$ (marginal likelihood) is simply the integral of the numerator over all real-valued fields $h$. Given $p$ and $B$ we plot $I=-\log {\cal P}(p,B,\sigma^2)/B$ as a function of $\sigma^2$. The general shape of $I$ is shown in Fig.~\ref{fig-shapeI}. The value of $\sigma^2$ minimizing $I$ is the most likely to have generated the data, and should be chosen on Bayesian grounds. 

For more than one spin calculating the marginal likelihood would be difficult. We thus need an alternative way of obtaining the best value for $\sigma^2$. The idea is to calculate $I$ through a saddle-point method, and include the Gaussian corrections which turn out to be crucial. This approach is correct when the size of the data set is large. A straightforward calculation leads to
\begin{equation} \label{approxi}
I \simeq \log \big(1+\exp( h^*)\big) -p\; h^* + \frac{\Gamma}2 \; (h^*)^2 + \frac 1{2B} \log\bigg[ 1 + \frac 1\Gamma\, \frac{\exp(- h^*)}{(1+\exp(- h^*))^2}\bigg]
\end{equation}
where $\Gamma=1/(B\sigma^2)$ and  $h^*$ denotes the root of $(1 + \exp( -h^*))^{-1}-p +\Gamma\; h^*=0$. $I$ decreases from $I(\sigma^2=0)=\log 2$ with a strong negative slope, $dI/d\sigma^2 (0)\simeq - B\; p^2$, and increases as $\log \sigma^2/(2B)$ for large values of the variance. Expression (\ref{approxi}) cannot be distinguished from the logarithm of the true marginal likelihood  $I$ shown in Fig.~\ref{fig-shapeI}.

\subsubsection{Case of $N\ge 2$ spins}

The above saddle-point approach can be generalized to any number $N$ of spins, with the result
\begin{equation} \label{isuite1}
I \simeq S_{Ising}[\{h_i,J_{ij}\}|{\bf p}] +\frac 1{2B} 
\log\ \mbox{det} \left( 1+ \frac{\bf H}{\Gamma} \right)
\end{equation}
where $S_{Ising}$ was defined in (\ref{s2}) and $\boldsymbol \chi$ is the $N+\frac 12N(N-1)= \frac 12 N(N+1)$-dimensional Hessian matrix composed of the second derivatives of $S_{Ising}$ with respect to the couplings and fields (\ref{defchi}). In principle $\boldsymbol\chi$ could be diagonalized and the expression (\ref{isuite1}) calculated. However this task would be time-consuming. As we have seen in the previous subsection we expect $I$ not to increase too quickly with $\sigma^2$ (for not too small variances) and approximate calculations of $I$ can be done under some data-dependent hypothesis. We now give an example of such an approximation, valid in the case of multi-electrode recordings of neural cell populations.

A simplification arises when the number $B$ of configurations and the frequency $p$ are such that: (a) each spin $i$ is active ($=1$) in a number of configurations much larger than 1 and much smaller than $B$ {\em i.e.} $1 \ll B\times p \ll B$; (b) the number $n_2$ of pairs of spins that are never active together is much larger than one and much smaller than $\frac {N(N-1)}2$. These assumptions are generically true for the applications to neurobiological data. For instance, the recording of the activity of $N=40$ salamander retinal ganglion cells in \cite{bialek} fulfills conditions (a) and (b) for a binning time  $\Delta t=5$~msec: a cell $i$ firing at least once in a time-bin corresponds to $\sigma_i=1$, while a silent cell is indicated by $\sigma_i=0$. More precisely: (a) the least and most active neurons respectively fire 891 and 17,163 times (among $B=636,000$ configurations); 
(b) $n_2=34$ pairs of cells (among $780$ pairs) are never active together. 

Condition (a) allows us to omit the presence of $\Gamma$ in the calculation of the fields, $h_i \simeq \log p_i$, to the first order of a large (negative) field expansion. Condition (b) forces us to introduce a non--zero $\Gamma$ to calculate the couplings, with the result that interactions between pairs $i,j$ of cells not active together are equal to $J_{ij} \simeq \log \Gamma + O(\log \log (1/\Gamma ) )$. Finally we obtain the asymptotic scaling of the entropy when $\Gamma \to 0$,
\begin{equation}
S_{Ising} \simeq n_2\ \frac{\Gamma}2 \  (\log \Gamma)^2 + O\Big(\Gamma
\log \Gamma \log\log \frac 1\Gamma \Big) \ .
\end{equation}
We are now left with the calculation of the determinant in (\ref{isuite1}). From assumption (b) the number of pairs of neurons not spiking together is small with respect to $N^2$, meaning that most of the eigenvalues $\lambda ^a$ of the Hessian matrix of $S_{Ising}$ are non zero. Hence,
\begin{equation}
 \log\ \mbox{det} \left( 1+\frac{\boldsymbol\chi}{\Gamma} \right) = \sum _{a=1}
^{N(N-1)/2} \log \left(1 + \frac{ \lambda ^a}{\Gamma} \right) \simeq
- \frac {N^2}2 \ \log \Gamma \ .
\end{equation} 
Putting both contributions to $I$ together we get
\begin{equation}
I (\Gamma ) \simeq   n_2\ \frac{\Gamma}2 \  (\log \Gamma)^2 - \frac
{N^2}{4 B} \ \log \Gamma \ .
\end{equation}
The optimal value for the variance $\sigma ^2$ is the root of
\begin{equation}
\frac{dI}{d\Gamma} (\Gamma ) =0 \simeq  \frac {n_2} 2 \  
(\log \Gamma)^2 - \frac {N^2}{4\;B \;\Gamma } \simeq \frac {n_2} 2 \  
(\log B)^2 - \frac{N^2}{4}\;\sigma^2\ .
\end{equation}
We finally deduce the optimal variance
\begin{equation}
\sigma^2 \simeq 2\; n_2\; \left(\frac{\log B}N\right) ^2 \ .
\end{equation}
For the data described above we find $\sigma^2\simeq 8$.

\section{Expression of the entropy of clusters with size $K=3$}
\label{entroK3}

In this Appendix, we give the analytical expression for the entropy of a cluster with $K=3$ spins. Using this expression instead of minimizing the cross-entropy (\ref{s2}) offers a valuable computational speed-up as there are $O(N^3)$ clusters of size $K=3$. We start with the definition of the entropy $P(\boldsymbol\sigma)$:
\begin{equation}
S_3=-\sum_{\substack{\sigma_1=0,1\\ \sigma_2=0,1\\ \sigma_3=0,1}} P(\sigma_1,\sigma_2,\sigma_3)\; \log P(\sigma_1,\sigma_2,\sigma_3) \ .
\label{entro3}
\end{equation}
We then replace the probabilities $P(\sigma_1,\sigma_2,\sigma_3)$ of the eight configurations of the three spins above with their expressions in terms of the probabilities $\{p_i,p_{kl}\}$ in the data, and of the probability $p_{123}$ that the three spins are equal to 1: 
\begin{eqnarray}
P(1,1,1)&=&p_{123} \nonumber \\
P(1,1,0)&=& p_{12}-p_{123} \nonumber \\
P(1,0,1)&=& p_{13}-p_{123} \nonumber \\
P(0,1,1)&=& p_{23}-p_{123} \nonumber \\
P(0,0,1)&=&p_3-p_{23}-p_{13} +p_{123} \nonumber \\
P(0,1,0)&=&p_2-p_{12}-p_{23} +p_{123} \nonumber \\
P(1,0,0)&=&p_1-p_{13}-p_{12} +p_{123} \nonumber \\
P(0,0,0)&=&1-p_1-p_2-p_3+p_{12}+p_{13}+p_{23}-p_{123} \nonumber \\
\end{eqnarray}  
The only unknown quantity (not available in ${\bf p}$) is the probability $p_{123}$. To determine $p_{123}$ we impose
\begin{equation}
\label{j30}
\frac{d S_3}{d p_{123} }=0 \ ,
\end{equation}
which means that the three-body coupling $J_{123}$ vanishes. Condition (\ref{j30}) gives a third degree equation on $p_{123}$,
\begin{equation}
p_{123}^3+\alpha\, p_{123}^2+\beta\, p_{123}+\gamma=0
\label{p123}
\end{equation}
with
\begin{eqnarray}
\alpha &=&p_1\,p_2+p_1\,p_3+p_2\,p_3-p_1\,p_{23}-p_2\,p_{13}-p_3\,p_{12}
-p_{12}-p_{23}-p_{13}\ , \nonumber \\
\beta &=&p_1\,p_{23}^2+p_2\,p_{13}^2+p_3\,p_{12}^2
-p_1\,p_2\,p_{23}-p_1\,p_2\,p_{13}- p_1\,p_3\,p_{12} 
-p_1\,p_3\,p_{23}- \nonumber \\
 & &p_2\,p_3\,p_{12} -p_2\,p_3\,p_{13}+2\,p_{12}\,p_{13}\,p_{23}+
p_{12}\,p_{13} + p_{12}\,p_{23}+p_{13}\,p_{23}+p_1\,p_2\,p_{3} \ , \nonumber \\
\gamma &=& p_1\,p_2+p_3\;(1-p_1-p_2-p_3+p_{12}+p_{13}+p_{23}) \ .
\end{eqnarray}
Upon substitution of $p_{123}$ in (\ref{entro3}) we obtain the desired cross-entropy $S_3$, as a function of the three average values $p_i$ and the three two-point averages $p_{kl}$. The expression of the cluster-entropy is given by,
\begin{equation}
\Delta S_{ (i,j,k)}=S_3(p_i,p_j,p_k,p_{ij},p_{ik},p_{jk}) -\Delta S_{(i,j)} - \Delta S_{(i,k)}-\Delta S_{(j,k)} -\Delta S_{(i)}-\Delta S_{(j)}-\Delta S_{(k)}\ ,
\end{equation}
according to (\ref{deltas3}). The expressions of the cluster-entropies for one and two spins are given by, respectively, (\ref{deltas1}) and (\ref{deltas2}).
Similarly, one obtains the expressions for the contributions of the 3-spin cluster to the values of the interactions parameters by differentiating $\Delta S$ with respect to the $p_i$'s and the $p_{kl}$'s.

\section{Leading diagrammatic contributions to small cluster-entropies}
\label{app-diag}

We analyze the dominant diagrams contributing to the cluster-entropies for the various values of the cluster sizes, $K$, in the limit of small connected correlations $c_{kl}$.

\subsection{Case $K=2$}

The entropy $\Delta S_{(i,j)}$ of a 2-spin cluster is the sum of all diagrammatic contributions containing two spins and an arbitrary number of links between them, corresponding to the power of the expansion parameter $M_{ij}=c_{ij}/(p_i\,(1-p_i)\,p_j\,(1-p_j))$ (Fig.~\ref{fig-diag}) and Section \ref{secexpdiag}. For small values of $M_{ij}$ the largest contribution to $\Delta S_{(i,j)}$ is the one with three links (cubic power of $M_{ij}$), if the reference entropy $S_0=S_{MF}$ removes the two-link loop diagram. The entropy contribution of this diagram was computed in  \cite{diag}, with the result
\begin{equation}
\label{hats2}
\Delta S^{(3)}_{i,j}=\alpha_{i,j} \, (c_{ij})^3\ ,
\end{equation}
where
\begin{equation}
\alpha_{i,j}^{(3)}= \frac {(2\, p_i-1)\,(2\, p_j-1)}{6 \;(p_i)^2\,(1-p_i)^2\;(p_j)^2\,(1-p_j)^2} \ .
\label{alpha}
\end{equation}
The superscript 3 refers to the power of the connected correlation.

\subsection{Case $K=3$}\label{secleading3}

For $K=3$  the leading term to $\Delta S_{(i,j,k})$ in powers of $M_{ij}$ was not derived analytically in \cite{diag}. Based on the studies of the unidimensional Ising model and the independent spin models (Appendix {\ref{app-ising1d}), we find that the leading diagrams are diagrams (a), (b), (c) in Fig.~\ref{diag3entrofig} (bold diagrams in Fig.~\ref{fig-diag}), whose sum is given by
\begin{equation}
\label{hats3}
\Delta S^{(5)}_{i,j,k}= \alpha^{(5)}_{ijk}\;  (c_{ij})^2\,(c_{jk})^2\,c_{ki} \,+\alpha^{(5)}_{jik} \;
(c_{ij})^2\,c_{jk}\,(c_{ki})^2 +  \alpha^{(5)}_{jki}\;c_{ij}\,(c_{jk})^2\,(c_{ki})^2\ ,
\end{equation}
with 
\begin{equation}\label{alpha3}
\alpha_{ijk} ^{(5)}=-\frac{(2 p_i-1)\,(2p_k-1)}{2\; (p_i)^2\,(1-p_i)^2\;(p_j)^2\,(1-p_j)^2\;
(p_k)^2\;(1-p_k)^2 }. 
\end{equation}
Note that $\alpha^{(5)}_{ijk}$ differs from $\alpha^{(5)}_{jik}$. We have also found the coefficients of the subsequent diagrams, of the order of $M^6$. These diagrams are labelled by (d), (e), (f) in Fig.~\ref{diag3entrofig}. Their total contribution to the cluster-entropy is
\begin{equation}
\label{hhats3}
\Delta S^{(6)}_{i,j,k}=  \alpha^{(6)}_{ijk}\;  (c_{ij})^3\,(c_{jk})^3+  \alpha^{(6)}_{jik}\;
(c_{ij})^3\,(c_{ki})^3 +   \alpha^{(6)}_{jki}\; (c_{jk})^3\,(c_{ki})^3 
\end{equation}
with 
\begin{equation}\label{alpha3p}
\alpha^{(6)}_{ijk} =\frac{(2 p_i-1)\,(2p_k-1)}{3\, (p_i)^2\,(1-p_i)^2\;(p_j)^3\,(1-p_j)^3\;
(p_k)^2\;(1-p_k)^2 }. 
\end{equation}

\begin{figure}
\begin{center}
\epsfig{file=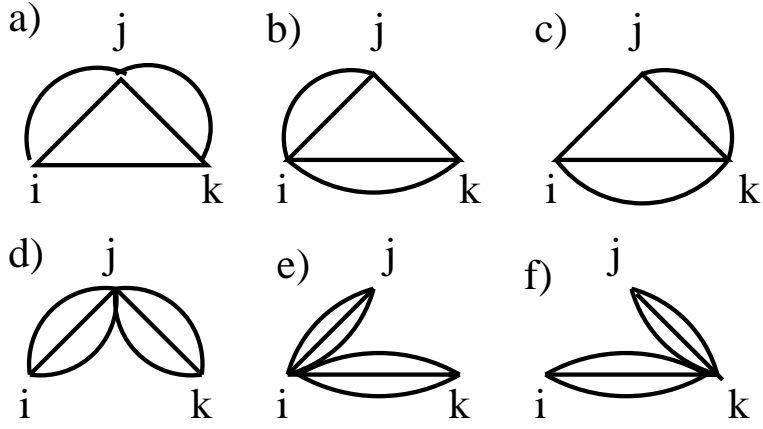,width=10cm}
\caption{Leading diagrams to the order $5$ (top) and $6$ (bottom) in the connected correlation for the entropy of 3-clusters.}
\label{diag3entrofig}
\end{center}
\end{figure}

\subsection{Generic case $K\ge 4$}

The above results for $K=3$ are easily generalized to any value of the cluster size $K\ge 4$. The diagrammatic expansion of  a $K$--spin cluster includes all circuits where pairs of spins are linked together. Each diagram with (one or two)  links between $i_l$ and $i_{l+1}$ ($l=1,\ldots, K-1$)  and (one or two) links  between $i_1$ and $i_K$ gives 
\begin{eqnarray}
\label{hatsk}
\Delta S^{(2K-1)}_{i_1,\ldots i_k}&=& \frac{(-1)^K}{2 \prod_{l=1}^{K} (p_{i_l})^2\,(1-p_{i_l} )^2}\;\bigg[
(2 p_{i_{k-1}}-1)\, (2 p_{i_{k}}-1)\, (c_{i_1,i_2})^2(c_{i_2,i_3})^2 \ldots (c_{i_{k-2},i_{k-1}})^2 c_{i_{k-1},i_{k}}  \nonumber \\
&+&  (2 p_{i_{k-2}}-1)\,(2 p_{i_{k-1}}-1)\, (c_{i_1,i_2})^2 \,(c_{i_2,i_3})^2 \ldots c_{i_{k-2},i_{k-1}} (c_{i_{k-1},i_{k}})^2+ \ldots  \nonumber \\
& +& (2 p_{i_1}-1)\,(2 p_{i_2}-1)\, c_{i_1,i_2}\,(c_{i_2,i_3})^2 \ldots (c_{i_{k-2},i_{k-1}})^2\, (c_{i_{k-1},i_{k}})^2  \bigg]\ .
\end{eqnarray}
At the next order in power of $M_{ij}$, each diagram with three links between $i_{l}$ and $i_{l+1}$  ($l=1, \ldots ,K-1$) gives a contribution
\begin{eqnarray}
\label{hatskk}
\Delta S_{i_1,\ldots i_k}^{(3K-3)}&=& \frac{(-1)^{K-1}\,(2 p_{i_1}-1)\;
(2 p_{i_k}-1)}{3 (p_{i_1})^2\,(1-p_{i_1} )^2 
\; (p_{i_k})^2\,(1-p_{i_k} )^2\prod_{l=2}^{K-1} (p_{i_l})^3\,(1-p_{i_l} )^3} \prod_{l=1}^{K-1} (c_{i_{l},i_{l+1}} )^3\ .
\end{eqnarray}

\section{Critical correlation length $\xi_c$ for the absolute convergence}
\label{app-bound}

In this Appendix, we briefly explain why the cluster-entropy series is absolutely convergent if and only if the correlation length $\xi$ is smaller than
\begin{equation}
\xi_c =Ê\frac \Omega{\log v} \ .
\end{equation}
Here, $\Omega =2$ when the reference entropy is $S_0=0$, and $\Omega =3$ when $S_0=S_{MF}$. Parameter $v$ denotes the number of neighbours of a site on the lattice, supposed to be uniform. For instance, $v=2D$ on a hypercubic lattice in dimension $D\ge 1$.

Consider a set of $K$ distinct points on the lattice. Let ${\cal N} (L)$ be the number of closed paths of length $L$ visiting all $K$ points. We obviously have ${\cal N} (L)\le v^L$. Hence, the series
\begin{equation}\label{series67}
\sum _L {\cal N} (L) \; \exp\Big(\! -\Omega\, L/\xi\Big) 
\end{equation}
is convergent if $\xi < \xi_c$. Reciprocally, let $L_{0}$ be the length of the shortest closed path ${\cal C}_0$ encircling the $K$ points. A closed path of length $L_1+L_0$ can be built from ${\cal C}_0$ by attaching a closed loop of length $L_1$ to any one of the sites in ${\cal C}_0$. Hence, for $L\ge L_0+2$, ${\cal N} (L)\ge L_0\;v^{L-L_0}$. We deduce that the series (\ref{series67}) is divergent if $\xi>\xi_c$.

\section{Distribution of cluster-entropies for the Independent Spin model}
\label{secisdominant}

We generate $B$ configurations of $N$ independent spins $\sigma_i$.  Spin $i$ is equal to $1$ with probability $p$ and to zero with probability $1-p$ (for simplicity we assume here that all the frequencies $p_i$ are equal to the same value $p$). The empirical connected correlations $c_{ij}$ computed from the $B$ sampled configurations of spins are generally non zero. The marginal distribution of $c_{ij}$ is a normal law, with zero mean and standard deviation (\ref{valuecb}). The largest values of the correlations are, for a system with $N$ spins, of the order of 
\begin{equation}
c_{ij}^{MAX}=c_B\;\sqrt{ 4 \log N} \ ,
\label{largec}
\end{equation}
according to extreme value theory.

\begin{figure} [t]
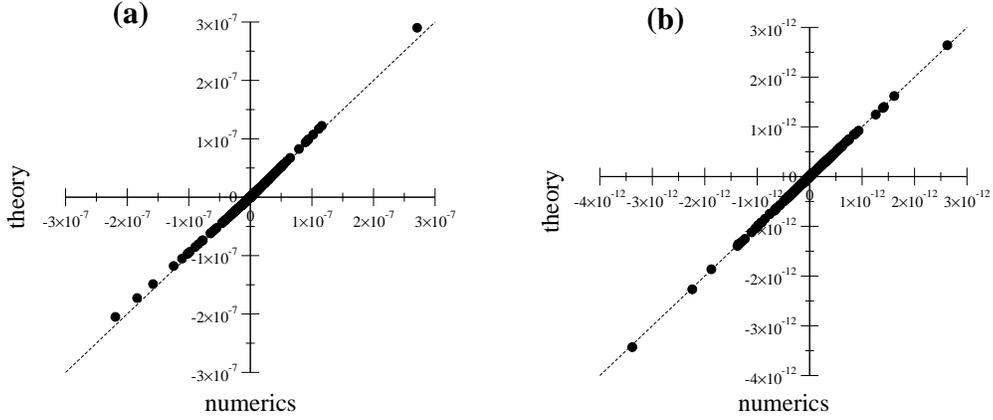

\begin{center}
\epsfig{file=fig34a.eps,width=6cm}
\hskip 1cm
\epsfig{file=fig34b.eps,width=6cm}
\caption{Comparison of the numerical ($x$-axis) and theoretical ($y$-axis) values for the entropies of clusters $(i,j)$ {\bf (a)} and $(i,j,k)$ {\bf (b)}. The system is made of $N=40$ independent spins, with the same $p_i$ as in the neural data of Ref. \cite{bialek}; the average value of the $p_i$'s is $p=0.0238$. Theoretical predictions correspond to (\ref{hats3}) and (\ref{hhats3}). The number of sampled configurations is $B=10^6$.}
\label{fig-entrocompalea}
\end{center}
\end{figure}

We compare in Fig.~\ref{fig-entrocompalea} formulas (\ref{hats3}) and (\ref{hhats3}) for, respectively, the cluster-entropies $\Delta S_{ij}$ and $\Delta S_{ijk}$ with numerics carried out from randomly sampled configurations. Each pair $(i,j)$ (Fig.~\ref{fig-entrocompalea}(a)) and triplet $(i,j,k)$ (Fig.~\ref{fig-entrocompalea}(b)) define a point, whose coordinates are the numerical and theoretical values of the entropy corresponding to the pair- or triplet-cluster. The agreement, for $B=10^6$ sampled configurations, is excellent due to the small value of $c_B\simeq 2\, 10^{-5}$.

\subsection{Distribution of cluster-entropies for $K=2$}

The distribution of the entropy of $K=2$-clusters for a set of $B=10^6$ configurations is shown in Fig.~\ref{fig-entro2alea}. To derive the analytical expression of the distribution in the $N\to\infty$ limit, we use the small-correlation formula (\ref{hats2}) for $\Delta S_{(1,2)}$, and the fact that the distribution of the connected correlation is Gaussian. As a result, approximating $\alpha_{1,2}$ with its average value $\alpha$ obtained by substituting $p_{1}$ and $p_2$ with $p$ in (\ref{alpha}), we obtain 
\begin{equation}
H_{IS}(\Delta S_{(1,2)})= \frac {\exp \Big(-\frac {(\Delta S_{(1,2)} )^{2/3}}{2 \,(c_B)^2\alpha^{2/3}}\Big)} {3\,\alpha^{1/3}\, \sqrt{2\pi (c_B)^2}\,(\Delta S_{(1,2)})^{2/3} } 
\label{pds}
\end{equation}
This distribution is a stretched exponential at infinity, and diverges in zero. Its standard deviation is
\begin{equation}
\sigma_{\Delta S_{(1,2)}} =\sqrt{15}\;\alpha\; (c_B)^3= 
\frac {\sqrt{15}\, (2p-1)^2}{6\, p(1-p)\, B^{3/2}}
\end{equation}
For $B=10^6$ and $p=0.0238$ we obtain that the standard deviation is $\simeq 2.7\, 10^{-8}$. Distribution (\ref{pds}) is  compared to the histogram obtains from numerics in Fig.~\ref{fig-entro2alea}. The standard deviation and the distribution at small entropies are in good agreement. 
Large values of the correlations (\ref{largec}) give rise to isolated values of $\Delta S_{(1,2)}$, of the order of 
\begin{equation}
\Delta S_{(1,2)} ^{MAX} \simeq (4 \log N )^{3/2}\; \Big( \langle (\Delta S_{(i,j)})^2\rangle\Big)^{1/2} \ ,
\label{dsind}
\end{equation}
approximately equal to $1.2\,10^{-6}$ for $N=40$. This value is about twice the largest cluster-entropy observed in Fig.~\ref{fig-entro2alea} for one particular realization of the sampled configurations.

\begin{figure}
\begin{center}
\epsfig{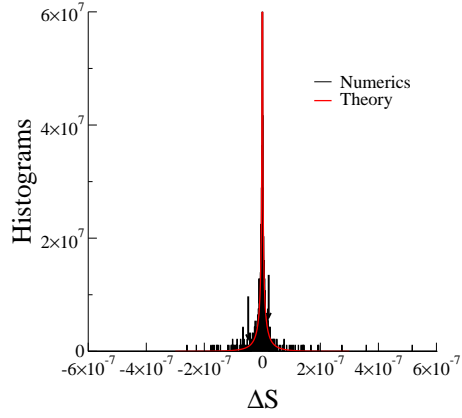}
\caption{Theoretical (red) and numerical (black) histograms $H_{IS}$  for the entropies $\Delta S$ of 2-spin clusters in a system of independent spins and $B=10^6$ configurations. Simulations were done with $N=40$ spins, with heterogeneous $p_i\simeq .0238$, see caption of Fig.~\ref{fig-entrocompalea}.}
\label{fig-entro2alea}
\end{center}
\end{figure}

\subsection{Distribution of the cluster-entropies for $K=3$}\label{dist3-calcul}

The leading order contribution to the entropy of a $3$-cluster is given by (\ref{hats3}). We want to calculate the distribution of $\Delta S_{(1,2,3)}$ when the connected correlations $c_{ij}$ are random Gaussian variables, of zero mean and variance $(c_B)^2$. We neglect the correlations between $c_{12},c_{13},c_{23}$, which is legitimate for large $B$. Let us call $x=\Delta S_{(1,2,3)}/ (\alpha (c_B)^5)$, with $\alpha$ given by (\ref{alpha3}), and let $P(x)$ be the probability density of $x$. Though we have not been able to find a closed expression for $P(x)$, the asymptotics behavior of $P$ for large or small arguments can characterized analytically.

\subsubsection{Large $x$ behaviour}

The Mellin transform  of $P$ \cite{flajolet} is
\begin{equation}\label{mellin}
\int _0^\infty dx P(x) x^\lambda = \left(\frac 2\pi\right)^{3/2}
\int _0^\infty dc_{12}\, dc_{13}\, dc_{23}\; e^{-F(c_{12},c_{13},c_{23})}
\end{equation}
where
\begin{equation}
F(c_{12},c_{13},c_{23}) =-\frac 12 \big( c_{12}^2+c_{13}^2+c_{13}^3) 
+\lambda \;
\log ( c_{12}c_{13}^2c_{23}^2+c_{12}^2c_{13}c_{23}^2+c_{12}^2c_{13}^2c_{23} ) \ .
\end{equation}
The tail of $P(x)$ at large $x$ can be studied by considering large values of $\lambda$. We expect the dominant contribution to the multiple integral on the right hand side of (\ref{mellin}) to come from large correlations. The location of the main contribution to the integral is the value of $(c_{12},c_{13},c_{23})$ which maximizes $F$. As $F$ is invariant under any permutation of its
arguments, we look for a maximum where $c_{12}=c_{13}=c_{23}\equiv c^*$. A straigthforward calculation shows that
\begin{equation}
c^*(\lambda)=\sqrt{\frac 53\, \lambda}, \quad F^*(\lambda)=\frac 52 \lambda \log\lambda +
\lambda \left(\log 3 +\frac 52 \log \frac 53 -\frac 52\right) \ .
\end{equation}
We now use the saddle-point method again, this time to estimate the integral on the left hand side of (\ref{mellin}). We obtain
\begin{equation}\label{legendre}
\max _x \big[ \log P(x) + \lambda \; \log x\big] =
F^*(\lambda) \ ,
\end{equation}
which is true when $\lambda$ is very large. Hence, $F^*(\lambda)$ is the Legendre transform of $\log P(x)$. Solving (\ref{legendre}) gives
\begin{equation}
\label{plargex}
\log P(x) \simeq - \frac 3 2 \, \left(\frac x3\right)^{2/5}
\end{equation}
at large $x$. The distribution of the cluster entropies $\Delta S_{(1,2,3)}$ thus follows a stretched exponential with exponent $\frac 25$. This decay is much slower than an exponential, and leads to large tails as can be seen from Fig.~\ref{fig-diag3vssoln40-alea}.

\subsubsection{Small $x$ behavior}

In order for the rescaled entropy $x$ to be small, at least one among the three correlations should be small according to (\ref{hats3}). Without restriction, we may assume that $c_{12}$ is the smallest of the three correlations. As $c_{12}$ appears once with power one, and twice with power two in (\ref{hats3}), we approximate $x \simeq c_{12}c_{13}^2c_{23}^2$. The Mellin transform of $P$ is, for negative $\lambda$,
\begin{equation}\label{melli2}
\int _0^\infty dx P(x) x^\lambda \simeq 3\left( \int _0^\infty dc \frac2{\sqrt{2\pi}}\;
c^\lambda \; e^{-c^2/2} \right)\;\left( \int _c^\infty dc \frac2{\sqrt{2\pi}}\;
c^{2\lambda} \; e^{-c^2/2} \right)^2 \ .
\end{equation}
The largest pole is located in $\lambda=-\frac 12$, and is of order $2$. According to standard results on the inversion of Mellin transforms \cite{flajolet}, we obtain a precise characterization of the divergence of the probability density at small $x$, 
\begin{equation}
P(x) \simeq C \; \frac{(-\log x)}{\sqrt{x}} \ ,
\end{equation}
where $C$ is a constant. 

\subsubsection{Typical value of $x$}

The typical value of the $x$ is defined through 
\begin{equation}
x_{typ} = \exp \left( \int _0^\infty dx \, P(x) \, \log x \right) \ .
\end{equation}
This quantity is less sensitive than the average value of $x$  to the presence of the
long tails in $P(x)$ at large $x$.  We write $x=(c_{12}c_{13}c_{23})^2 \,z$ where
\begin{equation}
z=\frac 1{c_{12}}+ \frac 1{c_{13}}+\frac 1{c_{23}} \ .
\end{equation}
Taking the logarithm, and averaging over the correlation, we obtain the following expression for the average value of the logarithm of $x$,
\begin{equation}
\langle\log  x \rangle = 6 \left(\int _0^\infty dc \,\frac{2\, \log c}{\sqrt{2\pi}}\; e^{-c^2/2} \right) + \langle 
\log z\rangle_z \ .
\end{equation}
The integral over $c$ in the above equation can be calculated numerically, with a value $\simeq -.63518$. To calculate the average value of $\log z$, we first use the identity
\begin{equation}
\log z = \int _{0}^\infty \frac{du}u \left( e^{-u} - e^{-u\, z}\right) 
\ .
\end{equation}
Taking the average on both sides, we have
\begin{equation}\label{integu}
\langle \log z\rangle _z = \int _{0}^\infty \frac{du}u \left( e^{-u} - \langle e^{-u\, z}
\rangle _z\right) 
\ .
\end{equation}
As $z$ is a sum of independent random variables its Laplace transform is the product of their Laplace transforms,
\begin{equation}
\langle e^{-u\, z}\rangle _z = \left(\int _0^\infty dc \,\frac{2}{\sqrt{2\pi}}\; e^{-c^2/2 - u/c} \right)^3 =
\left( \frac{\lambda}{2\pi\sqrt 2}\; G^{30}_{03} \left(\frac{\lambda^2}8 \bigg| \begin{array}{c}
-\\ -\frac 12 , 0, 0\end{array} \right) \right)^3 \ ,
\end{equation}
where $G$ is the Meijer-G function. we have calculated the integral (\ref{integu}) using the Mathematica software.  Some care must be taken for the numerical accuracy when $z\to 0$.  The outcome is $\langle \log z\rangle _z \simeq 2.09643$. Putting all contributions together we obtain $x_{typ}\simeq 0.18$. The corresponding values of $\Delta S_{(1,2,3)}$ are $3.5\times10^{-15}$ for $B=10^6$, and $1.1\times10^{-12}$ for $B=10^5$, in good agreement with the numerical value, respectively, $3\times10^{-15}$ and $9\times10^{-13}$.

\subsubsection{Standard deviation of x} 

We can easily evaluate the variance of each of the three terms of the sum in (\ref{hats3}) as the product of the variances of the three terms in the product, based on the approximation that the connected correlations $c_{ij}$ are independent stochastic variables. We obtain   
\begin{equation}\label{sigmads3}
\sigma_{\Delta S_{ijk}}
=\frac{ 3\,\sqrt{3} (2p-1)^2 \,(c_B)^5}{2 \,p^6\,(1-p)^6} =\frac{3\,\sqrt{3}\, (2\,p-1)^2}{2\,p\,(1-p)\,B^{5/2}} \ .
\end{equation}
With the values of $N$ and $p$ chosen in Fig.~\ref{fig-entro2alea}, we find that the standard deviation is of the order of $10^{-13}$ for $B=10^6$, and $2 \,10^{-11}$ for $B=10^5$, see Fig.~\ref{fig-diag3vssoln40-alea}. 

\subsection{Distribution of cluster-entropies for generic $K\ge 4$}

In general, for $K\ge 3$, the leading contribution to $\Delta S_{(i_1,i_2,\ldots i_K)}$ (\ref{hatsk}) contains the sum of $K\times (K-1)!/2$ terms, each one being the product of $K$ random variables, among which $(K-1)$ are elevated to power two, and 1 is elevated to power 1. The factor $K$ comes from the fact that there are $K$ way of choosing the single link in the circuits with $K$ spins. The factor $(K-1)!/2$ is the number of non equivalent circuits going through $K$ spins. We define the rescaled entropy $x$ through
\begin{equation}
x =|\Delta S_{(i_1,i_2,\ldots i_K)}|\times  \frac{2 \,(p(1-p))^{2K}}  {\sqrt{\frac{K!}2} \,(2p-1)^2\, (c_B)^{2K-1}}
\end{equation}
The approach followed in Section \ref{dist3-calcul} to calculate the asymptotic behaviour 
of the probability density $P$ of $x$ for $K=3$ can be extended without difficulty to any value of $K>3$. 
We find that $P(x)$ diverges when $x\to 0$, with
\begin{equation}
\label{divzero}
P(x) = C \; \frac{(-\log x)^{K-2}}{\sqrt x} \ .
\end{equation} 
where $C$ is a constant.
Hence the shape of the distribution of $x$ is, up to logarithmic terms, independent of $K$.
On the contrary, the tail of the distribution for large $x$ is very sensitive to $K$,
\begin{equation}\label{e22}
\log P(x) \simeq -\frac{K}{2(K-\frac 12)^2} \; \left(\frac xK\right)^{2/(2K-1)}
\end{equation}
As in the $K=3$ case, the  distribution of the cluster entropies $\Delta S$ 
follows a stretched exponential. The exponent of the stretched exponential decreases with  $K$. The variance of the distribution   can be easily evaluated, with the result
 \begin{equation}
\label{varianza}
\sigma_{\Delta S_{(i_1,i_2,\ldots i_K)}}=
\frac{\sqrt{K!/2} (\sqrt{3})^{K-1} (2p-1)^2 \,(c_B)^{2K-1}}{ 2 (p\,(1-p))^{2K}} \ .
\end{equation}

\section{Properties of the cluster-entropies of the one-dimensional Ising model}
\label{app-ising1d}

Consider the one-dimensional Ising model with nearest-neighbour couplings and periodic boundary conditions. The Hamiltonian of the model is
\begin{equation}\label{ei}
H=-h \sum_i \sigma_i- J\sum_i \sigma_i\,\sigma_{i+1}\ ,
\end{equation}
where the spins $\sigma_i$ take 0,1 values. The parameters of the model are the $N$ identical fields $h_i=h$, the $N$ couplings $J_{i,i+1}=J$ between neighbours and the remaining $N\times(N-3)/2$ zero couplings $J_{i,j}=0$ between non neighbours.

We recall a few elementary facts about the model. The transfer matrix is
\begin{equation}
T=\left(\begin{array}{c  c}
 e^{J+h}&e^{h/2} \\
e^{h/2} &1
\end{array} \right) \ .
\end{equation}
The eigenvalues are $\lambda_{\pm}=\frac1 2 \left(e^{J+h}+1\pm \sqrt{(e^{J+h}-1)^2+4\,e^h} \right)$, and the two components of the eigenvectors are, respectively,  $v_{\pm} (1)=-(1-\lambda_{\pm})/\sqrt{e^h+(1-\lambda_{\pm})^2}$ and $v_{\pm} (2)=e^{h/2}/\sqrt{e^h+(1-\lambda_{\pm})^2}$. The  probability that a spin is up is given by, in the $N\to \infty$ limit,
\begin{equation}
p=\langle \sigma _i\rangle_{\bf J} =\big(v_+(1)\big)^2 \ ,
\label{pising}
\end{equation}
and the connected correlation at distance $d$ is 
\begin{equation}
\label{cising}
c_{i,i+d}= \langle \sigma _i\,\sigma_{i+d}\rangle_{\bf J}-\langle \sigma _i\rangle_{\bf J}\langle \sigma _{i+d}\rangle _{\bf J}= p(1-p)\,\left(\frac{\lambda_-}{\lambda_+}\right)^{d}= p(1-p) \exp(-d/\xi)\ ,
\end{equation}
where the correlation length is given by $\xi=-1/\log(\lambda_-/\lambda_+)$.

\subsection{Calculation of the cluster-entropies and cancellation property}

In this Section, we show the exact cancellation property between the entropies of clusters with different sizes discussed in Section \ref{secpropdeltas}. We will see that this property is a direct consequence of the existence of a unique interaction path along the unidimensional chain.

\subsubsection{Case $S_0=0$}\label{sec39ising}

We first consider the case where the reference entropy is zero. Let $\Gamma=(i_1,i_2,\ldots ,i_K)$ be a cluster of size $K$, with $i_1<i_2<\ldots < i_K$. Due to the unidimensional nature of the interactions, the Gibbs distribution over the $K$-spin configurations $\boldsymbol\sigma$ obeys the chain rule,
\begin{equation}\label{chain}
P_{\bf J}[\boldsymbol\sigma] = P_{\bf J}(\sigma_{i_K}| \sigma_{i_{K-1}})\dots P_{\bf J}(\sigma_{i_4}| \sigma_{i_3})\, P_{\bf J}(\sigma_{i_3}| \sigma_{i_2}) \, P_{\bf J}(\sigma_{i_2},\sigma_{i_1}) \ ,
\end{equation}
where $P(\cdot , \cdot)$ and $P(\cdot|\cdot)$ denote, respectively, joint and conditional probabilites. Inserting the above formula into expression (\ref{s2bis}) for the cross-entropy, we obtain
\begin{eqnarray}
S_{Ising}({\bf J}|{\bf p}) &=& -\sum _{\boldsymbol\sigma} P_{obs}[\boldsymbol\sigma] \bigg( \sum _{l=2}^{K-1}  \log P_{\bf J}(\sigma_{i_{l+1}}|\sigma_{i_{l}}) + \log P_{\bf J}(\sigma_{i_2},\sigma_{i_1}) \bigg)  \nonumber \\&=& -\sum _{\boldsymbol\sigma} P_{obs}[\boldsymbol\sigma] \bigg( \sum _{l=1}^{K-1}  \log P_{\bf J}(\sigma_{i_{l+1}},\sigma_{i_{l}}) - \sum _{l=2}^{K-1} \log P_{\bf J}(\sigma_{i_l}) \bigg)  \label{appear}\\
&=& \sum _{l=1}^{K-1} S_{Ising} ( h^\rightarrow_{i_l},h^\leftarrow_{i_{l+1}},J_{i_l,i_{l+1}}|p_{i_l},p_{i_{l+1}},p_{i_l,i_{l+1}}) - \sum _{l=2}^{K-1}  S_{Ising}(h^0_{i_l}|p_{i_l}) \ .\nonumber
\end{eqnarray}
Each variable $\sigma_{i_l}$, with $l=2,\ldots ,K-1$, appears three times in (\ref{appear}), which explains the presence of three fields $h$ with the same index $i_l$. After optimization over ${\bf J}=(\{J_{i_l,i_{l+1}}\},\{h^\rightarrow_{i_l}\},\{h^\leftarrow_{i_l}\},\{h^0_{i_l}\})$ all these fields are equal, and we obtain
\begin{equation}\label{finalsdep}
S({\bf p}) =\sum _{l=1}^{K-1} S (p_{i_l},p_{i_{l+1}},p_{i_l,i_{l+1}}) - \sum _{l=2}^{K-1}  S(p_{i_l}) = \sum _{l=1}^{K-1} \Delta S_{(i_l,i_{l+1})}({\bf p})  +\sum _{l=2}^{K-1}  \Delta S_{(i_l)} ({\bf p}) \ .
\end{equation}
Hence the cross-entroy $S({\bf p})$ is the sum of the 1-cluster entropies and of the entropies of the 2-clusters made of adjacent sites. None of the other cluster-entropies appear, which proves that they cancel each other. To illustrate the cancellation mechanism, consider the case $K=3$. According to (\ref{finalsdep}), 
\begin{equation}\label{finalsdep3}
S({\bf p}) =\Delta S_{(i_1,i_2)}({\bf p})  +\Delta S_{(i_2,i_3)}({\bf p})+ \Delta S_{(i_1)} ({\bf p}) +\Delta S_{(i_2)} ({\bf p}) +\Delta S_{(i_3)} ({\bf p}) \ .
\end{equation}
Comparing with (\ref{deltas3}) we obtain
\begin{equation}
\Delta S_{(i_1,i_2,i_3)} ({\bf p})  = - \Delta S_{(i_1,i_3)} ({\bf p}) \ ,
\end{equation}
which shows that the entropy of a 3-cluster and the one of a 2-cluster with the same extremities $i_1,i_3$ are opposite to one another. By a recursive applications of (\ref{finalsdep}) this result can be immediately generalized to higher values of $K$. The entropy of a $K$-cluster is simply the entropy of the 2-cluster with the same extremities, multiplied by $(-1)^{K-2}$. Hence, identity (\ref{deltas_D1}) is established.
 
According to formula (\ref{cising}) for the connected correlation, the entropy of a two-site cluster is a function of the distance $d$ between the two sites:
\begin{equation}\label{deltas89}
\Delta S_{(i,i+d)} = F\Big(\exp(-d/\xi)\Big) \ ,
\end{equation}
where
\begin{eqnarray}\label{expF}
F(u ) &=& - 2p(1-p)(1-u)\log(1-u) - p\big(p+(1-p)u\big)  \log\Big(1+\frac{(1-p)u}p\Big) \nonumber \\ &-& (1-p)\big(1-p+p\,u\big)  \log\Big(1+\frac{p\,u}{1-p}\Big)\ .
\end{eqnarray}
To obtain the expression (\ref{expF}) for $F$, we have used formula (\ref{deltas2b}) for the 2-spin cluster-entropy, with $p_1=p_2=p$ and $p_{12}=p^2+c_{12}$, where the correlation $c_{12}$ is given by (\ref{cising}). Note that $F(u)=O(u^2)$ for small $u$, in agreement with scaling (\ref{asymptotics_deltas}).

\subsubsection{Case $S_0=S_{MF}$}

We now introduce the reference entropy $S_0=S_{MF}$. The matrix $M$ defined in (\ref{loopentro}) has elements
\begin{equation}
M_{ij} = \frac{c_{ij}}{\sqrt{p_i(1-p_i)p_j(1-p_j)}} = \exp(-|i-j|/\xi) \ .
\end{equation}
The inverse of $M$, $G=M^{-1}$, is a tridiagonal matrix, whose non zero elements are
\begin{equation}\label{tridiag}
G_{ii}= \frac{1+\exp(-2/\xi)}{1-\exp(-2/\xi)}\ , \quad G_{i,i\pm 1} = - \frac{\exp(-1/\xi)}{1-\exp(-2/\xi)} \ .
\end{equation}
Consider now the Gaussian model over $N$ real-values variables $\varphi_i$, whose energy function is given by
\begin{equation}\label{modelg}
E[\boldsymbol\varphi] = \frac 12 \sum _{i,j} G_{ij}\, \varphi_i\,\varphi_j \ .
\end{equation}
For this Gaussian model, the logarithm of the partition function is (up an irrelevant additional constant), $\log Z[{\bf G}]= -\frac 12 \log \text{det}\, G$. By construction, model (\ref{modelg}) is the solution of the inverse Gaussian problem, with data: $\langle \varphi_i \rangle =0, \langle \varphi_i \,\varphi_j\rangle =M_{ij}$. Hence, $S_0$ can be interpreted as the cross-entropy of Gaussian model (\ref{modelg}) under those data. A key feature of the Gaussian model above is that its interaction matrix $G_{ij}$ is tridiagonal. Only nearest neighbour variables are coupled to each other according to (\ref{tridiag}). We conclude that the Gaussian model is a one-dimensional model. Consequently, it obeys a chain rule similar to (\ref{chain}). This is the only requirement for the main conclusion of Section \ref{sec39ising} to hold: in the cluster expansion of $S_0$, the entropy of a $K$--cluster is simply equal to the entropy of the 2-cluster with the same extremities, multiplied by $(-1)^{K-2}$. As both the expansions of $S$ and the one of $S_0$ enjoy this property, so does the expansion of $S-S_0$.

We conclude this Section by the expression of the 2-cluster entropy $\Delta S_{(i,i+d)}$. In the presence of the reference entropy $S_0=S_{MF}$, we substract the following contribution to expression (\ref{deltas89}), see (\ref{loopentro}), 
\begin{equation}
(\Delta S_0\big) _{(i,i+d)} = \frac 12 \log \text{det} \, \left( \begin{array} {c c}
1& M_{i,i+d} \\M_{i,i+d}  & 1 \end{array}\right) \ .
\end{equation}
Hence, function $F(u)$ defined in (\ref{expF}) should be substracted $\frac 12 \log (1-u^2)$. It is a simple check that $F(u)- \frac 12 \log (1-u^2)= O(u^3)$, in agreement with scaling (\ref{deltas_D1_bis}).

\subsection{Examples and calculation of diagrammatic coefficients}

We now show the histograms of cluster-entropies for $K=2$ and $K=3$ for specific choices of $J,h$.  The averages $p_i$  and $p_{ij}$ were calculated exactly through formulas (\ref{pising}) and (\ref{cising}) (perfect sampling). Figure~\ref{fig-entroising}(a)  shows the histogram of entropies for clusters of the type $(i,i+d)$. Entropy values are discrete and labelled by the distance $d$. They  range from $10^{-2}$ (for nearest neighbours, distance $d=1$) to values smaller than $10^{-15}$ for $6<d<15$. All entropies smaller than the numerical accuracy $\simeq 10^{-15}$ are put in the peak at the origin. Expanding $F(u)$ to the lowest order in $u$ (for $S_0=S_{MF}$) we find the asymptotic formula for the 2-cluster entropy:
\begin{equation}
\label{ds2ising}
 \Delta S_{i,i+d} \simeq \frac{(2p-1)^2}{6\,p\,(1-p)}\; e^{-3\,d/\xi} \ ,
\end{equation}
in agreement with (\ref{hats2}). We have verified that this formula is in very good agreement with the numerics as soon as $d\ge 4$ for the parameters of Fig.~\ref{fig-entroising}. 

\begin{figure}[t]
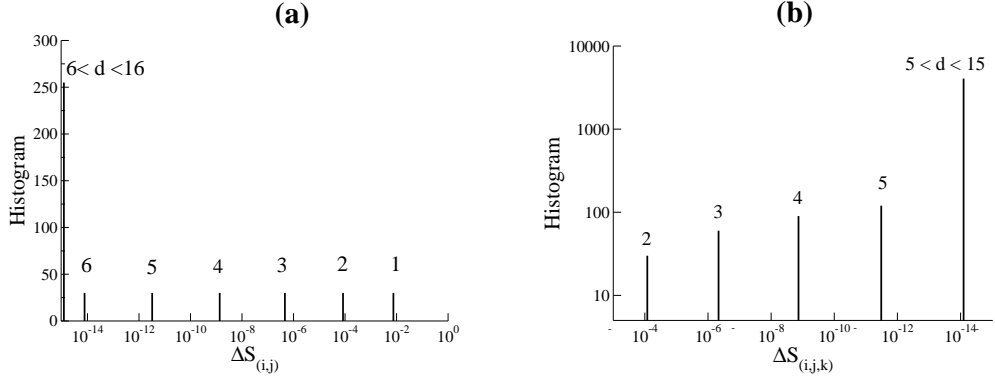

\begin{center}
\epsfig{file=fig36a.eps,width=6cm}
\hskip 1cm
\epsfig{file=fig36b.eps,width=6cm}
\caption{Histograms of the entropies for clusters of size 2 {\bf (a)} and 3 {\bf (b)}; in the latter case, entropies are negative. Data are generated from the unidimensional Ising model (\ref{ei}) with $N=30$ spins, and parameters $J=4$ and $h=-6$. Each peak is labelled by the distance $d$ between the extremities of the clusters. The reference entropy is $S_0=S_{MF}$.}
\label{fig-entroising}
\end{center}
\end{figure}

Figure~\ref{fig-entroising}(b) show the histogram of the entropies of 3-clusters $(i,j,k)$.  Let $d=k-i$ be the distance between the extremities. We observe that the entropies are gathered into peaks, and are exactly the opposite of the ones found in Fig.~\ref{fig-entroising}(a) as expected. Two differences are:
\begin{itemize}
\item  The peak in $d=1$ is not present because the minimal distance between three spins is $d=2$. The largest $3$-spins entropy thus corresponds to triplets of the type $(i,i+1,i+2)$.
\item The height of the peak (number of clusters) corresponding to distance $d$ is $(d-1)\, N$.  The degeneracy $(d-1)$ is the number of ways of choosing the location of site $i_2$ in between $i_1$ and $i_3$.
\end{itemize}

We now show how the value of the cluster entropy can be found back from the leading terms in the diagrammatic expansion calculated in Section \ref{secleading3}. Let us call $d'=j-i<d$ the distance between the first two sites in the cluster. For each diagram in Fig.~\ref{diag3entrofig} we give in Table~\ref{deltas3isingesatto} the sum of the distances of its links, {\em i.e.} the power of $\exp(-1/\xi)$.

\begin{center}
\label{deltas3isingesatto}
\begin{tabular}{|c|c|}
\hline
diagram & sum of distances\\
\hline
a) & $d+2d'+2(d-d')=3d$ \\
b) &  $2d+2d'+(d-d')=3d+d'$ \\
c) &  $2d+d'+2(d-d')=4d-d'$ \\
d) &  $3d'+3(d-d')=3d$ \\
e) &   $3d+3d'$ \\
f)  &  $3  d +3(d-d')=6d-3d'$\\
\hline
\end{tabular}
\end{center}

Interestingly, the lowest total distances are found in diagrams a) and d), while the latter diagram is of a higher power (6) in terms of the correlated function than the former (5). Hence, contrary to the case of independent spins (Section \ref{secisdominant}), diagrams a) and d) give the dominant contributions to the entropy. Summing the contributions of a) and d) we find
\begin{equation}
\label{ds3ising}
\Delta S_{(i,j,k)} = \alpha_{ijk} ^{(5)} \;  (c_{ij})^2\,(c_{jk})^2\,c_{ki}  +\alpha_{ijk}^{(6)}  \; (c_{ij})^3\,(c_{jk})^3 = \Big( \alpha_{ijk} ^{(5)} + \alpha_{ijk}^{(6)}\Big) \;\big(p(1-p)\, \exp(-1/\xi)\big)^{3d}\ .
\end{equation}
To derive the coefficients $\alpha^{(5)}$ and $\alpha^{(6)}$, we impose that $\Delta S_{(i,j,k)}$ is the opposite of (\ref{ds2ising}). We deduce that $\alpha^{(5)}$ and $\alpha^{(6)}$ are given by, respectively, (\ref{alpha3}) and (\ref{alpha3p}).

The exact cancellation property discussed above has important consequences for the inferred fields and couplings. Consider for instance the coupling $J_{i,i+2}$, which vanishes in the 1D-Ising model with nearest-neighbour interactions (\ref{ei}). As the connected correlation $c_{i,i+2}$ is not equal to zero, a contribution to the coupling will be collected from the cluster $(i,i+2)$ itself, equal to
\begin{equation}
\Delta J_{i,i+2;(i,i+2)}=- \frac{\partial \Delta S_{(i,i+2)}}{\partial p_{i,i+2}} \ .
\end{equation}
Other contributions will come from larger clusters. For instance the cluster $(i,i+1,i+2)$ will give an additional
\begin{equation}
\Delta J_{i,i+2;(i,i+1,i+2)}=- \frac{\partial \Delta S_{(i,i+1,i+2)}}{\partial p_{i,i+2}} \ .
\end{equation}
The sum of the two contributions above vanishes due to the cancellation property. It can be checked that the contributions coming from all the other clusters vanish, too, which makes the coupling $J_{i,i+2}=0$ as it should.

\section{Inverse susceptibility matrix for the unidimensional Ising model}
\label{app-chiquatre}

Hereafter, we want to invert the matrix $\boldsymbol\chi$, whose elements are given in (\ref{chi4}). The matrix is of dimension $\frac 12 N(N-1)$, and each element is labelled by two indices $(i,j)$ and $(k,l)$, with $i<j$ and $k<l$. Each index $(i,j)$ can be represented by a site of coordinates $i$ and $j$ on the half-grid of Fig.~\ref{fig-halfgrid}(a). We now show that the non-zero entries of the inverse susceptibility matrix, $\big(\chi^{-1}\big)_{ij,kl}$, are in one-to-one correspondence with the sites $(i,j)$ and $(k,l)$ that are either identical, or nearest neighbours, or diagonally opposed on the elementary mesh of the half-grid (Fig.~\ref{fig-halfgrid}(b,c,d)). Depending on the value of the difference $j-i$, the number of those sites is equal to 9, 8, or 6.

\begin{figure}
\begin{center}
\epsfig{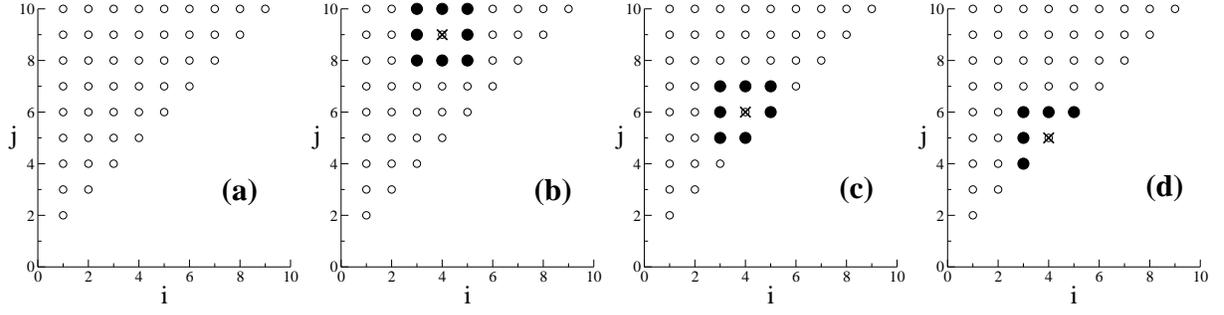}
\caption{Half grid representing the index $(i,j)$ of the entries of the inverse susceptibility matrix, with $i<j$ {\bf (a)}. Black circles locate the nearest-neighbours and the diagonally opposed sites $(k,l)$ of $(i,j)$ (cross), with $i=4$ and $j=9$ {\bf (b)}, $6$ {\bf (c)}, $5$ {\bf (d)}. }
\label{fig-halfgrid}
\end{center}
\end{figure}

We start with the case $j-i\ge  3$ (Fig.~\ref{fig-halfgrid}(b)). By symmetry, the nine unknown matrix elements $\big(\chi^{-1}\big)_{ij,kl}$ take only three independent values, denoted by $\gamma$ for $(k,l)=(i,j)$, $\beta$ for $(k,l)$ and $(i,j)$ nearest neighbours, and $\alpha$ for $(k,l)=(i\pm 1,j\pm 1)$. We now write the matrix inversion identity,
\begin{equation}\label{invio}
\sum _{k<l} \big(\chi^{-1}\big)_{ij,kl}\; \chi_{kl,mn} = \delta_{i,m}\, \delta_{j,n} \ ,
\end{equation}
for various values of $(m,n)$. Let $d=j-i$. For $m=i,n=j$, constraint (\ref{invio}) gives
\begin{equation}
\gamma ( 1- x^{2d}) + 2\beta \big( 2x-x^{2d-1}-x^{2d+1}\big)+\alpha \big( 4x^2-x^{2d-2}-x^{2d}-x^{2d+2}\big)= 1\ , 
\end{equation}
which should hold for all $d\ge 3$. We deduce two coupled equations for the three unknown variables: 
\begin{eqnarray}
\gamma+ 2\,\Big( x+\frac 1x\Big) \, \beta + 4\,\Big( x+\frac 1x\Big)^2 \,  \alpha &=& 0\ , \label{cv5} \\
\gamma+ 4\, x\, \beta + 4\, x^2\,  \alpha &=& 1\ .\label{cv6}
\end{eqnarray}
For $m=i+1,n=j$, constraint (\ref{invio}) is equivalent to
\begin{equation}
\gamma ( x- x^{2d-1}) + \beta \big(1+ 3x^2-x^{2d-2}-3x^{2d}\big)+\alpha \big( 2x+2x^3-x^{2d-3}-2x^{2d-1}-x^{2d+1}\big)=0\ .
\end{equation}
The $d$-dependent term in the equation above cancels by virtue of (\ref{cv5}). We are left with an additional equation over $\alpha,\beta,\gamma$:
\begin{equation}
\gamma \, x + \beta \, (1+ 3x^2)+2\, x\, (1+x^2)\, \alpha =0 \ . \label{cv7}
\end{equation}
By symmetry of the matrices $\boldsymbol\chi, \boldsymbol \chi^{-1}$, no new constraint is obtained when the values of $m,n$ are further varied. Solving (\ref{cv5}), (\ref{cv6}), (\ref{cv7}) we obtain 
\begin{equation}
 \alpha = \frac{x^2}{(1-x^2)^2}\ , \quad \beta =- \frac{x(1+x^2)}{(1-x^2)^2}
\ ,\quad \gamma= \frac{(1+x^2)^2}{(1-x^2)^2} \ .
\end{equation}

The analysis of the other cases $j=i+2$ (Fig.~\ref{fig-halfgrid}(c)) and $j=i+1$ (Fig.~\ref{fig-halfgrid}(d)) can be done along the same lines. We do not write the calculations in details, and simply report the results. The case $j=i+2$ is very similar to the previous case. There are 8 coefficients to be calculated, with three independent values, $\alpha',\beta',\gamma'$. It turns out that 
\begin{equation}
\alpha'=\alpha\ ,\quad \beta'=\beta\ ,\quad \gamma'=\gamma \ .
\end{equation}
As for the last case, $j=i+1$, we call $\alpha''$ the values of the entries of $\boldsymbol \chi^{-1}$ with $(k,l)=(i-1,j-1),(i-1,j+1),(i+1,j+1)$, $\beta''$ the values of the entries with $(k,l)=(i-1,j),(i,j+1)$, and $\gamma''$ the diagonal element corresponding to $(k,l)=(i,j)$. After some elementary algebra, we find
\begin{equation}
\alpha''=\alpha\ ,\quad \beta''=\beta\ ,\quad \gamma''=\frac{1+x^2+x^4}{(1-x^2)^2} \ .
\end{equation}
All those results are reported in (\ref{invchi4}).


\begin{thebibliography}{9999}
\bibitem{spin01}
In the present work where spins are equal to 0,1, the couplings ($J_{ij}$)
 and fields ($h_i$) are the one's in spin 0,1 ($\hat J_{ij},\hat{h_i}$)
  by the transformation: $J_{ij}=4 \hat J_{i,j}$ and
 $h_i=-\frac 1 2 \sum_{j\neq i} J_{ij} +\hat h_i.$
 The  numerical experiments of \cite{bento} where done with $\pm 1$ spins.
\end{thebibliography}
\end{document}